\documentclass[aps,prb,amsmath,amssymb,floatfix,twocolumn,amsmath,superscriptaddress,
twocolumn,nofootinbib,tighten,letterpaper]{revtex4-2}
\usepackage{lmodern}
\usepackage[T1]{fontenc}
\usepackage[latin9]{inputenc}
\usepackage{float}
\usepackage[normalem]{ulem}
\usepackage{amsmath, dsfont}
\usepackage{amssymb}
\usepackage{mathtools}
\usepackage{mathdots}

\usepackage{graphicx}
\usepackage{wasysym}
\usepackage{color, xcolor}
\usepackage{bm}
\usepackage{hyperref}
\hypersetup{pdfpagemode=UseNone}
\usepackage[many]{tcolorbox}
\newsavebox\MBox

\allowdisplaybreaks

\DeclareMathOperator{\Tr}{Tr}

\newcommand{\bsub}{\begin{subequations}}
\newcommand{\esub}{\end{subequations}}

\newcommand{\RR}[1]{{\bf r}_{#1}}

\newcommand{\ra}{\rangle}
\newcommand{\la}{\langle}

\newcommand{\tr}{{\rm \,tr\,}}
\newcommand{\str}{{\rm \,str\,}}

\newcommand{\bd}{b^{\dagger}}
\newcommand{\fd}{f^{\dagger}}
\newcommand{\bph}{b^{\vphantom{\dagger}}}
\newcommand{\fph}{f^{\vphantom{\dagger}}}

\newcommand{\up}{\uparrow}
\newcommand{\down}{\downarrow}

\newcommand{\bup}{\bph_\up}
\newcommand{\fup}{\fph_\up}
\newcommand{\bdup}{\bd_\up}
\newcommand{\fdup}{\fd_\up}
\newcommand{\bdown}{\bph_\down}
\newcommand{\fdown}{\fph_\down}
\newcommand{\bddown}{\bd_\down}
\newcommand{\fddown}{\fd_\down}

 \graphicspath{{Images/}}

\begin{document}

\title{Generalized multifractality at the spin quantum Hall transition: Percolation mapping and pure-scaling observables}

\author{Jonas F.~Karcher}
\affiliation{{Institute for Quantum Materials and Technologies, Karlsruhe Institute of Technology, 76021 Karlsruhe, Germany}}
\affiliation{{Institut f\"ur Theorie der Kondensierten Materie, Karlsruhe Institute of Technology, 76128 Karlsruhe, Germany}}

\author{Ilya A.~Gruzberg}
\affiliation{Ohio State University, Department of Physics, 191 West Woodruff Ave, Columbus OH, 43210, USA}

\author{Alexander D.~Mirlin}
\affiliation{{Institute for Quantum Materials and Technologies, Karlsruhe Institute of Technology, 76021 Karlsruhe, Germany}}
\affiliation{{Institut f\"ur Theorie der Kondensierten Materie, Karlsruhe Institute of Technology, 76128 Karlsruhe, Germany}}

\date{\today}

\begin{abstract}

This work extends the analysis of the generalized multifractality of critical eigenstates at the spin quantum Hall transition in two-dimensional disordered superconductors  [J.~F.~Karcher et al, Annals of Physics, {\bf 435}, 168584 (2021)]. A mapping to classical percolation is developed for a certain set of generalized-multifractality observables. In this way, exact analytical results for the corresponding exponents are obtained. Furthermore, a general construction of positive pure-scaling eigenfunction observables is presented, which permits a very efficient numerical determination of scaling exponents. In particular, all exponents corresponding to polynomial pure-scaling observables up to the order $q=5$ are found numerically.  For the observables for which the percolation mapping is derived, analytical and numerical results are in perfect agreement with each other.  The analytical and numerical  results unambiguously demonstrate that the generalized parabolicity (i.e., proportionality to eigenvalues of the quadratic Casimir operator) does not hold for the spectrum of generalized-multifractality exponents. This excludes Wess-Zumino-Novikov-Witten models, and, more generally, any theories with local conformal invariance, as candidates for the fixed-point theory of the spin quantum Hall transition. The observable construction developed in this work paves a way to investigation of generalized multifractality at Anderson-localization critical points of various symmetry classes.

\end{abstract}

\maketitle

\section{Introduction}
\label{sec:intro}

Despite a long history, Anderson localization remains a field of active theoretical and experimental research~\cite{50_years_of_localization}.  This concerns, in particular, investigation of Anderson transitions between the phases of localized and delocalized states and between localized phases characterized by distinct topological indices~\cite{evers08}. Developments of the symmetry classification of disordered systems as well as theoretical and experimental studies of topological insulators and superconductors have given an additional boost to the Anderson-localization research.

A remarkable property of eigenstates at critical points of Anderson-localization transitions is their multifractality. While the multifractality in the conventional sense refers to the scaling  of moments of eigenfunctions (or of the local density of states, LDOS)~\cite{evers08,rodriguez2011multifractal}, it is instructive to consider also a much broader class of observables characterizing the eigenstates  at criticality~\cite{gruzberg2013classification}. In the sigma-model formalism, which serves as a field theory of Anderson localization, these observables are represented by composite operators without gradients. In the preceding paper~\cite{karcher2021generalized}, we have introduced the term ``generalized multifractality'' for the scaling of the whole set of these observables.

Two-dimensonal (2D) Anderson-localization critical points attract a particular interest. Their realizations include 2D materials (such as, e.g., graphene) and interfaces (e.g., in semiconductor heterostructures), as well as surfaces of three-dimensional (3D) topological insulators and superconductors. A canonical example of such a critical point is provided by the integer quantum Hall (QH) plateau transition (the symmetry class A). This transition has its counterparts for superconducting systems: the spin quantum Hall (SQH) transition in the symmetry class C and the thermal quantum Hall (TQH) transition in the symmetry class D.  While the SQH transition~\cite{kagalovsky1999quantum, senthil1999spin} is a close cousin of the conventional QH transition, it has one distinct property: some of the SQH critical exponents can be determined exactly via a mapping to classical 2D percolation~{  \cite{gruzberg1999exact, beamond2002quantum}}. This includes, in particular, the exponent $\nu = 4/3$ of the localization length as well as the exponent $x_1 = 1/4$ characterizing the scaling of the disorder-averaged local density of states (LDOS) with the system size, $\langle \nu(\mathbf{r}) \rangle \sim L^{-x_1}$~\cite{gruzberg1999exact, beamond2002quantum}. Furthermore, it was shown~\cite{mirlin2003wavefunction, evers2003multifractality, subramaniam2008boundary} that the mapping to percolation yields exact values of the multifractal exponents $x_2 = 1/4$ and $x_3 = 0$ characterizing the scaling of the LDOS moments, $\langle \nu^q(\mathbf{r}) \rangle \sim L^{-x_q}$ with $q=2$ and $q=3$.

In Ref.~\cite{karcher2021generalized}, we have addressed the generalized multifractality at the SQH transition. Using the formalism of the non-linear sigma model (with the target space $G/K$ corresponding to the class C), we have determined composite operators $\cal{P}^C_\lambda$ that exhibit pure scaling at criticality, {  with the scaling dimension $x_\lambda$}. Here $\lambda$ is a multi-index, $\lambda = (q_1,q_2,\ldots,q_n)$, that labels representations (or, equivalently, heighest weights) of the Lie algebra of the group $G$. For operators that are polynomials of order $q$ with respect to the field $Q$ of the sigma model, all $q_i$ are integers satisfying $q_1 \ge q_2 \ge \ldots \ge q_n > 0$ and $q_1 + q_2 + \ldots + q_n = q \equiv |\lambda|$, so that $\lambda$ is an integer partition of $q$ or, equivalently, a Young diagram.  Further, we have found in Ref.~\cite{karcher2021generalized} wave-function observables corresponding to these pure-scaling operators. At this stage, a difficulty occurred: the found observables had indefinite signs before averaging, which required averaging over a very large number of realizations, thus seriously restricting the numerical determination of the critical exponents. We have found two ways to partly resolve this difficulty. Specifically, we constructed sets of observables that are strictly positive. While these observables were not of strictly pure-scaling nature, their dominant scaling yielded some of the sought generalized-mulifractality exponents.

One of central findings of Ref.~\cite{karcher2021generalized} was that the numerically obtained values of these exponents exhibited a strong violation of the generalized parabolicity. The latter property means that the exponents $x_\lambda$ are proportional to the eigenvalues of the quadratic Casimir operator, i.e., are uniquely determined by the symmetry of the sigma-model manifold, up to a single overall factor. Furthermore, it was shown in Ref.~\cite{karcher2021generalized} that a violation of the generalized parabolicity implies that the critical theory does not satisfy the local conformal invariance. A violation of the local conformal invariance at the SQH transition is quite remarkable since one usually assumes that 2D critical theories do satisfy local conformal invariance---even though (i) this can only be proven under certain rather restrictive conditions (that are not satisfied at critical points of Anderson localizations) and (ii) counterexamples are known.

{   Despite a very important progress in previous works with respect to the SQH transition, two major challenges remained. The first one is whether---and if yes, then how---one can extend the percolation mapping to calculate analytically a broader class of generalized-multifractalty exponents (beyond the three ``conventional-multifractality'' exponents $x_{(1)}$, $x_{(2)}$ and $x_{(3)}$). Exact analytical results for exponents characterizing localization transitions are very scarce, emphasizing the importance of this question. Furthermore, analytical knowledge of the exponents would greatly help in understanding physical properties of the fixed-point theory, also with respect to candidate theories. In particular, it can be expected to provide an analytical proof of the absence of generalized parabolicity and thus of local conformal invariance. The second challenge is to
derive a systematic construction of strictly positive wave-function observables corresponding to pure-scaling class-C composite operators $\cal{P}^C_\lambda$ with arbitrary $\lambda$. This would open the door to an efficient, high-precision numerical evaluation of  the generalized-multifractality exponents $x_\lambda$  in a broad domain of $\lambda$.
This work successfully solves both these challenging problems. More specifically, our key results here are as follows:
}

\begin{enumerate}
	
	\item
	In Sec.~\ref{sec:sqh_mapping}, we develop a percolation mapping for all SQH correlation functions up to the order $q=3$. Using these results, we obtain expressions for pure-scaling correlation functions from the representations $\lambda = (1,1)$, (2,1), and (1,1,1) {  in terms of classical percolation probabilities}.
	
	\item
	Using the above mapping and performing analysis of the obtained percolation correlation functions, in Sec.~\ref{sec:sqh_Pana} we determine analytically the corresponding SQH generalized-multifractality exponents $x_{(1)}, x_{(2)}, x_{(1,1)}, x_{(3)}, x_{(2,1)}$, and $x_{(1,1,1)}$ by relating them to the scaling dimensions $x_n^{\rm h}$ of $n$-hull operators from classical percolation. We verify the analysis  of percolation correlation functions by numerical simulations in {  Sec.~\ref{sec:sqh_Pnumerics}}.
	
	\item
	Furthermore, in Sec.~\ref {sec:susy} we use the supersymmetry approach that allows us to develop the percolation mapping also for the correlators of the type $\lambda = (1,1,1, ...,1) \equiv (1^q)$ ($q$ units) with arbitrary $q$ and to determine the corresponding scaling dimension $x_{(1^q)}$. By virtue of the Weyl symmetry relations satisfied by the generalized-multifractality exponents~\cite{gruzberg2011symmetries, gruzberg2013classification}, we also get the exponents $x_{(2,1^{q-1})} = x_{(1^q)}$  [and, in fact, further exponents generated by the Weyl orbit of $(1^q)$].

	\item
	We derive a general construction of positive wave-function observables that satisfy (upon disorder averaging) the pure scaling for a generic representation $\lambda = (q_1, q_2, \ldots, q_n)$, with arbitrary $q_i$ (that may be fractional and even complex). This is greatly beneficial for numerical determination of the exponents $x_\lambda$, as we also demonstrate in Sec.~\ref{sec:scaling}. In particular, we determine numerically (by using a network model of class C~\cite{kagalovsky1999quantum, senthil1999spin}) all exponents corresponding to polynomial pure-scaling observables up to the order $q=5$ in Sec.~\ref{sec:sqh_numerics}.	{  Our preliminary results indicate that the construction derived here applies also to critical points in symmetry classes AII, DIII, CI and CII, which further emphasizes its importance.}
	
	\item
	We find an excellent agreement between the analytical results (based on the percolation mapping) and the numerical results (based on the eigenstate observable construction) obtained in this work. These results---both analytical and numerical---unambiguously demonstrate that the generalized parabolicity does not hold at the SQH critical point, thus
	confirming the conclusion of Ref.~\cite{karcher2021generalized}. This strictly excludes Wess-Zumino-No\-vi\-kov-Wit\-ten models, and, more generally, any theories with local conformal invariance, as candidates for the fixed-point theory of the SQH transition.
	
\end{enumerate}

\section{Percolation mapping}
\label{sec:sqh_percolation}

The version of the Chalker-Coddington network model appropriate for the symmetry class C of the SQH transition has two channels on each link, and the group SU(2) acts on the corresponding wave-function spinors~\cite{kagalovsky1999quantum, senthil1999spin}. Certain observables in the SU(2) network model at criticality can be mapped to probabilities of the classical percolation problem. The SU(2) average turns the quantum-mechanical coherent sum over all amplitudes into a sum over configurations of classical percolation hulls weighted by the corresponding probabilities~\cite{gruzberg1999exact, beamond2002quantum}. This mapping exists for all products of $q=1$, 2, and 3 Green's functions, and  the scaling dimensions of the moments of the local density of states (LDOS) $x_{(1)}, x_{(2)}$ and $x_{(3)}$ can be computed this way~\cite{mirlin2003wavefunction}. Products of Green's functions with finite level broadening $\gamma$ also involve products of distinct wave functions that do not behave as pure LDOS powers but show subleading multifractality. Since the mapping to percolation is exact,
we should therefore have access to the exponents $x_{(1,1)},x_{(2,1)}$ and 
$x_{(1^3)}$
as well. We carry out this program in the present section.

We begin by introducing notations for SQH correlation functions in Sec.~\ref{sec:sqh_pnot}. Then in Sec.~\ref{sec:sqh_mapping} the mapping to percolation as developed in Ref.~\cite{mirlin2003wavefunction} is presented for all SQH correlation functiions with $q=1$, 2, and 3. Subsequently, in Sec.~\ref{sec:sqh_Pana},  analytical properties of classical percolation probabilities are discussed and implications for the SQH correlation functions are analyzed. Nume\-ri\-cal simulations of classical percolation are performed in Sec.~\ref{sec:sqh_Pnumerics} in order to verify the analytical predictions based on the percolation theory. The key result of Sec.~\ref{sec:sqh_percolation} are relations between the multifractal scaling dimensions $x_\lambda$ at the SQH critical point and scaling dimensions of hull operators $x_n^{\rm h}$ in classical percolation.

\subsection{SQH correlation functions}
\label{sec:sqh_pnot}

\subsubsection{Notations}

A network model is characterized by an evolution operator $\mathcal{U}$ (which describes the evolution of the system over a unit time step).
The evolution operator is unitary, so that its eigenvalues lie on the unit circle in the complex plane:
\begin{align}
	\mathcal{U}\psi_i &= e^{i\epsilon_i}\psi_i.
\end{align}
The SQH network model in class C is characterized by the particle-hole symmetry which implies that
every quasienergy $\epsilon_i$ has a partner $-\epsilon_i$. The wave function $\psi_{i\alpha}({\bf r})$
(eigenfunction of $\mathcal{U}$) lives on links ${\bf r}$ of the network and carries a spin index $\alpha = \uparrow, \downarrow$.

The Green's function associated with the operator $\mathcal{U}$ (a $2 \times 2$ matrix in the spin space) is defined as
\begin{align}
	G(\RR2,\RR1;z)&= \langle \RR2|(1-z\mathcal{U})^{-1}|\RR1\rangle,
	\label{eq:fgreen}
\end{align}
with a complex argument $z$.
Since the evolution operator $\mathcal{U}$ is unitary, the poles of this Green's function lie on the unit circle $|z|=1$. Inside the unit disc, $|z|<1$, the expansion of this expression in powers of $z$ is absolutely convergent. Outside of the unit disc, i.e. for $|z| >  1$, one can use the identity
\begin{align}
	\left[G(\RR1,\RR2;z^{-1})\right]^\dagger &= \mathds{1}-G(\RR2,\RR1;z^*),
	\label{eq:inv}
\end{align}
which generates a convergent expression in powers of $(z^*)^{-1}$.
The class-C particle-hole symmetry implies the following property of the Green's function:
\begin{align}
	\left[G(\RR1,\RR2;z)\right]^\dagger &= -(i\sigma_2) \left[G(\RR1,\RR2;z)\right]^T (i\sigma_2),
	\label{eq:Csym}
\end{align}
where  $\sigma_2$ is the second Pauli matrix in the spin space.

We are interested in SQH correlation functions that are (linear combinations of) disorder-averaged products of $2q$ wave function amplitudes. As pointed out above,  in
Sec.~\ref{sec:sqh_percolation} we focus on the cases $q=1$, 2, and 3. To explain notations, we are now going to consider the case $q=2$ in detail.

In the order $q=2$, there are two independent correlation functions that we denote $\mathcal{D}_{(2)}$ and $\mathcal{D}_{(1,1)}$, respectively:
\onecolumngrid
\begin{align}
	\mathcal{D}_{(2)}(\RR1,\RR2;\epsilon_1,\epsilon_2)
	&= \Big\langle \sum_{ij \alpha\beta} \psi^*_{i\alpha}(\RR1) \psi_{j\alpha}(\RR1) \psi_{i\beta}(\RR2) \psi^*_{j\beta}(\RR2) \delta(\epsilon_1-\epsilon_i)\delta(\epsilon_2-\epsilon_j) \Big\rangle,\\
	\mathcal{D}_{(1,1)}(\RR1,\RR2;\epsilon_1,\epsilon_2)
	&= \Big\langle\sum_{ij\alpha\beta} |\psi_{i\alpha}(\RR1)|^2 |\psi_{j\beta}(\RR2)|^2 \delta(\epsilon_1-\epsilon_i) \delta(\epsilon_2-\epsilon_j) \Big\rangle.
	\label{eq:D-HF}
\end{align}
\twocolumngrid
\noindent
These functions describe correlations between amplitudes of two eigenstates $\psi_i$ and $\psi_j$ with energies $\epsilon_1$ and $\epsilon_2$ at two links $\RR1,\RR2$ of the network. Spin indices are denoted by Greek letters $\alpha, \beta$.  The angular brackets in Eq.~\eqref{eq:D-HF} denote averaging over disorder (i.e., over random
SU(2) matrices on the links of
the network). We will refer to $\mathcal{D}_{(1,1)}$ and $\mathcal{D}_{(2)}$ as the Hartree and the Fock correlation functions, respectively. The origin of this terminology is rather clear: exactly these structures arise when one calculates the Hartree and the Fock contributions to the energy of an interacting system.

It is straightforward to express the correlation functions $\mathcal{D}_{(2)}$ and $\mathcal{D}_{(1,1)}$ in terms of the Green's function~\eqref{eq:fgreen}. For further analysis, it turns out to be useful to perform an analytical continuation of the arguments $e^{i\epsilon_1}=z=e^{-\gamma}$ and $e^{i\epsilon_2}=w=e^{-\delta}$ to the real axis. All correlators $\mathcal{D}$ of wave functions contain the difference of retarded and advanced Green's functions
\begin{align}
	\Delta G(\mathbf{r}_1, \mathbf{r}_2;z)
	&\equiv G(\mathbf{r}_1, \mathbf{r}_2; z) - G(\mathbf{r}_1, \mathbf{r}_2; z^{-1}),
	\label{DeltaG}
\end{align}
and we get
\begin{align}
	&(2\pi)^{2} \mathcal{D}_{(2)}(\RR1,\RR2;\gamma,\delta)
	\nonumber \\
	&= \big\langle  \Tr \{ \Delta G(\mathbf{r}_1, \mathbf{r}_2;z) \Delta G(\mathbf{r}_2, \mathbf{r}_1;w) \}
	\big\rangle,
	\label{eq:F}
	\\
	&(2\pi)^{2} \mathcal{D}_{(1,1)}(\RR1,\RR2;\gamma,\delta)
	\nonumber \\
	&= \big\langle  \Tr \Delta G(\mathbf{r}_1, \mathbf{r}_1;z)
	\Tr \Delta G(\mathbf{r}_2, \mathbf{r}_2;w) \big\rangle.
	\label{eq:H}
\end{align}
Here traces go over the spin space. The subscripts in notations $\mathcal{D}_{(2)}$ and $\mathcal{D}_{(1,1)}$ indicate that the correlation function~\eqref{eq:F} involves a loop ($\RR1 \leftarrow \RR2 \leftarrow \RR1$) formed by two Green's functions, while Eq.~\eqref{eq:H} involves a product of two loops ($\RR1 \leftarrow \RR1$ and $\RR2 \leftarrow \RR2$), each of them formed by a single Green's function. These notations are straightforwardly extended to higher-order correlation functions $\mathcal{D}_\lambda$ with integer partitions  $\lambda= (q_1, \ldots, q_n)$.

In order to extract the scaling behavior of the correlation functions~\eqref{eq:F}  and~\eqref{eq:H}, it is convenient to choose $\gamma=\delta$ (i.e., $z=w$).  This choice is beneficial for performing the percolation mapping following Ref.~\cite{mirlin2003wavefunction}. We are thus left with the correlation functions
\onecolumngrid
\begin{align}
	(2\pi)^2\mathcal{D}_{(2)}(\RR1,\RR2;\gamma)
	&= \Big\langle \sum_{ij\alpha\beta} \psi^*_{i\alpha}(\RR1) \psi_{j\alpha}(\RR1) \psi_{i\beta}(\RR2) \psi^*_{j\beta}(\RR2) \delta_\gamma(\epsilon_i) \delta_\gamma(\epsilon_j) \Big\rangle,
	\nonumber \\
	(2\pi)^2\mathcal{D}_{(1,1)}(\RR1,\RR2;\gamma)
	&= \Big\langle \sum_{ij\alpha\beta} |\psi_{i\alpha}(\RR1)|^2 |\psi_{j\beta}(\RR2)|^2 \delta_\gamma(\epsilon_i) \delta_\gamma(\epsilon_j) \Big\rangle,
	\label{eq:dg2}
\end{align}
\twocolumngrid
\noindent
where
\begin{align}
	\delta_\gamma(\epsilon_i)& = \mathrm{Im}\frac{1}{\epsilon_i - i\gamma}
\end{align}
is the $\delta$ function (times $\pi$) broadened by $\gamma$. We define $\mathcal{D}_\lambda(\RR1,\RR2, \ldots;\gamma)$ for an arbitrary integer partition $\lambda$ in an analogous fashion.
In everything that follows there is only one broadening parameter $\gamma$, and we will suppress the corresponding variable $z$ in the difference $\Delta G$, Eq.~\eqref{DeltaG}.

\subsubsection{Pure-scaling combinations}

A system at the critical point of the SQH transition (or, more generally, at any Anderson-transition critical point) is characterized by generalized multifractality.
This means that there are composite operators labeled by the multi-index $\lambda = (q_1, \ldots, q_n)$ that exhibit power-law scaling with scaling dimensions $x_\lambda$.
The dimensions $x_\lambda$ are related to the anomalous dimensions $\Delta_\lambda$ describing eigenstate correlations via
\begin{align}
	\Delta_\lambda &= x_\lambda-|\lambda| x_{(1)},
	&
	|\lambda| = q \equiv q_1 + \ldots + q_n,
	\label{eq:Delta_lambda}
\end{align}
where $x_{(1)}$ is the LDOS scaling dimension.
For each $\lambda$, one can construct appropriate wave-function observables $O_\lambda[\psi]$ that reveal the corresponding power-law scaling:
\begin{align}
	L^{2q}\langle O_\lambda[\psi](\RR1,\ldots,\RR{q}) \rangle  &\sim (r/L)^{\Delta_\lambda}.
\end{align}
Here the points $\RR1,\ldots,\RR{q}$ are separated by a distance $\sim r$ from each other, with $r \ll L$. In Sec.~\ref{sec:scaling}, we explicitly construct such pure scaling operators $O_\lambda[\psi](\RR1,\ldots,\RR{q})$ for all $\lambda$ for class C.
Here we discuss how these scaling combinations for $q=2$ appear in the correlation functions $\mathcal {D}_{(1,1)}$ and $\mathcal {D}_{(2)}$ introduced above.

The double sums over $i$ and $j$ in Eq.~\eqref{eq:dg2} contain two contributions: the diagonal one ($i = j$) and the one with $i \ne j$:
\begin{align}
	&(2\pi)^2\mathcal{D}_{(2)}(\RR1,\RR2;\gamma)
	= \Big\langle \sum_{i,\alpha\beta} |\psi_{i\alpha}(\RR1)|^2 |\psi_{i\beta}(\RR2)|^2 \left(\delta_\gamma(\epsilon_i)\right)^2
	\nonumber\\
	& + \sum_{i\neq j; \,\alpha\beta} \psi^*_{i\alpha}(\RR1) \psi_{j\alpha}(\RR1) \psi_{i\beta}(\RR2) \psi^*_{j\beta}(\RR2) \delta_\gamma(\epsilon_i) \delta_\gamma(\epsilon_j)\Big\rangle,
	\nonumber\\
	&(2\pi)^2\mathcal{D}_{(1,1)}(\RR2,\RR1;\gamma)
	= \Big\langle \sum_{i,\alpha\beta} |\psi_{i\alpha}(\RR1)|^2 |\psi_{i\beta}(\RR2)|^2 \left(\delta_\gamma(\epsilon_i)\right)^2
	\nonumber\\
	& + \sum_{i\neq j;\, \alpha\beta} |\psi_{i\alpha}(\RR1)|^2 |\psi_{j\beta}(\RR2)|^2 \delta_\gamma(\epsilon_i) \delta_\gamma(\epsilon_j)\Big\rangle.
\end{align}
The terms with $i=j$, i.e., with a summation over one eigenstate $i$, yield pure $\Delta_{(2)}$ scaling:
\begin{align}
	&\Big\langle \sum_{i,\alpha\beta} |\psi_{i\alpha}(\RR1)|^2 |\psi_{i\beta}(\RR2)|^2 \left(\delta_\gamma(\epsilon_i)\right)^2 \Big\rangle
	\nonumber\\
	& = L^2\int d \epsilon \, \nu(\epsilon) \left(\delta_\gamma(\epsilon)\right)^2
	\Big\langle \sum_{\alpha\beta}|\psi_{i\alpha}(\RR1)|^2|\psi_{i\beta}(\RR2)|^2 \Big\rangle \Big|_{\epsilon_i = \epsilon}
	\nonumber\\
	& \qquad \sim L^{-2}\int d \epsilon\,  \nu(\epsilon) \left(\delta_\gamma(\epsilon)\right)^2  (r/\xi_\gamma)^{\Delta_{(2)}}
	\nonumber\\
	& \qquad \sim L^{-2} \nu(\gamma)\, \gamma^{-1}(r/\xi_\gamma)^{\Delta_{(2)}},
	\label{eq:scaling-q2-diagonal}
\end{align}
where
\begin{align}
	\nu(\epsilon) &= \rho_0\epsilon^{-1/7},
	&
	\xi_\gamma &\sim \gamma^{-4/7}
\end{align}
are the average density of states and the localization length, respectively,
and we have used
\begin{align}
	\langle |\psi_{i\alpha}(\RR1)|^2|\psi_{i\beta}(\RR2)|^2\rangle|_{\epsilon_i
		= \epsilon}  \sim L^{-4}(r/\xi_\epsilon)^{\Delta_{(2)}}.
\end{align}
The formula~\eqref{eq:scaling-q2-diagonal} and analogous scaling formulas below hold for $r \ll \xi_\gamma$.
In the thermodynamic limit $L\rightarrow \infty$ and for fixed $\gamma$, these diagonal contribution are suppressed in comparison with those coming from two different eigenstates ({  the double} sum over $i \ne j$).

\begin{table*}[t]
	\centering
	\caption{Probabilities involved in the mapping to classical percolation. Here $\RR1,\RR2,\RR3$ are links on a percolation network and $N,N', N''$ are lengths (of segments) of paths through the links. }
	\label{tab:perc}
	\begin{tabular}{c|c}
		\hline\hline
		probability & description  \\
		\hline
		$p(\RR1;N)$ & loop of length $N$ running through link $\RR1$\\
		\hline
		$p(\RR1,\RR2;N)$ & loop of length $N$ running through links $\RR1\leftarrow \RR2$ \\
		$p_1(\RR1,\RR2;N)$ & loop running through links $\RR1\leftarrow \RR2$, where the path $\RR1\leftarrow \RR2$ has length $N_{\RR1\RR2}=N$\\
		$p(\RR1,\RR2;N_{\RR1\RR2} = N, N_{\RR2\RR1}= N')$ & loop running through links $\RR1\leftarrow \RR2$, where the path $\RR1 \leftarrow \RR2$ has length $N_{\RR1\RR2} = N$\\& and the path $\RR2\leftarrow \RR1$ has length $N_{\RR2\RR1}= N'$\\
		$p(\RR1;N|\RR2;N')$ & loops of lengths $N,N'$ running through links $\RR1,\RR2$ respectively\\
		\hline
		$p(\RR1,\RR2,\RR3;N)$ & loop of length $N$ running through links $\RR1\leftarrow \RR2\leftarrow \RR3$\\
		$p_1(\RR1,\RR2,\RR3;N)$ & loop running through links $\RR1\leftarrow \RR2\leftarrow \RR3$, where $N_{\RR1\RR3}=N$\\
		$p(\RR1,\RR2;N|\RR3;N')$ & loop of length $N$ running through links $\RR1\leftarrow \RR2$\\
		& and loop of length $N'$ running through link $\RR3$ \\
		$p_1(\RR1,\RR2;N|\RR3;N')$ & loop of length $N$ running through links $\RR1\leftarrow \RR2$,\\
		& where $N_{\RR1\RR2}=N$ and loop of length $N'$ running through link $\RR3$ \\
		$p(\RR1;N|\RR2;N'|\RR3;N')$ & loops of lengths $N,N',N''$ running through links $\RR1,\RR2,\RR3$ respectively\\
		\hline\hline
	\end{tabular}
\end{table*}

For two-eigenstate correlation functions ($i \ne j$), we know from Ref.~\cite{karcher2021generalized} that pure-scaling observables
$\mathcal{P}^C_{(1,1)}$ and $\mathcal{P}^C_{(2)}$
are given by the following linear combinations of $\mathcal {D}_{(1,1)}$ and $\mathcal {D}_{(2)}$ (see Eq.~\eqref{eq:rg_C_result} in Appendix~\ref{appendix:rg} where we briefly summarize the relevant results of Ref.~\cite{karcher2021generalized}):
\begin{align}
	\begin{pmatrix}
		\mathcal{P}^C_{(1,1)}\\
		\mathcal{P}^C_{(2)}
	\end{pmatrix}
	& = \begin{pmatrix*}[r]
		1& -2\\
		1& 1
	\end{pmatrix*}
	\begin{pmatrix}
		\mathcal{D}_{(1,1)}(\RR1,\RR2;z) \\
		\mathcal{D}_{(2)}(\RR1,\RR2;z)
	\end{pmatrix}.
	\label{rg:perc}
\end{align}
This result (and its extension to higher $q$) was inferred in Ref.~\cite{karcher2021generalized} from the renormalization-group (RG) analysis of the class-C non-linear sigma model (NL$\sigma$M).  This implies the following scaling properties of $\mathcal {D}_{(1,1)}$ and $\mathcal {D}_{(2)}$ (for $r \ll \xi_\lambda$):
\begin{align}
	&\mathcal{D}_{(2)}(\RR2,\RR1;\gamma)
	\nonumber\\
	& \simeq c \nu^2(\gamma)L^4
	\Big\langle  \sum_{\alpha\beta} \psi^*_{i\alpha}(\RR1) \psi_{j\alpha}(\RR1) \psi_{i\beta}(\RR2) \psi^*_{j\beta}(\RR2) \Big\rangle
	\Big|_{\epsilon_i,\epsilon_j
		\sim \gamma}
	\nonumber\\
	& \qquad \simeq c' \nu^2(\gamma) \left[ (r/\xi_\gamma)^{\Delta_{(2)}}
	- (r/\xi_\gamma)^{\Delta_{(1,1)}}\right],
	\nonumber\\
	&\mathcal{D}_{(1,1)}(\RR2,\RR1;\gamma)
	\nonumber\\
	& \qquad \simeq c \nu^2(\gamma)L^4 \Big\langle  \sum_{\alpha\beta} |\psi_{i\alpha}(\RR1)|^2 |\psi_{j\beta}(\RR2)|^2 \Big\rangle \Big|_{\epsilon_i,\epsilon_j\sim\gamma}
	\nonumber\\
	& \qquad \simeq c' \nu^2(\gamma)\left[2(r/\xi_\gamma)^{\Delta_{(2)}}
	+ (r/\xi_\gamma)^{\Delta_{(1,1)}}\right],
	\label{eq:MF-scaling-q2}
\end{align}
where $c$ and $c'$ are constants.
At small $r/\xi_\gamma$, one thus has $2\mathcal{D}_{(2)}(\RR2,\RR1;\gamma)\approx \mathcal{D}_{(1,1)}(\RR2,\RR1;\gamma)$, so that the difference $\mathcal{D}_{(1,1)}-2\mathcal{D}_{(2)}$ reveals the subleading $\Delta_{(1,1)}$ scaling.

In the following, it is convenient to switch to the variable $z=e^{-\gamma}<1$  to
characterize the broadening $\gamma > 0$.

\subsection{Mapping to percolation}
\label{sec:sqh_mapping}

As shown in Ref.~\cite{mirlin2003wavefunction},  the SU(2) averaging reduces quantum-mechanical sums over amplitudes that enter
SQH correlation functions involving products of two or three Green's functions to classical sums over non-intersecting paths. (For the average of a single Green's function, this approach was implemented in Ref.~\cite{beamond2002quantum}.) As a result, disorder-averaged quantum-mechanical correlation functions with $q \le 3$ are expressed in terms of probabilities of classical percolation hulls (external perimeters of the percolation clusters). In brief, this mapping obeys the following rules:
\begin{itemize}
	\item Each quantum-mechanical Green's function $G(\RR{i},\RR{j}; z)$  is given by a sum over paths {  (segments of percolation hulls)} from $\RR{j}$ to $\RR{i}$.
	Upon averaging of a product of Green's functions, only those contributions remain, where each link is traversed exactly 0 or 2 times by the paths.
	\item Paths of length $2N$ are weighted with a factor $z^{2N}$.	
	\item Each spin trace gives a factor $-1$. The negative sign originates from the SU(2) average, since $\langle U^k\rangle_{\mathrm{SU}(2)} =c_k \mathds{1}$, with $c_2=-\frac12<0$ and with all other $c_k=0$ for $k>0$.	
\end{itemize}

The sum over all resulting classical paths traversing these links can then be interpreted as a sum over hulls of the classical percolation problem. The full quantum mechanical problem is then re-expressed in terms of the classical percolation probabilities. The relevant probabilities are defined in Table~\ref{tab:perc}. These are probabilities to find loops (percolation hulls) of given lengths $N,N',\ldots$ running through given links $\RR1\leftarrow \RR2 \leftarrow \ldots$ in a given order. For some of the probabilities, the length of a segment  between two links  (e.g., $N_{\RR1\RR2}$) is specified. For example, $p(\RR1;N)$ is a probability that the link $\RR1$ belongs to a loop of the length $N$; $p(\RR1,\RR2;N)$ is a probability that the links $\RR1$ and $\RR2$ belong to the same loop of the length $N$;  $p(\RR1;N|\RR2;N')$ is a probability that the link $\RR1$ belongs to a loop of length $N$, while the link $\RR2$ belongs to a different loop of length $N'$, and so on, see description in Table~\ref{tab:perc}.

The probabilities introduced in Table~\ref{tab:perc} satisfy a number of identities that follow directly from their definitions. In particular,
\begin{align}
	\sum_N p(\RR1,\RR2;N) = \sum_N p_1(\RR1,\RR2;N)&= p(\RR1,\RR2),
	\label{eq:sum_rule1}
	\\
	p(\RR1,\RR2;N) + \sum_{N'}p(\RR1;N|\RR2;N') &= p(\RR1;N),
	\label{eq:sum_rule2}
	\\
	\sum_N p(\RR1;N)&= 1.
	\label{eq:sum_rule3}
\end{align}
Analogous identities hold for probabilities with three (or more) spatial arguments.

In the following, the mapping is applied to all correlation functions $\mathcal{D}_\lambda$ with $q=1,2,3$ Green's functions (or, equivalently, $2q=2,4,6$ wave function amplitudes). We recall that the correlation function are labeled by integer partitions $\lambda$, with $|\lambda|=q$.
{  For $\lambda = (1)$ the resulting form of $\mathcal{D}_\lambda$ was obtained within this approach in  Ref.~\cite{beamond2002quantum}, and for $\lambda =   (2), (1,1), (3)$ in Ref.~\cite{mirlin2003wavefunction}.  For reader's convenience, we present the derivation here for all $|\lambda|=3$ (i.e. including also the previously known results) within the notations introduced in Tab.~\ref{tab:perc}. Most of the technical details of calculations are given in Appendix~\ref{appendix-mapping_}. We omit the general proof of validity of the mapping for $|\lambda|  \le 3$ (see the rules formulated above), referring the reader to  Ref.~\cite{mirlin2003wavefunction}.}

\subsubsection{One Green's function: $q=1$}
\label{sec:mapping-q1}

\begin{figure}
	\noindent\begin{center}
		\includegraphics[width=0.48\textwidth]{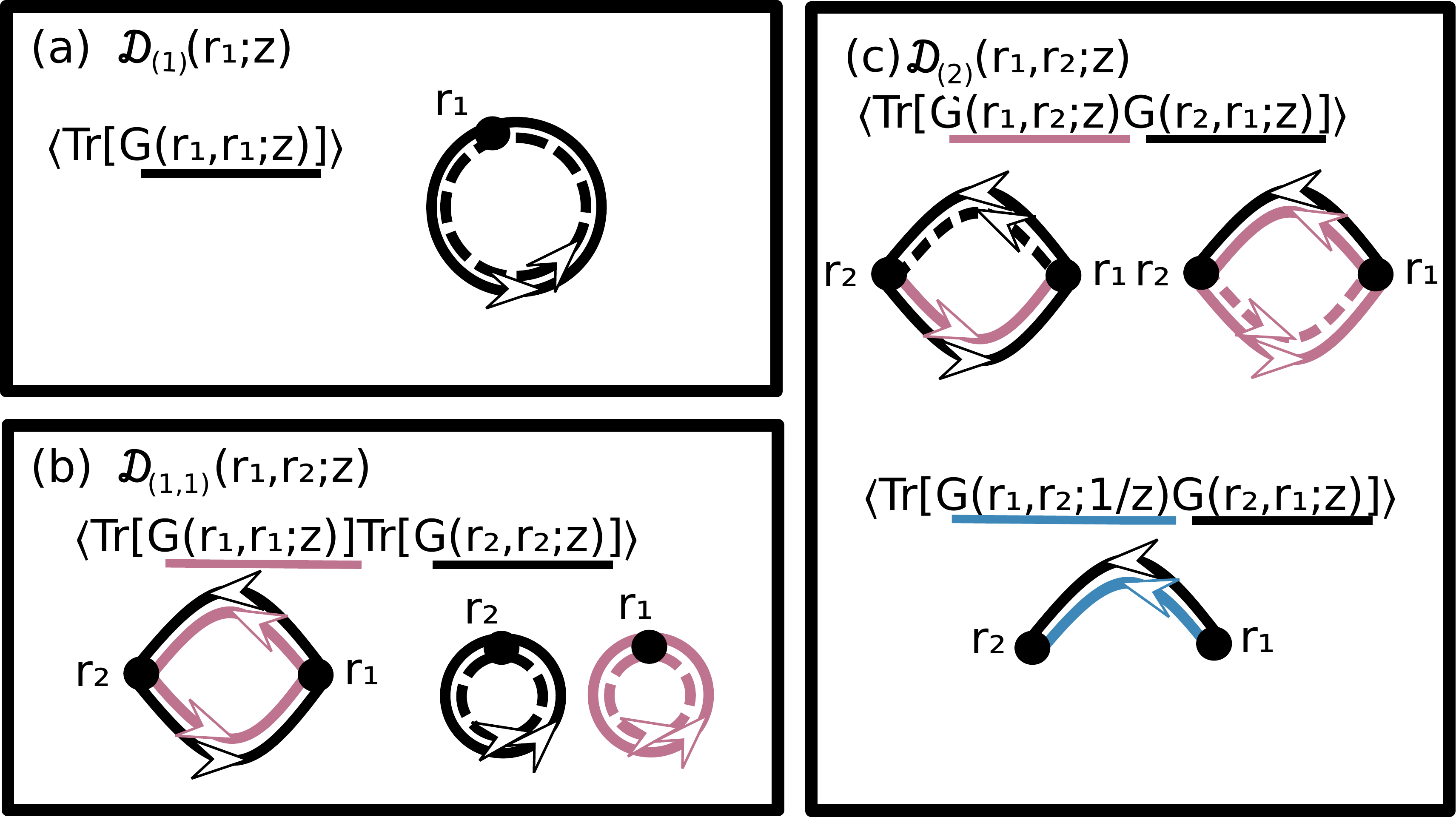}
	\end{center}
	\caption{Schematic representation of path configurations yielding individual contributions within the percolation mapping for the $q=1$ and $q=2$ correlation functions presented in Sec.~\ref{sec:mapping-q1} and~\ref{sec:mapping-q2}. Paths corresponding to each Green's function are shown by the color corresponding to the line underlining this Green's function. When a path traverses a segment of the loop for the second time, it is shown by a dashed line.
	}
	\label{fig:paths-12}
\end{figure}

In this order, there is only one SU(2)-invariant expression, $\Tr G(\RR1,\RR1;z)$, corresponding to the average LDOS $\langle \nu(\RR1; \gamma)\rangle$ (which is the same as the average global density of states $\rho(\gamma)$, since disorder averaging restores translational invariance). In our notations, this corresponds to $\mathcal{D}_\lambda$ with $\lambda = (1)$.  We have
\begin{align}
	(2\pi)\mathcal{D}_{(1)}(\RR1;z)
	&= \rho(\gamma) \Big\langle \sum_\alpha|\psi_\alpha(\RR1)|^2 \Big\rangle
	\nonumber\\
	& = \big\langle \Tr \Delta G(\RR1,\RR1)
	\big\rangle.
	\label{eq:D1}
\end{align}
The expectation values of the Green's functions can be turned into sums over percolation probabilities:
\begin{align}
	\big\langle \Tr G(\RR1,\RR1;z)\big\rangle
	&= 2-\sum_{N=1}^\infty p(\RR1;N)z^{2N},
	\nonumber\\
	\big\langle \Tr G(\RR1,\RR1;z^{-1})\big\rangle
	&= \sum_{N=1}^\infty p(\RR1;N)z^{2N}.
	\label{eq:q1-G}
\end{align}
Non-vanishing contributions to $\langle \Tr G(\RR1,\RR1;z)\rangle $ are given by loops containing the link $\RR1$ and traversed exactly twice, see
Fig.~\ref{fig:paths-12}a. The contribution 2 in the first line originates from the spin trace  of the identity matrix (the zeroth order in the expansion over powers of $\mathcal{U}$).  The second line is obtained using Eq.~\eqref{eq:inv}.

Substituting Eq.~\eqref{eq:q1-G} into Eq.~\eqref{eq:D1} and using the normalization of the probability, one gets the percolation expression for the average LDOS
\begin{align}
	(2\pi)\mathcal{D}_{(1)}(\RR1;z) &= 2\sum_{N=1}^\infty (1-z^{2N}) p(\RR1;N).
	\label{DOS}
\end{align}
This expression does not depend on the link $\RR1$ and reveals, upon evaluation of the percolation sum, the well-known scaling
$\langle \nu(\mathbf{r}) \rangle \sim L^{-x_1}$ with $x_1=\frac14$.

\subsubsection{Two Green's functions: $q=2$}
\label{sec:mapping-q2}

In the case of averaged products of two Green's functions (or, equivalently, four wave functions), the two SU(2)-invariant combinations are labeled by $\lambda=(2)$ and $\lambda=(1,1)$. These correspond to the Fock and the Hartree terms, respectively, and have been introduced in Eqs.~\eqref{eq:F} and~\eqref{eq:H}.

\paragraph{Correlator $\mathcal{D}_{(2)}$.}
{  The calculations are carried out in detail in Appendix~\ref{appendix-mapping2-2}. Combining together the individual contributions~\eqref{eq:D2-term1},~\eqref{eq:D2-term2}, and~\eqref{eq:tr_Gz_Gz-1} derived there according to the mapping rules, we obtain}
\begin{align}
	&(2\pi)^{2}\mathcal{D}_{(2)}(\RR1,\RR2;z)
	= 4\sum_{N=1}^\infty (1-z^{2N})
	\nonumber\\
	& \quad \times \big[p(\RR1,\RR2;N) - p^{(s)}_1(\RR1,\RR2;N)\big].
	\label{eq:D2}
\end{align}
Here we used the sum rule~\eqref{eq:sum_rule1}. The superscript $(s)$ denotes symmetrization with respect to the spatial arguments $\RR1, \RR2$; in particular,
\begin{align}
	2 p^{(s)}_1(\RR1,\RR2;N) = p_1(\RR1,\RR2;N)+p_1(\RR2,\RR1;N).
\end{align}
Let us emphasize that, here and below, $(s)$ means symmetrization with respect to all $q$ arguments of the considered correlation function $\mathcal{D}_\lambda$ (in the present case, $q=2$).

\paragraph{Correlator $\mathcal{D}_{(1,1)}$.}

We turn now to the Hartree correlator, Eq.~\eqref{eq:H} (with $z=w$).
{ 
Individual contributions are derived in Appendix~\ref{appendix-mapping2-11}, see Eq.~\eqref{eq:D11-perc-terms}. Combining them and
using the sum rule~\eqref{eq:sum_rule2}, we can rewrite the Hartree correlation function in the form}
\begin{align}
	&(2\pi)^{2}\mathcal{D}_{(1,1)}(\RR1,\RR2;z)
	= 4 \sum_{N=1}^\infty (1-z^{2N})
	\nonumber\\
	& \quad \times \big[p(\RR1,\RR2;N)+p_-(\RR1,\RR2;N) \big]
	\label{eq:D11},
\end{align}
where the auxiliary function
\begin{align}
	& p_-(\RR1,\RR2;N)=\sum_M \big[p(\RR1;M|\RR2;N)
	\nonumber\\
	& + p(\RR1;N|\RR2;M)-p(\RR1;N-M|\RR2;M)\big]
	\label{eq:pm}
\end{align}
is a linear combination of percolation probabilities, and it can assume negative values. {  Note that the correlator $\mathcal{D}_{(1,1)}$ is by definition symmetric with respect to the interchange $\RR1 \leftrightarrow \RR2$. The result~\eqref{eq:D11} explicitly obeys this symmetry.}

\subsubsection{Three Green's functions: $q=3$}
\label{sec:perc-mapping-q3}

There are three SU(2) invariant correlation function in this order
that can be labeled with multi-indices $\lambda =(3)$, $(2,1)$ and 
$(1^3)$:
\begin{align}
	& (2\pi)^3 \mathcal{D}_{(3)}(\RR1,\RR2,\RR3;z)
	\nonumber \\
	&= \big\langle \Tr \big\{
	\Delta G(\RR1,\RR2) \Delta G(\RR2,\RR3) \Delta G(\RR3,\RR1)
	\big\} \big\rangle,
	\label{eq:dg3}
	\\
	& (2\pi)^3 \mathcal{D}_{(2,1)}(\RR1,\RR2|\RR3;z)
	\nonumber \\
	&= \big\langle \Tr \big\{
	\Delta G(\RR1,\RR2) \Delta G(\RR2,\RR1)
	\big\}
	\Tr \Delta G(\RR3,\RR3)
	\big\rangle,
	\label{eq:dg21}
	\\
    & (2\pi)^3 \mathcal{D}_{(1^3)}(\RR1,\RR2,\RR3;z)
	\nonumber \\
	&=  \big\langle
	\Tr \Delta G(\RR1,\RR1) \Tr \Delta G(\RR2,\RR2) \Tr \Delta G(\RR3,\RR3)
	\big\rangle.
	\label{eq:dg111}
\end{align}
We recall that the notation $1^q$ stands for $1,1,\ldots, 1$ ($q$ units).
The averaged correlation function 
$\mathcal{D}_{(1^3)}(\RR1,\RR2,\RR3;z)$
is manifestly invariant under permutations of links $\RR1,\RR1,\RR3$. For $\mathcal{D}_{(3)}(\RR1,\RR2,\RR3;z)$, this invariance is emergent. In the case of $\mathcal{D}_{(2,1)}(\RR1,\RR2|\RR3;z)$, we need to explicitly symmetrize over the permutations of $\RR1,\RR2,\RR3$ in order to obtain the correlator  that is invariant under all permutations of the links $\RR1,\RR1,\RR3$; we will denote it as $\mathcal{D}^{(s)}_{(2,1)}(\RR1,\RR2,\RR3;z)$.

The mapping to percolation is performed for $q=3$ correlators in analogy with the case $q=2$.   Here we present the final results for  $\mathcal{D}_{(3)}$, $\mathcal{D}_{(2,1)}$, and 
$\mathcal{D}_{(1^3)}$.
Percolation expressions for individual contributions to these correlators are presented in Appendix~\ref{appendix-mapping}. Percolation probabilities appearing in these expressions are listed in Table~\ref{tab:perc}.

\paragraph{Correlator $\mathcal{D}_{(3)}$.}

The results for the percolation mapping applied to the individual terms entering Eq.~\eqref{eq:dg3} are presented in Eq.~\eqref{eq:d3ic} of Appendix~\ref{appendix-mapping3-3}.
Adding up these contributions, we get the following result for the correlation function:
\begin{align}
	&(2\pi)^{3}\mathcal{D}_{(3)}(\RR1,\RR2,\RR3;z)
	= \sum_{N=1}^\infty \big[-8p^{(s)}(\RR1,\RR2,\RR3;N)
	\nonumber\\
	& \quad  + 12 p^{(s)}_1(\RR1,\RR2,\RR3;N) \big] z^{2N}.
	\label{eq:D3-perc}
\end{align}
We recall that the superscript $(s)$ denotes full symmetrization with respect to all $q=3$ spatial arguments ($\RR1$, $\RR2$, $\RR3$) of the considered correlation function ($\mathcal{D}_3$).
For $z=1$, the expression ~\eqref{eq:D3-perc} simplifies to:
\begin{align}
	(2\pi)^{3}\mathcal{D}_{(3)}(\RR1,\RR2,\RR3;z=1) &= 4p^{(s)}(\RR1,\RR2,\RR3),
\end{align}
since both $p(\RR1,\RR2,\RR3;N)$ and $p_1(\RR1,\RR2,\RR3;N)$ give $p(\RR1,\RR2,\RR3)$ when summed over all loop lengths $N$. This implies the following decomposition of
$\mathcal{D}_{(3)}$ into a $z$-independent term and a term vanishing at $z=1$:
\onecolumngrid
\begin{align}
	&(2\pi)^{3}\mathcal{D}_{(3)}(\RR1,\RR2,\RR3;z) = 4p^{(s)}(\RR1,\RR2,\RR3)
	+\sum_{N=1}^\infty\big[8p^{(s)}(\RR1,\RR2,\RR3;N)
	-12p^{(s)}_1(\RR1,\RR2,\RR3;N)\big](1-z^{2N}).
	\label{eq:D3n}
\end{align}
This decomposition is very useful for the purposes of numerical simulations (see below).
Note that the first term on the right-hand side of Eq.~\eqref{eq:D3n} does not depend on the energy $\gamma$ (we recall that $z = e^{-\gamma}$).

For the analytical investigation of the scaling of the correlation function (see below), it is useful to rewrite the obtained percolation expression for $\mathcal{D}_{(3)}$ in the following form
\begin{align}
	(2\pi)^{3}\mathcal{D}_{(3)}(\RR1,\RR2,\RR3;z)
	&= 4p^{(s)}(\RR1,\RR2,\RR3)
	-12 \sum_{N, N'} (1-z^{2N})(1-z^{2N'}) \,
	p^{(s)}(\RR1,\RR2,\RR3;N_{\RR1\RR2}=N, N_{\RR2\RR3}=N')
	\nonumber\\
	& + 8 \sum_{N, N',N''} \!\! (1-z^{2N})(1-z^{2N'})(1-z^{2N''}) \,
	p^{(s)}(\RR1,\RR2,\RR3;N_{\RR1\RR2}=N, N_{\RR2\RR3}=N', N_{\RR3\RR1}=N'').
	\label{eq:D3c}
\end{align}

\paragraph{Correlator $\mathcal{D}_{(2,1)}$.}

For the correlation function  ~\eqref{eq:dg21}, the individual terms are mapped onto percolation expressions as given in Eq.~\eqref{eq:d21ic} of
Appendix~\ref{appendix-mapping3-21}.
Combining the individual contributions from Eq.~\eqref{eq:d21ic}, we obtain
\begin{align}
	(2\pi)^{3}\mathcal{D}_{(2,1)}(\RR1,\RR2|\RR3;z)
	&= 8 \sum_{N=1}^\infty p^{(s_2)}(\RR1,\RR2,\RR3;N) z^{2N}
	- 8\sum_{N=1}^\infty [p(\RR1,\RR2;N)-p_1^{(s_2)}(\RR1,\RR2;N)]z^{2N}
	\nonumber\\&
	+ 8\sum_{N,N^{\prime\prime}=1}^\infty [p(\RR1,\RR2;N|\RR3;N^{\prime\prime})
	- p_1^{(s_2)}(\RR1,\RR2;N|\RR3;N^{\prime\prime})] z^{2N} z^{2N''}.
	\label{eq:D21-perc}
\end{align}
\twocolumngrid
\noindent
Here the superscript $(s_2)$ denotes the symmetrization with respect to the permutation of two variables $\RR1,\RR2$ only. Note that, by its definition, the correlation function $\mathcal{D}_{(2,1)}(\RR1,\RR2|\RR3;z)$ is invariant with respect to this permutation; however, it does not possess the full invariance with respect to the permutations of $\RR1,\RR2,\RR3$.  It is convenient to perform the corresponding symmetrization; we denote the fully symmetrized form of this correlator by $\mathcal{D}_{(2,1)}^{(s)} (\RR1,\RR2,\RR3;z)$.

For $z=1$, the percolation expression for $\mathcal{D}_{(2,1)}^{(s)}$ simplifies to
\begin{align}
	(2\pi)^{3}\mathcal{D}^{(s)}_{(2,1)}(\RR1,\RR2,\RR3;z=1) &= 8p^{(s)}(\RR1,\RR2,\RR3),
\end{align}
since $p(\ldots;N)$ and $p_1(\ldots;N)$ both give $p(\ldots)$, when summed over the loop length $N$.

In analogy with Eq.~\eqref{eq:D3c} for $\mathcal{D}_{(3)}$, it is useful to rewrite $\mathcal{D}_{(2,1)}^{(s)}$ in the following form that is especially convenient for  the analytical investigation of scaling properties:
\onecolumngrid
\begin{align}
	&(2\pi)^{3}\mathcal{D}_{(2,1)}^{(s)}(\RR1,\RR2,\RR3;z)
	= 8p^{(s)}(\RR1,\RR2,\RR3)
	\nonumber\\ &
	- 4\sum_{N, N'} (1-z^{2N})(1-z^{2N'}) \big[ 4 p^{(s)}(\RR1,\RR2,\RR3;N_{\RR1\RR2}=N, N_{\RR2\RR3}=N')
	- p^{(s)}(\RR1,\RR2;N|\RR3;N') \big]
	\nonumber\\
	&- 4 \sum_{N, N',N''} (1-z^{2N})(1-z^{2N'})(1-z^{2N''})
	\nonumber \\
	&\qquad \times [p^{(s)}(\RR1,\RR2;N_{\RR1\RR2}=N, N_{\RR2\RR1}=N'|\RR3;N'')
	- 2 p^{(s)}(\RR1,\RR2,\RR3;N_{\RR1\RR2}=N, N_{\RR2\RR3}=N', N_{\RR3\RR1}=N'')].
	\label{eq:D21c}
\end{align}

\paragraph{Correlator 
$\mathcal{D}_{(1^3)}$.
}

For the correlation function  ~\eqref{eq:dg111}, results of the percolation mapping for individual terms are presented in Eq.~\eqref{eq:tr111} of
Appendix~\ref{appendix-mapping3-111}.
Adding up these individual contributions,  we obtain
\begin{align}
	& 
    (2\pi)^3\mathcal{D}_{(1^3)}(\RR1,\RR2,\RR3;z)
	= 8 - 24 \sum_N p^{(s)}(\RR1,\RR2;N)z^{2N}
	+ 24 \sum_{NN'} p^{(s)}(\RR1;N|\RR2;N')z^{2N}z^{2N'}
	+ 24 \sum_{N} p^{(s)}(\RR1;N)z^{2N}
	\nonumber\\
	& \qquad -8\sum_{NN'N^{\prime\prime}} p^{(s)}(\RR1;N|\RR2;N'|\RR3;N^{\prime\prime}) z^{2N}z^{2N'}z^{2N^{\prime\prime}}
	-24\sum_{NN'}p^{(s)}(\RR1,\RR2;N|\RR3;N')z^{2N}z^{2N'}.
	\label{eq:D111-perc}
\end{align}
Using the identity
\begin{align}
	&\sum_{N'N^{\prime\prime}}p(\RR1;N|\RR2;N'|\RR3;N^{\prime\prime})
	+ \sum_{N'} [p(\RR1,\RR2;N|\RR3;N')
	+ p(\RR1,\RR3;N|\RR2;N')+ p(\RR2,\RR3;N|\RR1;N')]
	\nonumber\\
	& \qquad + p(\RR1,\RR2,\RR3;N) + p(\RR1,\RR3,\RR2;N) = p(\RR1;N),
\end{align}
which states that a loop running through $\RR1$ of length $N$ can either run also through two other points $\RR2,\RR3$, or through one of them, or just through $\RR1$, one finds for $z=1$:
\begin{align}
    (2\pi)^3\mathcal{D}_{(1^3)}(\RR1,\RR2,\RR3; z=1)
	&= 16p^{(s)}(\RR1,\RR2,\RR3).
\end{align}
Further, we can rewrite the percolation expression for 
$\mathcal{D}_{(1^3)}$
in the form analogous to Eqs.~\eqref{eq:D3c} and~\eqref{eq:D21c}:
\begin{align}
	(2\pi)^{3}\mathcal{D}_{(1^3)}(\RR1,\RR2,\RR3;z)
&= 16 p^{(s)}(\RR1,\RR2,\RR3)
	+ 24 \sum_{N, N'} (1-z^{2N})(1-z^{2N'}) \, p^{(s)}(\RR1,\RR2;N |\RR3;N')
	\nonumber\\
	& \quad + 8 \sum_{N, N',N''} \!\! (1-z^{2N})(1-z^{2N'})(1-z^{2N''}) \, p^{(s)}(\RR1;N|\RR2;N'|\RR3;N''). \label{eq:D111c}
\end{align}
\twocolumngrid

\subsection{Analytical determination of SQH generalized-multifractality exponents}
\label{sec:sqh_Pana}

The mapping of the SQH correlation functions with $q \le 3$ onto percolation probabilities is very useful for the analytical study of the problem. Indeed, various critical exponents for the 2D percolation problem are known analytically.  As we discuss below in this section, the mapping allows us to relate the SQH generalized-multifractality exponents with $q \le 3$ to the scaling dimensions $x_n^{\rm h}$ of hull operators~\cite{saleur1987exact} in classical percolation theory. {  This extends the earlier obtained results for several exponents characterizing the LDOS moments, $x_{(1)}=x_{(2)}=x_1^{\rm h} = \frac14$ and $x_{(3)}=0$, see Sec.~\ref{sec:intro}.}
 In Sec.~\ref{sec:susy}, we will use the supersymmetry approach to {  further generalize} the connection to the most subleading  SQH exponents $x_{(1^q)}$ of arbitrary integer order $q$.

\subsubsection{Scaling of percolation probabilities}
\label{sec:scaling-perc-prob}

In Ref.~\cite{saleur1987exact} the scaling dimensions $x_n^{\rm h}$ of the $n$-hull operators were introduced. They determine the probability $P_n$ that $n$ segments of the hull of the infinite percolation cluster come close (within a distance $r \sim 1$) at a given point of the system:
\begin{equation}
	P_n\sim (p-p_c)^{\nu x_n^{\rm h}},
	\label{eq:Pn-hull}
\end{equation}
where $p-p_c$ is the detuning from the percolation threshold $p_c$ and $\nu = 4/3$ is the critical exponent of the correlation length. More generally, one can consider the probability that $n$ hull segments come close at two points separated by a distance $r$:
\begin{equation}
	P_n(r)\sim r^{-x_n^{\rm h}} (p-p_c)^{\nu x_n^{\rm h}},
	\qquad r \lesssim \xi,
	\label{eq:Pn-hull-r}
\end{equation}
where $\xi \sim (p-p_c)^{-\nu}$ is the correlation length. As was shown in Ref.~\cite{saleur1987exact}, the $n$-hull exponents are given by
\begin{align}
	x_n^{\rm h} &= \dfrac{4n^2-1}{12},
	\qquad n=1, 2, \ldots.
\end{align}
In particular, the values of the $n$-hull exponents for $n=1$, 2, and 3 are
\begin{equation}
	x_1^{\rm h}=\frac{1}{4},
	\qquad x_2^{\rm h}=\frac{5}{4},
	\qquad x_3^{\rm h}=\frac{35}{12}.
\end{equation}

To determine the SQH generalized-multifractality dimensions, we need the scaling of percolation probabilities from Table~\ref{tab:perc}. Importantly, scaling properties of all of them can be expressed in terms of the the $n$-hull exponents $x_n^{\rm h}$, as we are going to explain.

Let us start with the simplest case $q=1$, when we have a single percolation probability $p(\RR1;N)$.  Integrating $p(\RR1;N')$ over $N'$ from $N$ to infinity, we get a probability
$\sim N p(\RR1;N)$ that the point point $\RR1$ belongs to a loop of the length $\ge N$ at criticality.  (Since the dependence is of power-law type, integration reduces to multiplication by $N$, up to a numerical constant.) This can be identified with Eq.~\eqref{eq:Pn-hull} where we should make a replacement of the correlation length $\xi \sim (p-p_c)^{-\nu}$ by the length scale $\xi_N\sim N^{\frac47}$ associated with the loop length $N$.  Here we used the fractal dimension
\begin{equation}
	D_H = 2  - x_1^{\rm h} = \frac{7}{4}
\end{equation}
of the percolation hull.  Thus, we find
\begin{equation}
	p(\RR1;N) \sim N^{-1- \frac{4}{7} x_1^{\rm h} } = N^{-8/7}.
	\label{eq:prN-scaling}
\end{equation}

Now we extend these arguments to percolation probabilities with a larger number of spatial variables, keeping first a single $N$ variable (i.e., a single loop). The simplest probability of this type is $p(\RR1,\RR2;N)$. Its scaling with $N$ and with $r$  (distance between $\RR1$ and $\RR2$) can be obtained from Eq.~\eqref{eq:Pn-hull-r} following the above line of arguments. We get, for $N \gtrsim r^{7/4}$,
\begin{align}
	p(\RR1, \RR2;N)
	\sim N^{-1- \frac{4}{7} x_1^{\rm h} } r^{-x_1^{\rm h}}
	= N^{-8/7} r^{-1/4}.
	\label{eq:pr1r2N-scaling}
\end{align}
Note that summation over $N$ (that can be replaced by integration) yields the probability that the points $\RR1$ and $\RR2$ belong to the same loop:
\begin{align}
	p(\RR1, \RR2) \equiv \int_N dN' p(\RR1, \RR2;N') \sim r^{-2x_1^{\rm h}}.
	\label{eq:pr1r2-scaling}
\end{align}

A more accurate analysis using operator fusion algebra allows us to find also a subleading correction to Eq.~\eqref{eq:pr1r2N-scaling}. Specifically, since we have a correlation function with two spatial arguments, there is a correction coming from the 2-hull operator:
\begin{align}
	p(\RR1, \RR2;N)
	&\sim \frac{1}{N} \big( N^{- \frac{4}{7} x_1^{\rm h} } r^{-x_1^{\rm h}}
	+ c N^{- \frac{4}{7} x_2^{\rm h} } r^{x_2^{\rm h} -2x_1^{\rm h}} \big),
\end{align}
where $N\gtrsim r^{7/4}$ and $c \sim 1$.  Below we focus on the leading behavior of the percolation probabilities and do not write down such corrections.  The correlation function
$p_1(\RR1, \RR2;N)$ has the same scaling properties as $p(\RR1, \RR2;N)$.

The result~\eqref{eq:pr1r2N-scaling} can be straightforwardly extended to percolation probabilities
$p(\RR1,\RR2,\ldots, \RR{q};N)$ corresponding to a larger number of spatial arguments $\RR{j}$ belonging to the same loop of the length $N$. Their scaling with $N$ is the same as in Eq.~\eqref{eq:pr1r2N-scaling}. Further, we assume that all distances between the points $\RR{j}$ are of the same order $\sim r$.  Integration over $N$ should give
\begin{align}
	p(\RR1, \RR2, \ldots, \RR{q})
	\equiv \int_N dN' p(\RR1, \RR2, \ldots, \RR{q};N')
	\sim r^{-qx_1^{\rm h}},
	\label{eq:pr1r2rn-scaling}
\end{align}
in analogy with Eq.~\eqref{eq:pr1r2-scaling}. Thus, we find
\begin{align}
	& p(\RR1, \RR2, \ldots, \RR{q};N)
	\sim N^{-1- \frac{4}{7} x_1^{\rm h} } r^{(1-q)x_1^{\rm h}}
	\nonumber \\
	& \qquad \qquad = N^{-8/7} r^{(1-q)/4},
	\qquad \qquad N \gtrsim r^{7/4}.
	\label{eq:pr1r2rnN-scaling}
\end{align}

We further discuss the extension to probabilities with a larger number of $N$ arguments. Consider first the correlation functions with two such arguments, $p(\RR1,\RR2;N_{\RR1\RR2}=N,N_{\RR2\RR1}=N')$ and $p(\RR1;N|\RR2;N')$. The analysis below applies to both of them equally. Since they share the same scaling properties in the range of interest, $N,N' \gtrsim r^{7/4}$, we will use here a short notation $p(r;N,N')$ for either of these probabilities (with $r$ being the distance between $\RR1$ and $\RR2$ as before). Without restricting the generality, we can assume $N \gtrsim N'$. Let us fix $N$ and consider two limiting cases with respect to $N'$:  (i) $N' \sim r^{7/4}$ and (ii) $N' \sim N$.

To find the behavior at $N' \sim r^{7/4}$, we consider an integral of $p(\RR1,\RR2;N_{\RR1\RR2}=N,N_{\RR2\RR1}=N')$ over $N'$.  This integral is governed by the lower limit, $N' \sim r^{7/4}$, and, from the point of view of the scaling, the integration is equivalent to multiplication by $N'$.  The result is exactly the probability $p_1(\RR1,\RR2;N)$.
Using its scaling~\eqref{eq:pr1r2N-scaling}, we find
\begin{align}
	p(r;N,N') &\sim \frac{1}{NN'} N^{- \frac{4}{7} x_1^{\rm h} } r^{-x_1^{\rm h}}
	\nonumber \\
	&= \frac{r^{-2x_1^{\rm h}}}{NN'} \big( N^{- \frac{4}{7} }r \big)^{x_1^{\rm h}},
	\qquad N' \sim r^{7/4}.
	\label{prNN-low}
\end{align}

On the other hand, for $N' \sim N \gg r^{7/4}$, we have a geometry of two long hull segments that come close in the region including the points $\RR1$ and $\RR2$.  The scaling with $N$ is thus governed by the two-hull exponent:  $p(r, N, N' \sim N) \propto N^{- \frac{4}{7} x_2^{\rm h}}$.  To determine the scaling with $r$, we consider an integral of $p(\RR1,\RR2;N_{\RR1\RR2}=N,N_{\RR2\RR1}=N')$ over $N$ and $N'$, which scales as $NN' p(r,N,N')$ with $N \sim N' \sim r^{7/4}$. However, this integral is nothing but
$p(\RR1,\RR2) \sim r^{ -2x_1^{\rm h}}$.  Therefore,
\begin{align}
	p(r;N,N') &\sim \frac{r^{-2x_1^{\rm h}}}{NN'}
	\big(N^{- \frac{4}{7} }r \big)^{ x_2^{\rm h}},
	\qquad N' \sim N.
	\label{prNN-up}
\end{align}
Since we expect a power-law scaling with respect to $N'$, we can now interpolate between Eqs.~\eqref{prNN-low} and~\eqref{prNN-up} to restore the behavior of the function $p(r,N,N')$ in the whole range $r^{7/4} \lesssim N' \lesssim N$:
\begin{align}
	p(r;N,N') &\sim \dfrac{r^{-2x_1^{\rm h}}}{NN'}  \big(N^{-\frac47} r\big)^{x_1^{\rm h}}
	\big((N')^{-\frac47} r\big)^{x_2^{\rm h}-x_1^{\rm h}}.
	\label{eq:perc-scaling-q2}
\end{align}

In the same way, we can analyze the scaling of percolation probabilities [that we denote for brevity as $p(r;N_1,\ldots,N_q)$] containing $q$ spatial arguments $\RR1, \ldots, \RR{q}$ (separated by distances $|\RR{i}-\RR{j}|\sim r$) and an equal number of hull-length arguments $N_1, \ldots, N_q$. Assuming the hierarchy $N_1\gtrsim \ldots \gtrsim N_q$, we find
\begin{align}
	&p(r;N_1,\ldots,N_q)
	\sim \prod_{i=1}^q \bigg[ \dfrac{r^{-x_1^{\rm h}}}{N_i}
	\big(N_i^{-\frac47} r\big)^{x_i^{\rm h}-x_{i-1}^{\rm h}}\bigg],
	\label{eq:scaling}
\end{align}
with the definition $x_{0}^{\rm h}\equiv 0$. The scaling of percolation probabilities with a number $m$ of $N_j$ arguments smaller than the number $q$ of $\RR{j}$ arguments can be obtained from Eq.~\eqref{eq:scaling} by integrating over $N_q, N_{q-1}, \ldots, N_{m+1}$.  In particular, for $m=1$ we reproduce in this way  Eq.~\eqref{eq:pr1r2rnN-scaling}.

\begin{figure}
	\centering
	\includegraphics[width=.48\textwidth]{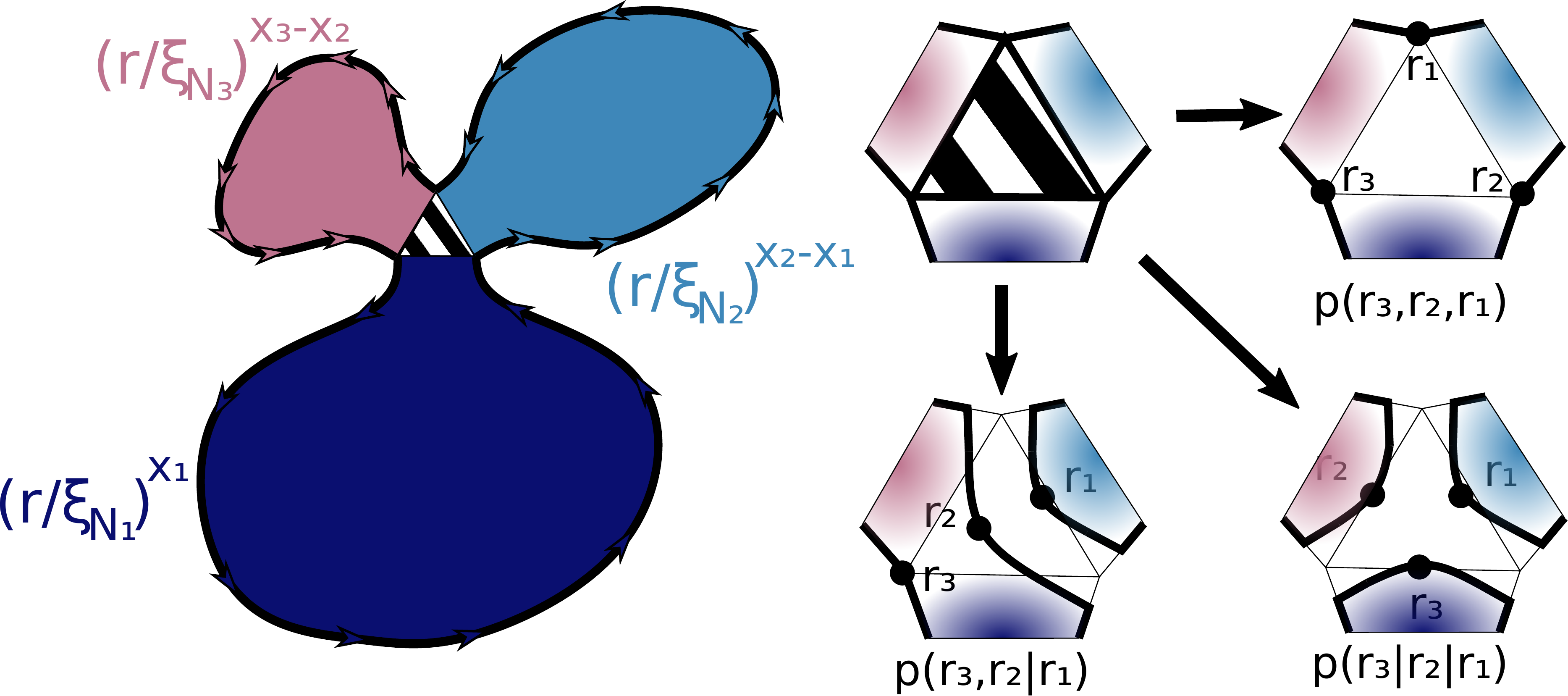}	
	\caption{Schematic illustration of $q=3$ percolation probabilities of the type~\eqref{eq:scaling}, with three spatial arguments $\RR1, \RR2, \RR3$ and three hull-length arguments $N_1, N_2, N_3$. The hierarchy of hull lengths $N_1 > N_2 > N_3$ is assumed, and the factors $\big(N_i^{-\frac47} r\big)^{x_i^{\rm h}-x_{i-1}^{\rm h}}$ in Eq.~\eqref{eq:scaling} that depend on this hierarchy are indicated, with the hull exponents $x_i^{\rm h}$ abbreviated as $x_i$. On the right-hand side of the figure, different possible connections between the paths are shown. They yield different percolation probabilities that exhibit the same scaling~\eqref{eq:scaling}.
	}
	\label{fig:three-hulls}
\end{figure}

In Fig.~\ref{fig:three-hulls} we illustrate the geometry corresponding to the percolation probabilties of the type~\eqref{eq:scaling} with $q=3$. They are governed by three hull loops (or ``nearly loops'') coming close to the region around the points $\RR1$, $\RR2$, and $\RR3$. Depending on the way these ``nearly loops'' are connected in the central region, one gets a probability with one, two, or three loops. All these probabilities are characterized by the same scaling, Eq.~\eqref{eq:scaling}.

We are now ready to determine the scaling of SQH observables by combining their mapping to percolation in Sec.~\ref{sec:sqh_mapping} with the above results for the scaling of percolation probabilities.

\subsubsection{SQH generalized multifractality: $q=1$}
\label{sec:perc-analyt-q1}

We start with the percolation expression for the averaged LDOS, Eq.~\eqref{DOS}, and use Eq.~\eqref{eq:prN-scaling} for the scaling of the percolation probability $p(\RR1;N)$.
The sum over $N$ is estimated as (we recall that $x_1^{\rm h}=\frac14$)
\begin{align}
	& \sum_{N} (1-z^{2N}) N^{-1 - \frac{4}{7} x_1^{\rm h}}
	\sim \int_{1}^{\infty} \!\! \frac{dN}{N} (1-z^{2N}) N^{- \frac{4}{7} x_1^{\rm h}}
	\nonumber\\
	&\sim \int_{\gamma^{-1}}^{\infty} \!\! \frac{dN}{N} N^{- \frac{4}{7}x_1^{\rm h}}
	\sim \gamma^{\frac{4}{7}x_1^{\rm h} } = \gamma^{\frac{1}{7}}.
	\label{eq:perc_sum1}
\end{align}
Hence
\begin{align}
	\mathcal{D}_{(1)} (\RR1;z)  \sim (a/\xi_\gamma)^{x_1^{\rm h}} \sim \gamma^{\frac17},
	\label{eq:D1-scaling}
\end{align}
where $\xi_\gamma \sim \gamma^{-\frac47}$ is the correlation length assosiated with the energy scale $\gamma$, and $a$ is the lattice constant. This yields  the relation
\begin{equation}
	x_{(1)}=x_1^{\rm h} = \frac14
	\label{eq:x1-result}
\end{equation}
between the SQH scaling exponent $x_{(1)}$ (which characterizes the average LDOS) and the scaling dimension $x_1^{\rm h}$ of the one-hull operator in the classical percolation.

\subsubsection{SQH generalized multifractality: $q=2$}
\label{sec:perc-analyt-q2}

Percolation expression for the SQH correlation functions $\mathcal{D}_{(2)}$ and $\mathcal{D}_{(1,1)}$ were obtained in Eqs.~\eqref{eq:D2} and~\eqref{eq:D11}. They can be cast in the following form
\begin{align}
	&\mathcal{D}_\lambda(\RR1,\RR2;z) = \sum_{N}^\infty (1-z^{2N}) \mathcal{D}^{(N)}_\lambda(\RR1,\RR2)
	\nonumber\\
	&+ \sum_{N,N'=1}^\infty (1-z^{2N})(1-z^{2N'}) \mathcal{D}^{(N,N')}_\lambda(\RR1,\RR2),
	\label{eq:D2d}
\end{align}
with
\begin{align}
	(2\pi)^2\mathcal{D}^{(N)}_{(2)}(\RR1,\RR2) &=2p(\RR1,\RR2;N), \nonumber\\
	(2\pi)^2\mathcal{D}^{(N,N')}_{(2)}(\RR1,\RR2)  &=-2p(\RR1,\RR2;N_{\RR1\RR2} \! = \! N,N_{\RR2\RR1} \! = \! N'), \nonumber\\
	(2\pi)^2\mathcal{D}^{(N)}_{(1,1)}(\RR1,\RR2) &=4p(\RR1,\RR2;N), \nonumber\\
	(2\pi)^2\mathcal{D}^{(N,N')}_{(1,1)}(\RR1,\RR2) &=4p(\RR1;N|\RR2;N').
	\label{eq:D2d-coeff}
\end{align}
The advantage of such presentation is that each sum over the length $N$ of a (segment of a) hull is accompanied by the factor $(1-z^{2N})$ that suppresses contributions from $N \ll \gamma^{-1}$ in the sum.
%
%
%
Performing the transformation~\eqref{rg:perc} to pure-scaling $q=2$ observables $\mathcal{P}_{(2)}^C$ and $\mathcal{P}_{(1,1)}^C$, we get for them representations analogous to Eq.~\eqref{eq:D2d} with
\begin{align}
	(2\pi)^2\mathcal{P}^{(N)}_{(2)}(\RR1,\RR2) &=6p(\RR1,\RR2;N), \nonumber\\
	(2\pi)^2\mathcal{P}^{(N,N')}_{(2)}(\RR1,\RR2) &=4p(\RR1;N|\RR2;N')\nonumber\\&-2p(\RR1,\RR2;N_{\RR1\RR2} \!= \! N,N_{\RR2\RR1} \!= \! N'), \nonumber\\
	(2\pi)^2\mathcal{P}^{(N)}_{(1,1)}(\RR1,\RR2) &=0, \nonumber\\
	(2\pi)^2\mathcal{P}^{(N,N')}_{(1,1)}(\RR1,\RR2) &=4p(\RR1,\RR2;N_{\RR1\RR2} \!= \! N,N_{\RR2\RR1} \!= \! N')\nonumber\\&+4p(\RR1;N|\RR2;N').
\end{align}

We estimate now the single and double sums entering Eq.~\eqref{eq:D2d} and the analogous formulas for $\mathcal{P}_\lambda^C$.
For the single sum, we use Eq.~\eqref{eq:pr1r2N-scaling} for the scaling of the corresponding percolation probabilities [that we abbreviate as $p(r,N)$] and get, in similartity with
Eq.~\eqref{eq:perc_sum1},
\begin{align}
	& \sum_N (1-z^{2N}) p(r,N)
	\sim \int_{1}^{\infty} \!\! \frac{dN}{N} (1-z^{2N})
	N^{- \frac{4}{7} x_1^{\rm h}} r^{-x_1^{\rm h}}
	\nonumber\\
	&\sim \int_{\gamma^{-1}}^{\infty} \!\! \frac{dN}{N}
	N^{- \frac{4}{7}x_1^{\rm h}} r^{-x_1^{\rm h}}
	\sim \gamma^{\frac{4}{7}x_1^{\rm h} }  r^{-x_1^{\rm h}} = \gamma^{\frac{1}{7}} r^{-\frac14}.
	\label{eq:single-N-sum}
\end{align}
As in all analogous formulas, we have assumed here $r < \xi_\gamma \sim \gamma^{-4/7}$, which is the condition for criticality.

We turn now to the analysis of the double sum. Using the scaling form~\eqref{eq:perc-scaling-q2} of the corresponding percolation probabilities, we obtain
\onecolumngrid
\begin{align}
	&\sum_{N,N'} (1-z^{2N})(1-z^{2N'}) p(r;N,N')
	\sim r^{-2x_1^{\rm h}} \int_1^\infty \!\! \frac{dN}{N} (1-z^{2N})
	\big(r N^{-\frac47} \big)^{x_1^{\rm h}}
	\int_1^N \!\! \frac{dN'}{N'} (1-z^{2N'})
	\big(r (N')^{-\frac47} \big)^{x_2^{\rm h}-x_1^{\rm h}}
	\nonumber\\
	& \qquad \qquad \sim r^{-2x_1^{\rm h}} \int_{\gamma^{-1}}^\infty \!\! \frac{dN}{N}
	\big(r N^{-\frac47} \big)^{x_1^{\rm h}}
	\int_{\gamma^{-1}}^N \!\! \frac{dN'}{N'}
	\big(r (N')^{-\frac47} \big)^{x_2^{\rm h}-x_1^{\rm h}}
	\sim r^{- 2x_1^{\rm h}} \big( r\gamma^{\frac47}\big)^{x_2^{\rm h}}
	= r^{\frac34} \gamma^{\frac57}.
	\label{eq:double-N-sum}
\end{align}
\twocolumngrid

Comparing Eqs.~\eqref{eq:single-N-sum} and~\eqref{eq:double-N-sum}, we see that the double-$N$ sum is much smaller than the single-$N$ sum, which is related to an obvious inequality
$x_2^{\rm h} > x_1^{\rm h}$.  The scaling of the correlation function $\mathcal{P}_{(2)}^C$ is thus determined by that of a single sum,
\begin{equation}
	\mathcal{P}_{(2)}^C (\RR1,\RR2;z)  \sim \gamma^{\frac{4}{7}x_1^{\rm h} }  r^{-x_1^{\rm h}} \sim  \gamma^{\frac{1}{7}} r^{-\frac14}.
	\label{eq:P2-scaling}
\end{equation}
On the other hand, for the correlation function $\mathcal{P}_{(1,1)}^C$ there is no contribution of a single sum,  since $\mathcal{P}^{(N)}_{(1,1)}(\RR1,\RR2)=0$.  Let us emphasize the importance of this result, which is a manifestation of a very non-trivial character and of a full consistency of our percolation analysis. Thus, the scaling of $\mathcal{P}_{(1,1)}^C$ is governed by that of the double-$N$ sum,
\begin{equation}
	\mathcal{P}_{(1,1)}^C (\RR1,\RR2;z)  \sim  r^{-2x_1^{\rm h}} \big(r\gamma^{\frac47}\big)^{x_2^{\rm h}}
	= r^{\frac34} \gamma^{\frac57}.
	\label{eq:P11-scaling}
\end{equation}
Comparing the results~\eqref{eq:P2-scaling} and~\eqref{eq:P11-scaling} with Eq.~\eqref{eq:MF-scaling-q2}, we read off the values of the $q=2$ scaling exponents characterizing the SQH critical point:
\begin{align}
	x_{(2)} &= x_1^{\rm h} = \frac14,  & \Delta_{(2)} &=  x_{(2)} - 2x_{(1)} = - \frac14,
	\label{eq:x2-result} \\
	x_{(1,1)} &= x_2^{\rm h} = \frac54, & \Delta_{(1,1)} &=  x_{(1,1)} - 2x_{(1)} = \frac34.
	\label{eq:x11-result}
\end{align}

In terms of the correlation functions $\mathcal{D}_{(2)}(\RR1,\RR2;z)$ and $\mathcal{D}_{(1,1)}(\RR1,\RR2;z)$, the results of the percolation mapping yield
\begin{align}
	\mathcal{D}_{(2)}(\RR1,\RR2;z) & = \xi_\gamma^{-\frac12} \big(c_0(r/\xi_\gamma)^{-\frac14}+c_1( r/\xi_\gamma)^{\frac34}\big),
	\label{D2-scaling-perc} \\
	\mathcal{D}_{(1,1)}(\RR1,\RR2;z) & = \xi_\gamma^{-\frac12} \big(2c_0(r/\xi_\gamma)^{-\frac14}+c_2( r/\xi_\gamma)^{\frac34}\big).
	\label{D11-scaling-perc}
\end{align}
Since we know from the RG analysis of the SQH problem  that $\mathcal{P}_{(2)}^C = \mathcal{D}_{(2)} + \mathcal{D}_{(1,1)}$ is a pure-scaling operator [see Eq.~\eqref{rg:perc}], the coefficients $c_1$ and $c_2$ in Eqs.~\eqref{D2-scaling-perc} and~\eqref{D11-scaling-perc} should satisfy the relation
\begin{equation}
	c_2 = - c_1 \,.
\end{equation}
(We do not have a proof of this relation that would be based solely on the percolation mapping.) Furthermore,
since polynomial operators of the order $q=2$ of the SQH problem are characterized by only two exponents, $x_{(2)}$ and $x_{(1,1)}$, there is no further subleading corrections
to Eqs.~\eqref{D2-scaling-perc} and~\eqref{D11-scaling-perc}. This observation is very non-trivial from the point of view of the percolation mapping.

Indeed, while the dominant contributions to the sums in Eq.~\eqref{eq:D2d} come from $N \sim N' \sim \gamma^{-1}$ as discussed above, one could also expect contributions from the lower limit in the integrals over $N$ and $N'$, i.e., $N \sim N'\sim r^{\frac74}$. In this region $\gamma N \ll 1$, and the factor $1-z^{2N} = 1 - e^{-2\gamma N}$ can be expanded in powers of $N$. A naive estimate of the corresponding contributions yields terms characterized by scaling exponents $\frac74 k$ and $x_1^{\rm h} + \frac74 k$, where $k=1,2, \ldots$. These exponents are larger than $x_{(2)} = x_1^{\rm h}$  and $x_{(1,1)} = x_2^{\rm h}$, so that the corresponding contributions would be subleading. Remarkably, the RG analysis of the SQH problem implies that these subleading corrections are in fact identically zero, as explained above.

This may look as just a curious observation for $q=2$ (and also for $q=3$), since the contributions under discussion would be of minor importance anyway. We will show, however, in Sec.~\ref{sec:susy} that the percolation mapping works for a particular type of correlation functions for $q \ge 4$ as well. In that case, the contributions from the lower limit, $N \sim r^{\frac74}$, might become dominant if they existed. It is natural to expect, however, that, since this contribution is absent for $q=2,3$, it is also absent for larger $q$.  This expectation is fully confirmed by the numerical determination of the generalized-multifractality exponents in Sec.~\ref{sec:sqh_numerics}.



\subsubsection{SQH generalized multifractality: $q=3$}
\label{sec:perc-analyt-q3}

In the case of $q=2$, we have seen that the presentation of the correlation function in the form~\eqref{eq:D2d} is very useful for the analysis of the scaling behavior. Analogous formulas  will also serve as a convenient starting point for the analysis of the $q=3$ correlators. The required formulas have been already derived above, see  Eqs.~\eqref{eq:D3c}, ~\eqref{eq:D21c}, and~\eqref{eq:D111c} for the correlation functions $\mathcal{D}_{(3)}$, $\mathcal{D}_{(2,1)}$, and  
$\mathcal{D}_{(1^3)}$,
respectively.
They all have the form
\begin{align}
	&\mathcal{D}_\lambda(\RR1,\RR2,\RR3;z)
	= \mathcal{D}_\lambda(\RR1,\RR2,\RR3;z=1)
	\nonumber\\
	&+ \sum_{N,N'=1}^\infty (1-z^{2N})(1-z^{2N'}) \mathcal{D}^{(N,N')}_\lambda(\RR1,\RR2,\RR3) \nonumber\\
	&+ \sum_{N,N',N''=1}^\infty (1-z^{2N})(1-z^{2N'})(1-z^{2N''})
	\nonumber \\
	& \qquad\qquad \times \mathcal{D}^{(N,N',N'')}_\lambda(\RR1,\RR2,\RR3),
	\label{eq:D3d}
\end{align}
with $\lambda = (3)$, (2,1), and 
$(1^3)$.
Here the first term is energy-independent and given by
\begin{align}
	(2\pi)^3 \mathcal{D}_\lambda(\RR1,\RR2,\RR3;z=1) &= b_\lambda p^{(s)}(\RR1,\RR2,\RR3),
\end{align}
with $b_{(3)} = 4$, $b_{(2,1)} = 8$, and 
$b_{(1^3)} = 16$.
The coefficients $\mathcal{D}^{(N,N')}_\lambda(\RR1,\RR2,\RR3)$ and $\mathcal{D}^{(N,N',N'')}_\lambda(\RR1,\RR2,\RR3)$ of the double and triple sums can be directly read off from Eqs.~\eqref{eq:D3c},~\eqref{eq:D21c}, and~\eqref{eq:D111c}, so that we do not repeat them here. Remarkably, there is no term with a single-$N$ sum in Eq.~\eqref{eq:D3d}. (Equivalently, we could write such a term and state that the corresponding coefficients vanishes: $\mathcal{D}^{(N)}_\lambda = 0$.)

In Ref.~\cite{karcher2021generalized}, the $q=3$ pure-scaling operators were determined by means of the sigma-model RG analysis (see Appendix~\ref{appendix:rg}  and in particular Eq.~\eqref{eq:rg_C_result} of the present work):
\begin{align}
	\begin{pmatrix}
        \mathcal{P}^C_{(1^3)} \\
		\mathcal{P}^C_{(2,1)} \\
		\mathcal{P}^C_{(3)}
	\end{pmatrix} &=
	\begin{pmatrix*}[r]
		1 & -6 & 8 \\
		1 & -1 & -2 \\
		1 & 3 & 2
	\end{pmatrix*}
	\begin{pmatrix}
        \mathcal{D}_{(1^3)} \\
		\mathcal{D}_{(2,1)} \\
		\mathcal{D}_{(3)}
	\end{pmatrix}.
	\label{eq:q3-pure-scaling-matrix}
\end{align}
Substituting Eq.~\eqref{eq:D3d} into Eq.~\eqref{eq:q3-pure-scaling-matrix} and keeping only the leading terms (the ones with the minimal number of summations) for the
functions $\mathcal{P}^C_{(3)}$ and $\mathcal{P}^C_{(2,1)}$, we find:
\onecolumngrid
\begin{align}
	\mathcal{P}^C_{(3)}
	&\simeq 48 p^{(s)}(\RR1,\RR2,\RR3),
	\nonumber\\
	\mathcal{P}^C_{(2,1)}
	&\simeq 20 \sum_{N,N'} (1-z^{2N})(1-z^{2N'})
	\big[ p^{(s)}(\RR1,\RR2;N|\RR3;N')
	+ 2p^{(s)}(\RR1,\RR2,\RR3;N_{\RR1\RR2}=N,N_{\RR2\RR1}=N')\big],
	\nonumber\\
    \mathcal{P}^C_{(1^3)}
	&= 8 \!\! \sum_{N,N',N''} \!\!\! (1-z^{2N})(1-z^{2N'})(1-z^{2N''})
	\big[ p^{(s)}(\RR1;N|\RR2;N'|\RR3;N'')
	+
	3p^{(s)}(\RR1,\RR2;N_{\RR1\RR2}=N,N_{\RR2\RR1}=N'|\RR3;N'')
	\nonumber\\
	& \quad +
	2p^{(s)}(\RR1,\RR2,\RR3;N_{\RR1\RR2}=N,N_{\RR2\RR3}=N',N_{\RR3\RR1}=N'')\big].
	\label{eq:q3-P-perc-formulas}
\end{align}
\twocolumngrid
\noindent
Let us emphasize that the terms without summations cancels in both $\mathcal{P}^C_{(2,1)}$ and 
$\mathcal{P}^C_{(1^3)}$
and, furthermore, the terms with a double sum cancel in
$\mathcal{P}^C_{(1^3)}$.
This is one more remarkable manifestation of the full consistency of our treatment that combines the SQH RG theory and the percolation analysis.
The scaling of the double and triple sums in Eq.~\eqref{eq:q3-P-perc-formulas} is determined in the same way as was done in Sec.~\ref{sec:perc-analyt-q2} for the single and double sums for the case $q=2$. Via the same token, the scaling of the triple sum (governed by the region $N\sim N'\sim N''\sim \gamma^{-1}$) involves the three-hull exponent $x_3^{\rm h}$.  The results read:
\begin{align}
	\mathcal{P}^C_{(3)} (\RR1, \RR2, \RR3; z) &\sim r^{-3x_1^{\rm h}},
	\label{eq:q3-P-perc-scaling3} \\
	\mathcal{P}^C_{(2,1)} (\RR1, \RR2, \RR3; z) &\sim r^{-3x_1^{\rm h}} (r/\xi_\gamma)^{x_2^{\rm h}},
	\label{eq:q3-P-perc-scaling21}  \\
    \mathcal{P}^C_{(1^3)}(\RR1, \RR2, \RR3; z)
&\sim r^{-3x_1^{\rm h}} (r/\xi_\gamma)^{x_3^{\rm h}}.
	\label{eq:q3-P-perc-scaling111}
\end{align}
We thus find the values of the $q=3$ scaling exponents at the SQH critical point:
\begin{align}
	x_{(3)} &= 0,
	& \Delta_{(3)} &=  x_{(3)} - 3x_{(1)} = - \frac34,
	\\
	x_{(2,1)} &= x_2^{\rm h} = \frac54,
	& \Delta_{(2,1)} &=  x_{(2,1)} - 3x_{(1)} = \frac12,
	\\
    x_{(1^3)}
    &= x_3^{\rm h} = \frac{35}{12},
	& 
    \Delta_{(1^3)}
    &=  
    x_{(1^3)}
    - 3x_{(1)} = \frac{13}{6}.
\end{align}
Using the numerical values of the hull exponents, we get the following scaling for the $q=3$ pure-scaling operators:
\begin{align}
	\mathcal{P}^C_{(3)} &\sim \xi_\gamma^{-\frac34}(r/\xi_\gamma)^{-\frac34},
	\nonumber\\
	\mathcal{P}^C_{(2,1)} &\sim \xi_\gamma^{-\frac34}(r/\xi_\gamma)^{\frac12},
	\nonumber\\
    \mathcal{P}^C_{(1^3)}
    &\sim \xi_\gamma^{-\frac34}(r/\xi_\gamma)^{\frac{13}6}.
\end{align}

This completes the analytical study of SQH genera\-lized-multifractality critical exponents with $q \le 3$. The obtained analytical results for the critical exponents are summarized in Table~\ref{tab:lC}.  This table includes also further results obtained in the rest of the paper, including numerical results of classical percolation simulations and of quantum (network-model) simulations as well as analytical results for some specific exponents with $q \ge 4$.

Two comments are in order at this point:

\begin{enumerate}
	
	\item[(i)]
	The analytical results obtained above satisfy the Weyl-symmetry relations: $x_{(3)}=x_{(0)} \equiv 0$,  $x_{(2)}=x_{(1)}$ as well as $x_{(2,1)}=x_{(1,1)}$. The Weyl symmetry is by no means explicit in the percolation mapping, and it is impressive to see how it emerges out of the percolation analysis.
	
	\item[(ii)]   This comment largely reiterates what was said in the end of Sec.~\ref{sec:perc-analyt-q2} for the case $q=2$. We know from the class-C RG that  $\mathcal{P}^C_{(3)}$, $\mathcal{P}^C_{(2,1)}$, and 
$\mathcal{P}^C_{(1^3)}$
are pure-scaling correlation functions. This means that there are no corrections to the formulas~\eqref{eq:q3-P-perc-scaling3},~\eqref{eq:q3-P-perc-scaling21}, and~\eqref{eq:q3-P-perc-scaling111} of the same type with other exponents $x$. In particular, there are no corrections to Eq.~\eqref{eq:q3-P-perc-scaling3} of the type~\eqref{eq:q3-P-perc-scaling21} and~\eqref{eq:q3-P-perc-scaling111}, and there is no corrections to  Eq.~\eqref{eq:q3-P-perc-scaling21} of the type~\eqref{eq:q3-P-perc-scaling111}.  Furthermore, there should be no corrections to these formulas with further subleading exponents. We do not have a direct proof of these statements based solely on the percolation formulas.
	
\end{enumerate}

{ 
\subsection{Numerical study of percolation expressions for SQH correlators}
\label{sec:sqh_Pnumerics}}

In Sec.~\ref{sec:sqh_Pana} we have performed an analytical investigation of the scaling of percolation expressions for the SQH correlation functions derived in Sec.~\ref{sec:sqh_mapping}. This has allowed us to obtain analytical results for the SQH generalized-multifractality exponents with $q \le 3$. Here we supplement these analytical results by numerical simulations of the corresponding percolation probabilities. Let us emphasize that these are classical simulations that are relevant to the SQH problem only for those observables for which the percolation mapping exists.

In Sec.~\ref{sec:sqh_numerics} we will perform direct quantum simulations (within the class-C version of the Chalker-Coddington network model) that can be implemented for wave-function observables of any representation $\lambda$. Further, we will see that the accuracy of the numerical determination of the exponents $x_\lambda$ is considerably higher in the case of quantum simulations. This is related to strong statistical fluctuations in percolation simulations; getting high-precision results for averaged correlators in this way would require a much larger statistical ensemble and thus a much longer computational time.

Since we have highly efficient quantum simulations (Sec.\ref{sec:sqh_numerics}) that fully support our analytical predictions, we find it unnecessary to invest such extreme efforts in classical simulations. On the other hand, we find it very instructive to present here results of classical simulations. Despite a modest accuracy of the corresponding numerical values of the exponents, these simulations nicely illustrate and corroborate the analytical study of the percolation probabilities in Sec.~\ref{sec:sqh_Pana}.

For the purpose of these  percolation simulations, we use the classical counterpart of the Chalker-Coddington network model of linear size $L=32768$ with periodic boundary conditions. The averaging over $10000$ random configurations is performed.
Each configuration is described by $L\times L$ binary degrees of freedom (``black'' or ``white'') that are associated with the nodes of the network. For black (white) nodes, the paths running through this node turn right (respectively, left). Any given realization of disorder thus generates a decomposition of  all $2L^2$ links of the network into a set of closed loops.
(Note that the disorder in the classical percolation problem, as described above, is very different form the link SU(2) randomness in the quantum network model.)
The probabilities from Table~\ref{tab:perc} are then obtained by counting the number of loops satisfying the desired properties and by averaging over disorder.

{  Before presenting details, we briefly announce main results of these percolation simulations:
\begin{itemize}
\item[(i)] We verify that the percolation probabilities entering the expressions for the SQH correlation functions $\mathcal{D}_\lambda$ with $|\lambda| \le 3$ exhibit a power-law scaling with $N$ as expected.
\item[(ii)] We demonstrate numerically leading-order cancellations between different terms whenever this is predicted analytically (Sec.~\ref{sec:sqh_Pana}).
\item[(iii)] The numerical results confirm the analytically predicted values of the exponents $x_{(1)} = x_{(2)} = x_1^h = \frac14$ and $x_{(1,1)} = x_{(2,1)} = x_2^h = \frac54$.
\end{itemize}
We turn now do a detailed exposition of the results of our percolation numerics for the SQH correlation functions.
}

\subsubsection{The case $q=1$}

\begin{figure}
	\centering
	\includegraphics[width=.48\textwidth]{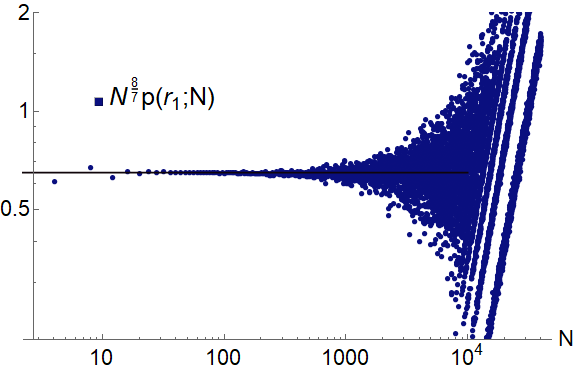}	
	\caption{Numerical simulations of the percolation probability $p(\RR1;N)$ entering Eq.~\eqref{DOS} for the $q=1$ SQH correlation function $\mathcal{D}_{(1)}(\RR1; z)$. The data fully confirm the analytical prediction $p(\RR1;N) \sim N^{-1- \frac{4}{7} x_1^{\rm h} } = N^{-8/7}$, see Eq.~\eqref{eq:prN-scaling}. In combination with Eq.~\eqref{DOS}, it yields the exponent $x_{(1)}=x_1^{\rm h} = \frac14$. For $ N > 1000$, strong statistical fluctuations are observed, which are related to insufficient averaging.
	}
	\label{fig:DP1}
\end{figure}

For $q=1$, there is a single pure-scaling correlation function $\mathcal{D}_{(1)}(\RR1; z)$ given by Eq.~\eqref{DOS}. We choose not to perform the $N$-summation numerically, since strong statistical fluctuations (due to insufficient averaging) at large $N$ make difficult to do this in reliable way. Instead, we plot in Fig.~\ref{fig:DP1} the function $p(\RR1, N)$ that enters the sum in  Eq.~\eqref{DOS}. The scaling of $p(\RR1, N)$ with $N$ is straightforwardly translated in scaling of $\mathcal{D}_{(1)}(\RR1; z)$ with $\gamma$, as discussed in Sec.~\ref{sec:perc-analyt-q1}.  The analytical prediction for the scaling of $p(\RR1, N)$ is given by Eq.~\eqref{eq:prN-scaling}, which results
in Eq.~\eqref{eq:D1-scaling} for the correlation function $\mathcal{D}_{(1)}$ and thus in the value $x_{(1)}=x_1^{\rm h} = \frac14$ of the exponent, see Eq.~\eqref{eq:x1-result}.

Figure~\ref{fig:DP1} provides a perfect confirmation of this analytical prediction. We plot there $N^{1+ \frac 47 x_1^{\rm h}} p(\RR1, N) = N^{\frac 87} p(\RR1, N)  $ and observe that it is indeed $N$-independent in the range between $N \approx 10$ and $N \approx 1000$, in full agreement with Eq.~\eqref{eq:prN-scaling}.

\subsubsection{The case $q=2$}

In this order, there are two correlation functions, $\mathcal{D}_{(2)}(\RR1,\RR2; z)$  and $\mathcal{D}_{(1,1)}(\RR1,\RR2; z)$.   We use here the representations~\eqref{eq:D2} and~\eqref{eq:D11} for them. These formulas have the form
\begin{align}
	\pi^2\mathcal{D}_{\lambda}(\RR1,\RR2; z)
	= \sum_N (1-z^{2N}) \mathcal{\tilde{D}}^{(N)}_\lambda (\RR1,\RR2),
	\label{eq:q2-D-sum_N}
\end{align}
with
\begin{align}
	\mathcal{\tilde{D}}^{(N)}_{(2)} (\RR1,\RR2) & = p^{(s)}_1(\RR1,\RR2;N)-p(\RR1,\RR2;N), \nonumber \\
	\mathcal{\tilde{D}}^{(N)}_{(1,1)} (\RR1,\RR2) & = p(\RR1,\RR2;N)+p_-(\RR1,\RR2;N).
	\label{eq:q2-D-perc-coeff}
\end{align}
Note that this representation is somewhat different from that in Eq.~\eqref{eq:D2d} where a double sum over path lengths was separated for the convenience of analytical investigation. In the numerical analysis, studying the dependence on two (or three) length variables would require much more statistics, which is why we use the representation~\eqref{eq:q2-D-sum_N} here. The coefficient functions $\mathcal{\tilde{D}}^{(N)}_\lambda (\RR1,\RR2)$ in Eq.~\eqref{eq:q2-D-sum_N} are related to the functions
\eqref{eq:D2d-coeff} entering the representation~\eqref{eq:D2d} as follows:
\begin{align}
	\mathcal{\tilde{D}}^{(N)}_\lambda = \mathcal{D}^{(N)}_\lambda + \sum_{N'} \left[ \mathcal{D}^{(N,N')}_\lambda + \mathcal{D}^{(N',N)}_\lambda + \mathcal{D}^{(N-N',N')}_\lambda \right].
	\label{eq:q2-tilde-D}
\end{align}

\begin{figure*}
	\noindent\begin{center}
		\includegraphics[width=.40\linewidth]{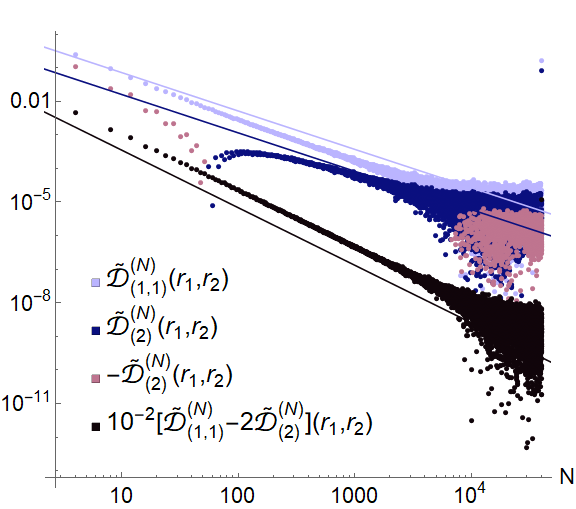}
		\includegraphics[width=.40\linewidth]{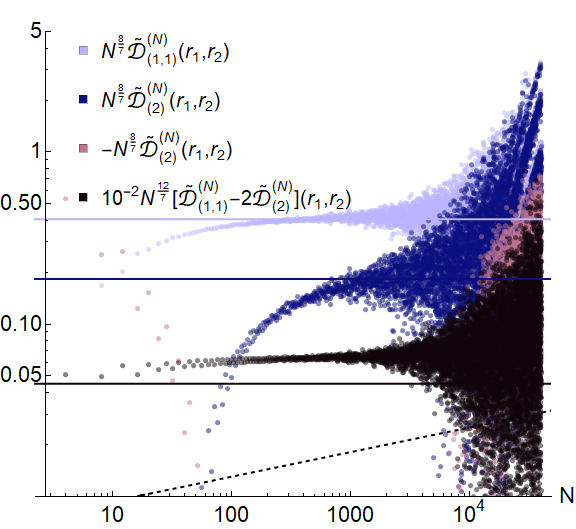}	
	\end{center}
	\caption{
		Numerically determined coefficient functions $\mathcal{\tilde{D}}^{(N)}_{(1,1)} (\RR1,\RR2)$  and  $\mathcal{\tilde{D}}^{(N)}_{(2)} (\RR1,\RR2)$ in the percolation representation~\eqref{eq:q2-D-sum_N},~\eqref{eq:q2-D-perc-coeff} of the $q=2$ SQH correlation functions. The links $\RR1,\RR2$ are chosen to be either horizontal or vertical nearest neighbors.  \textit{Left panel:} The Hartree coefficient function $\mathcal{\tilde{D}}^{(N)}_{(1,1)} (\RR1,\RR2)$ is shown by light blue symbols, while the Fock coefficient function $\mathcal{\tilde{D}}^{(N)}_{(2)} (\RR1,\RR2)$ by dark blue symbols. (When the Fock function is negative, its absolute value is shown by red symbols.) The results agree with the analytically predicted scaling $\mathcal{\tilde{D}}^{(N)}_{(1,1)}, \mathcal{\tilde{D}}^{(N)}_{(2)} \propto N^{-1-\frac47 x_1^{\rm h}} = N^{-\frac87}$, see Eq.~\eqref{eq:pr1r2N-scaling}, as shown by light blue and dark blue straight lines.  The black symbols show the combination $\mathcal{\tilde{D}}^{(N)}_{(1,1)} - 2 \mathcal{\tilde{D}}^{(N)}_{(2)}$ (multiplied by $10^{-2}$ for clarity)  corresponding to the pure-scaling correlation function  $\mathcal{P}^C_{(1,1)}$. It is seen that the leading terms $ \propto N^{-\frac87}$ cancel and the scaling follows the law $\mathcal{\tilde{D}}^{(N)}_{(1,1)} - 2\mathcal{\tilde{D}}^{(N)}_{(2)} \propto N^{-1-\frac47 x_2^{\rm h}} = N^{-\frac{12}{7}}$ (straight black line), in agreement with analytical prediction. {  \textit{Right panel:}}  Same data normalized to the analytically predicted power laws. In this representation, the predicted scaling corresponds to $N$-independence of the plotted functions (horizontal straight lines). To demonstrate violation of the generalized parabolicity, we also included the dashed line corresponding to $x_{(1,1)} = 1$ which would hold for generalized parabolicity.
	}
	\label{fig:HF}
\end{figure*}

The numerical simulations data shown in Fig.~\ref{fig:HF} fully support the analytically predicted scaling $\mathcal{\tilde{D}}^{(N)}_{(1,1)}, \mathcal{\tilde{D}}^{(N)}_{(2)} \propto N^{-1-\frac47 x_1^{\rm h}} = N^{-\frac87}$, see Eq.~\eqref{eq:pr1r2N-scaling}. Furthermore, we show in this figure the combination $\mathcal{\tilde{D}}^{(N)}_{(1,1)} -2 \mathcal{\tilde{D}}^{(N)}_{(2)}$ that corresponds to the pure-scaling correlation function  $\mathcal{P}^C_{(1,1)}$, see Eq.~\eqref{rg:perc}.
It is clearly seen that the terms $ \propto N^{-\frac87}$ cancel in this combination as expected. Furthermore, the resulting scaling is in agreement with the analytically predicted scaling $\mathcal{\tilde{D}}^{(N)}_{(1,1)} - 2\mathcal{\tilde{D}}^{(N)}_{(2)} \propto N^{-1-\frac47 x_2^{\rm h}} = N^{-\frac{12}{7}}$. Therefore, our $q=2$ numerical data are in a very good agreement with analytical predictions, leading to  the results $x_{(2)} = x_1^{\rm h} = \frac14$ and $ x_{(1,1)} = x_2^{\rm h} = \frac54$ for the critical exponents, see Eqs.~\eqref{eq:x2-result} and~\eqref{eq:x11-result}.


\subsubsection{The case $q=3$}

As we have seen in Sec.~\ref{sec:perc-mapping-q3}, all $q=3$ correlation functions $\mathcal{D}_{\lambda}^{(s)}(\RR1,\RR2,\RR3;z)$ can be written in the form
\begin{align}
	&(2\pi)^{3}\mathcal{D}^{(s)}_{\lambda}(\RR1,\RR2,\RR3;z) =(2\pi)^{3}\mathcal{D}^{(s)}_{\lambda}(\RR1,\RR2,\RR3;z=1)
	\nonumber\\
	& \quad + \sum_{N=1}^\infty (1-z^{2N}) \mathcal{\tilde{D}}^{(N)}_{\lambda}(\RR1,\RR2,\RR3).
	\label{eq:q3-D-decomp-numerics}
\end{align}
Expressions for the coefficient functions $\mathcal{\tilde{D}}^{(N)}_{\lambda}(\RR1,\RR2,\RR3)$ in this representations in terms of the functions $\mathcal{D}^{(N,N')}_{\lambda}(\RR1,\RR2,\RR3)$  and $\mathcal{D}^{(N,N',N'')}_{\lambda}(\RR1,\RR2,\RR3)$ in the representation~\eqref{eq:D3d} can be straightforwardly obtained in analogy with Eq.~\eqref{eq:q2-tilde-D}.  The first term in the right-hand side of Eq.~\eqref{eq:q3-D-decomp-numerics} is $z$-independent and yields the leading scaling with the exponent $x_{(3)}=0$. The second term, involving the sum over $N$ includes subleading corrections scaling with the exponents $x_{(2,1)}$ and 
$x_{(1^3)}$.
We consider now this term, focussing on its leading behavior that is governed by $x_{(2,1)}$.

The coefficient functions $\mathcal{\tilde{D}}^{(N)}_{\lambda}(\RR1,\RR2,\RR3)$ are linear combinations of percolation probabilities that are each of the type $p(\RR1,\RR2,\RR3;N)$, $p_1(\RR1,\RR2,\RR3;N)$, etc (i.e., depends on three spatial points and a single loop-length $N$). The scaling of such probabilities is given by
Eq.~\eqref{eq:pr1r2rnN-scaling} with $q=3$:
\begin{align}
	p(\RR1, \RR2, \RR3;N)
	\sim N^{-1- \frac{4}{7} x_1^{\rm h} } r^{-2x_1^{\rm h}}
	= N^{-\frac87} r^{-\frac12}.
\end{align}
A contribution of each such probability to the sum over $N$ in Eq.~\eqref{eq:q3-D-decomp-numerics} would yield a scaling with the exponent $x_1^{\rm h}$, in analogy with the
leading behavior of a similar sum~\eqref{eq:q2-D-sum_N} in the $q=2$ case. This might suggest that the exponent $x_{(2,1)}$ is given by $x_1^{\rm h}$. Our analytical study showed, however, that this is not correct. The fact that we have a representation~\eqref{eq:D3d} without a single sum over $N$ (i.e., with only double and triple sums) implies that the terms showing the $x_1^{\rm h}$ scaling should cancel and the leading scaling of the sum over $N$ in Eq.~\eqref{eq:q3-D-decomp-numerics}  is controlled by the exponent
$x_2^{\rm h}$, so that $x_{(2,1)}= x_2^{\rm h}$. Thus, we should have
\begin{align}
	\mathcal{\tilde{D}}^{(N)}_{\lambda}(\RR1,\RR2,\RR3)
	\sim N^{-1- \frac{4}{7} x_2^{\rm h} } r^{x_2^{\rm h} - 3x_1^{\rm h}}
	= N^{-\frac{12}{7}} r^{\frac12}.
\end{align}
%

\begin{figure*}
	\centering
	\noindent\begin{center}
		\vspace*{-0.5cm}
		\includegraphics[width=.40\linewidth]{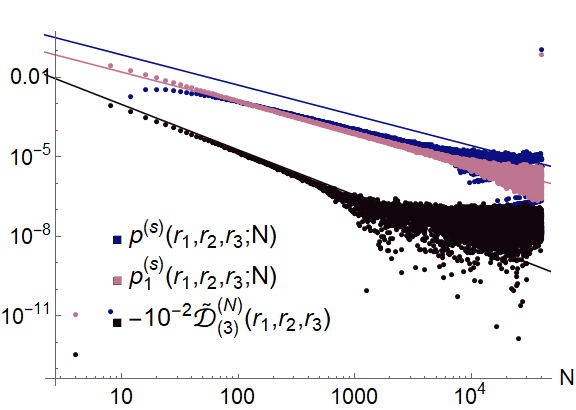}	\hspace{0.07\linewidth}
		\includegraphics[width=.40\linewidth]{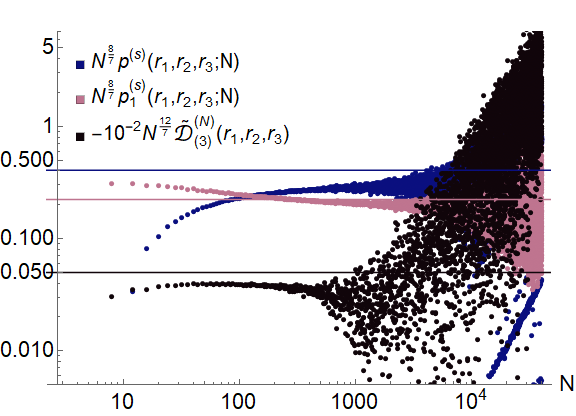}
	\end{center}
	\vspace*{-0.5cm}
	\noindent\begin{center}
		\includegraphics[width=.40\linewidth]{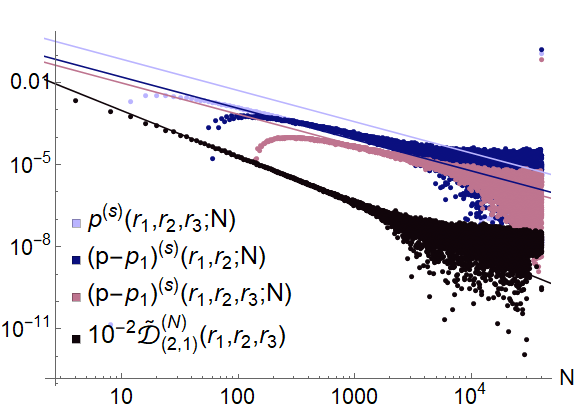}	\hspace{0.07\linewidth}
		\includegraphics[width=.40\linewidth]{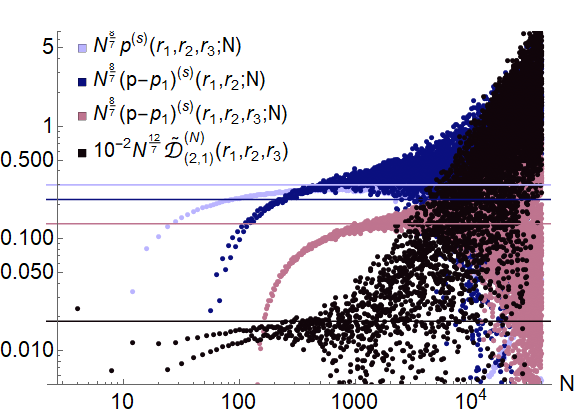}
	\end{center}
	\vspace*{-0.3cm}
	\noindent\begin{center}
		\includegraphics[width=.40\linewidth]{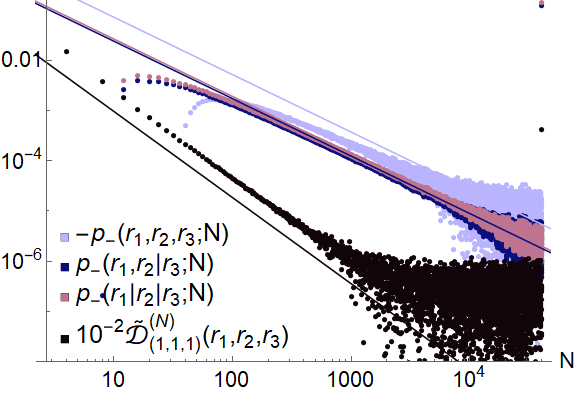}	 \hspace{0.07\linewidth}
		\includegraphics[width=.40\linewidth]{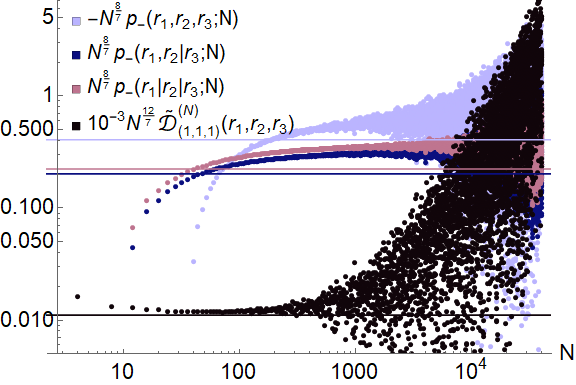}
	\end{center}
	\caption{ Numerically determined coefficient functions $\mathcal{\tilde{D}}^{(N)}_\lambda (\RR1,\RR2,\RR3)$  in the percolation representation~\eqref{eq:q3-D-decomp-numerics} of the $q=3$ SQH correlation functions. The links $\RR1,\RR2,\RR3$ are chosen to be either horizontal or vertical (next) nearest neighbors.
		\textit{Top:} $\mathcal{\tilde{D}}^{(N)}_{(3)} (\RR1,\RR2,\RR3)$   as given by Eq.~\eqref{eq:Dn3} (black) and individual
		probabilities  $p^{(s)}(\RR1,\RR2,\RR3;N)$ (blue) and $p_1^{(s)}(\RR1,\RR2,\RR3;N)$ (red) entering this formula.
		\textit{Middle:}  $\mathcal{\tilde{D}}^{(N)}_{(2,1)} (\RR1,\RR2,\RR3)$  as given by Eq.~\eqref{eq:Dn21} (black) and individual
		terms entering this formula (light blue, dark blue, red).
		\textit{Bottom:}  
$\mathcal{\tilde{D}}^{(N)}_{(1^3)} (\RR1,\RR2,\RR3)$,
Eq.~\eqref{eq:Dn111} (black), and the individual terms in this formula given by Eq.~\eqref{eq:Dn111-terms} (light blue, dark blue, and red).  The results for the individual terms agree with the analytically predicted scaling $\propto N^{-1-\frac47 x_1^{\rm h}} = N^{-\frac87}$ as shown by straight lines of the corresponding colors.  At the same time, the total expressions for the coefficient functions $\mathcal{\tilde{D}}^{(N)}_\lambda (\RR1,\RR2,\RR3)$  exhibit the analytically predicted scaling $\propto N^{-1-\frac47 x_2^{\rm h}} = N^{-\frac{12}{7}}$ (black straight lines).
		\textit{Right panels:}  Same data as in the respective left panels normalized to the analytically predicted power laws. In this representation, the predicted scaling corresponds to $N$-independence of the plotted functions (horizontal straight lines).
	}
	\label{fig:D3}
\end{figure*}

The results of the percolation simulations shown in Fig.~\ref{fig:D3} nicely confirm these predictions for the $N$ scaling. In particular, in the top left panel of this figure the data  for the coefficient function $\mathcal{D}^{(N)}_{(3)}(\RR1,\RR2,\RR3)$ are presented. Rewriting Eq.~\eqref{eq:D3c} in the form~\eqref{eq:q3-D-decomp-numerics}, we get
\begin{align}
	\mathcal{\tilde{D}}^{(N)}_{(3)}(\RR1,\RR2,\RR3) &= 8p^{(s)}(\RR1,\RR2,\RR3;N)\nonumber\\&-12p^{(s)}_1(\RR1,\RR2,\RR3;N).
	\label{eq:Dn3}
\end{align}
The blue and red symbols in the top left panel of Fig.~\ref{fig:D3} show the individual terms of Eq.~\eqref{eq:Dn3}, and the black symbols show the total expression.
The numerical data for the individual terms agree with the analytically predicted scaling $\propto N^{-1-\frac47 x_1^{\rm h}} = N^{-\frac87}$ presented by straight lines of the corresponding colors.  At the same time,  the scaling of for the total coefficient function~\eqref{eq:Dn3} is in full agreement with the anaytical prediction  $\propto N^{-1-\frac47 x_2^{\rm h}} = N^{-\frac{12}{7}}$ that is presented by a black line.  In the top right panel, the same data are shown normalized to the respective analytically predicted power laws.

In the same way, the middle and bottom panels of Fig.~\ref{fig:D3} display numerical data for $\mathcal{\tilde{D}}^{(N)}_{(2,1)}(\RR1,\RR2,\RR3)$ and 
$\mathcal{\tilde{D}}^{(N)}_{(1^3)}(\RR1,\RR2,\RR3)$,
respectively. Transforming Eq.~\eqref{eq:D21c} to the form~\eqref{eq:q3-D-decomp-numerics}, we obtain
\onecolumngrid
\begin{align}
	\mathcal{\tilde{D}}^{(N)}_{(2,1)}(\RR1,\RR2,\RR3)
	&= 8 \big[p^{(s)}(\RR1,\RR2;N)-p^{(s)}_1(\RR1,\RR2;N) \big]
	-4 p^{(s)}(\RR1,\RR2,\RR3;N)
	\nonumber\\&
	-8\sum_M \big[p^{(s)}(\RR1,\RR2;N-M|\RR3;M)
	- p_{1}^{(s)}(\RR1,\RR2;N-M|\RR3;M)\big].
	\label{eq:Dn21}
\end{align}
\twocolumngrid
\noindent
Similarly, Eq.~\eqref{eq:D111c} can be brought to the form~\eqref{eq:q3-D-decomp-numerics} with
\begin{align}
    \mathcal{\tilde{D}}^{(N)}_{(1^3)}(\RR1,\RR2,\RR3)
	&= p_-(\RR1,\RR2,\RR3;N)+ p_-(\RR1,\RR2|\RR3;N)
	\nonumber \\
	& \quad + p_-(\RR1|\RR2|\RR3;N),
	\label{eq:Dn111}
\end{align}
where
\begin{align}
	p_-(\RR1,\RR2,\RR3;N) &=  24 \sum_M \big[p^{(s)}(\RR1;M|\RR2;N)
	\nonumber\\
	& \quad - p^{(s)}(\RR1;N-M|\RR2;M)\big],
	\nonumber\\
	p_-(\RR1,\RR2|\RR3;N) &= 24 \sum_Mp^{(s)}(\RR1,\RR2;N-M|\RR3;M),\
	\nonumber\\
	p_-(\RR1|\RR2|\RR3;N) &= 8 \sum_{MK} p^{(s)}(\RR1;N-M-K|\RR2;M|\RR3;K).
	\label{eq:Dn111-terms}
\end{align}
It is clearly seen in Fig.~\ref{fig:D3} that the the individual terms in Eqs.~\eqref{eq:Dn21} and~\eqref{eq:Dn111} show the $N^{-\frac87}$ scaling, while the total functions
$\mathcal{\tilde{D}}^{(N)}_{(2,1)}(\RR1,\RR2,\RR3)$ and 
$\mathcal{\tilde{D}}^{(N)}_{(1^3)}(\RR1,\RR2,\RR3)$
scale as $N^{-\frac{12}7}$, as predicted analytically.

The percolation numerics thus nicely confirms the analytical prediction $x_{(2,1)} = x_2^{\rm h} = \frac54$.  In principle, by studying the combination corresponding to the most subleading correlation function  
$\mathcal{P}^C_{(1^3)}$
as given by  Eq.~\eqref{eq:q3-pure-scaling-matrix}, one could also verify in this way the prediction 
$x_{(1^3)} = x_3^{\rm h} = \frac{35}{12}$.
However, it turns out that statistical fluctuations prevent a reliable analysis of this correlation function by means of classical percolation numerics. In Sec.~\ref{sec:sqh_numerics} we will present an alternative numerical analysis---starting directly from the quantum formulation of the problem---that will allow us to get accurate numerical results for all exponents with $q \le 5$ and, in particular, to confirm all analytical predictions.

\section{Supersymmetry mapping}
\label{sec:susy}

\subsection{Basic ideas of the supersymmetry method}

The original mapping of the network model of the SQH transition in class C onto classical percolation was obtained with the help of the supersymmetry (SUSY) method~\cite{Read-1991, Gruzberg-Scaling-1997} in Ref.~\cite{gruzberg1999exact} where the exponents $\nu$ and $x_{(1)}$ characterizing the scaling of the localization length and of the average LDOS were computed. The SUSY method was later used~\cite{subramaniam2008boundary} to reproduce the leading multifractal exponents $x_{(q)}$ for $q \le 3$ from Ref.~\cite{mirlin2003wavefunction} and to extend their derivation to the multifractal wave functions near straight boundaries and corners. Here we will show that the SUSY mapping allows us to reproduce our results for all generalized-multifractality correlation functions from Sec.~\ref{sec:sqh_percolation} of this paper, and also to obtain exact results for the most irrelevant scaling operators $\mathcal{P}^C_{(1^q)}$ of arbitrary order $q$.

The basic idea of the SUSY method is to rewrite Green's functions and their products as quantum expectation values of bosonic or fermionic creation and annihilation operators in a Fock space. The Fock space factorizes into local Fock spaces associated with every link in the network. In the minimal version of this representation, which is sufficient to write down all products of no more than three Green's functions, and some other special combinations of arbitrary numbers of Green's functions, there is only one boson and one fermion per lattice link per spin direction: $\bup(\mathbf{r})$, $\bdown(\mathbf{r})$, $\fup(\mathbf{r})$, $\fdown(\mathbf{r})$.

The average over disorder independently on each link projects the minimal Fock space onto the fundamental three-dimensional representation of sl$(2|1)$ Lie superalgebra~\cite{gruzberg1999exact}. The states in this respresentation are the singlets under the random ``gauge'' SU(2) on the links. The SU(2)-singlet bilinear combinations of bosons and fermions that survive the disorder average are the eight generators of sl$(2|1)$:
\begin{align}
	\label{up-gen}
	\hat{B} &= \frac{1}{2}(\bdup \bup + \bddown \bdown + 1),
	&
	\hat{Q}_3 &= \frac{1}{2}(\fdup \fup + \fddown \fdown - 1),
	\nonumber \\
	\hat{Q}_+ &= \fdup \fddown,
	&
	\hat{Q}_- &= \fdown \fup,
	\nonumber \\
	\hat{V}_+ &= \frac{1}{\sqrt 2} (\bdup \fddown - \bddown \fdup),
	&
	\hat{W}_- &=
	\frac{1}{\sqrt 2} (\bup \fdown - \bdown \fup),
	\nonumber \\
	\hat{V}_- &= - \frac{1}{\sqrt 2} (\bdup \fup + \bddown \fdown),
	& \hat{W}_+ &=
	\frac{1}{\sqrt 2} (\fdup \bup + \fddown \bdown).
\end{align}
In the three-dimensional representation with the basis states chosen as
$|0\ra$, $|1\ra = \hat{V}_+ |0\ra$, and $|2\ra = \hat{Q}_+ |0\ra$,
these operators act as matrices
\begin{align}
	B &= \begin{pmatrix*}[r] 1/2 & 0 & 0 \\ 0 & 1 & 0 \\ 0 & 0 & 1/2 \end{pmatrix*},
	&
	Q_3 &= \begin{pmatrix*}[r] -1/2 & 0 & 0 \\ 0 & 0 & 0 \\ 0 & 0 & 1/2 \end{pmatrix*},
	\nonumber \\
	Q_+ &= \begin{pmatrix*}[r] 0 & 0 & 0 \\ 0 & 0 & 0 \\ 1 & 0 & 0 \end{pmatrix*},
	&
	Q_- &= \begin{pmatrix*}[r] 0 & 0 & 1 \\ 0 & 0 & 0 \\ 0 & 0 & 0 \end{pmatrix*},
	\nonumber \\
	V_+ &= \begin{pmatrix*}[r] 0 & 0 & 0 \\ 1 & 0 & 0 \\ 0 & 0 & 0 \end{pmatrix*},
	&
	W_- &= \begin{pmatrix*}[r] 0 & 1 & 0 \\ 0 & 0 & 0 \\ 0 & 0 & 0 \end{pmatrix*},
	\nonumber \\
	V_- &= \begin{pmatrix*}[r] 0 & 0 & 0 \\ 0 & 0 & -1 \\ 0 & 0 & 0 \end{pmatrix*},
	&
	W_+ &= \begin{pmatrix*}[r] 0 & 0 & 0 \\ 0 & 0 & 0 \\ 0 & 1 & 0 \end{pmatrix*}.
	\label{rep-3}
\end{align}

The mapping to percolation then goes as follows. Disorder-averaged products of Green's functions are written as expectation values of strings of sl$(2|1)$ generators inserted at various points. In the percolation picture, contributions to these expectation values are given in terms of loops (percolation hulls) that go through the insertion points. Each sl$(2|1)$ generator is represented by the corresponding matrix in the fundamental representation~\eqref{rep-3}, and each element of a loop by the diagonal ``attenuation'' matrix
\begin{align}
	Z = \text{diag}(1,z^2,z^2).
\end{align}
This matrix gives weights to the states according to how many bosons and fermions propagate in them along the loop. Then we take the supertrace in the fundamental representation. Finally, we need to take into account the classical probabilities of the loops in percolation that come from the product of the individual factors for turning left or right at each node of the network model.

Let us illustrate how this works in the simplest example of a single (retarded) Green's function:
\begin{align}
	G_{\alpha \beta}(\mathbf{r}_a, \mathbf{r}_b; z) &= \langle \bph_\alpha(\mathbf{r}_a) \bd_\beta(\mathbf{r}_b) \rangle_\text{q},
	\label{G-b}
\end{align}
where $\la \ldots \ra_\text{q}$ stands for the quantum average in the Fock space. For $\mathbf{r}_a \neq \mathbf{r}_b$ the Green's function is not SU(2) invariant, and it vanishes upon averaging over the disorder. However, for $\mathbf{r}_a = \mathbf{r}_b = \mathbf{r}$, we get (the angular brackets without subscripts stand for disorder average, as before)
\begin{align}
	\la G_{\alpha \beta}(\mathbf{r}, \mathbf{r}; z) \ra
	&= \frac{1}{2} \delta_{\alpha \beta}  \la \la 2\hat{B}(\mathbf{r}) + 1 \ra_\text{q} \ra.
\end{align}

Using the symmetry~\eqref{eq:inv} and subscripts to denote spatial points [$G_{12}(z) \equiv G(\mathbf{r}_1, \mathbf{r}_2; z)$], we can write
\begin{align}
	\Delta G_{12} &= G_{12}(z) - G_{12}(z^{-1})
	\nonumber \\
	& = G_{12}(z) - G_{21}(z) + \Tr G_{21}(z) - \delta_{12}.
\end{align}
For example, the average LDOS is obtained as
\begin{align}
	\nu(\mathbf{r}) &= \mathcal{D}_{(1)}(\mathbf{r}) = \frac{1}{2\pi} \big\la \Tr \Delta G(\mathbf{r}, \mathbf{r}) \big\ra
	\nonumber \\
	&= \frac{1}{\pi} \big\la \Tr G(\mathbf{r}, \mathbf{r}; z) - 1 \big\ra
	= \frac{2}{\pi} \la \la \hat{B}(\mathbf{r}) \ra_\text{q} \ra.
	\label{D1}
\end{align}
In the percolation picture this average is obtained by summing over all possible percolation hulls going through the point $\mathbf{r}$:
\begin{align}
	\nu(\mathbf{r}) &= \frac{2}{\pi} \sum_N \str [B Z^N] p(\mathbf{r},N)
	\nonumber \\
	&= \frac{1}{\pi} \sum_N \str
	\left[
	\begin{pmatrix*}[r]
		1 & 0 & 0\\ 0 & 2 & 0\\ 0 & 0 & 1
	\end{pmatrix*}
	\begin{pmatrix*}[l]
		1 & 0 & 0\\ 0 & z^{2N} & 0\\ 0 & 0 & z^{2N}
	\end{pmatrix*}
	\right]
	p(\mathbf{r},N)
	\nonumber \\
	&= \frac{1}{\pi} \sum_N \big(1 - z^{2N}\big) p(\mathbf{r}, N),
	\label{eq:nu-SUSY-perc}
\end{align}
which is exactly the known result~\eqref{DOS}. In Eq.~\eqref{eq:nu-SUSY-perc} we introduced the symbol ``str'', which means the supertrace, i.e., the trace over bosonic states
(here $|0\ra$ and $|2\ra$) minus trace over fermionic states  ($|1\ra$).

Notice that the Green's function in Eq.~\eqref{G-b} can be also expressed in terms of fermions:
\begin{align}
	G_{\alpha \beta}(\mathbf{r}_a, \mathbf{r}_b; z) &=
	\langle \fph_\alpha(\mathbf{r}_a) \fd_\beta(\mathbf{r}_b) \rangle_\text{q},
	\nonumber \\
	\la G_{\alpha \beta}(\mathbf{r}, \mathbf{r}; z) \ra
	&= \frac{1}{2} \delta_{\alpha \beta} \la \la 1 - 2\hat{Q}_3(\mathbf{r}) \ra_\text{q} \ra.
	\label{G-f}
\end{align}
Evaluating this expression in the percolation picture gives the same answer~\eqref{DOS}. It is this flexibility to use either bosons or fermions to represent Green's functions that allows us to generalize the SUSY mapping to arbitrary products of retarded and advanced Green's functions with $q \leq 3$.

This calculation has been extended in Ref.~\cite{subramaniam2008boundary} to the correlation functions $\mathcal{D}_{(1,1)}$ and $\mathcal{D}_{(1,1,1)}$.
We briefly reproduce the calculation of $\mathcal{D}_{(1,1)}$ here; an extension to $\mathcal{D}_{(1,1,1)}$ is quite straightforward. We have
\begin{align}
	(2\pi)^2 \mathcal{D}_{(1,1)}(\mathbf{r}_1, \mathbf{r}_2)
	&= \big\la \Tr \Delta G(\mathbf{r}_1, \mathbf{r}_1) \Tr \Delta  G(\mathbf{r}_2, \mathbf{r}_2) \big\ra
	\nonumber \\
	&= -16 \la \la \hat{B}(\mathbf{r}_1)\hat{Q}_3(\mathbf{r}_2)\ra_\text{q} \ra.
	\label{cal-D1-BQ}
\end{align}
The computation of this correlator proceeds analogously to the case of $\mathcal{D}_{(1)}$, Eq.~\eqref{D1}. There are two different kinds of loop configurations which contribute to the correlator: (i) a single loop passes through both points $\mathbf{r}_1$ and $\mathbf{r}_2$, (ii) a distinct loop passes through each of these two points. These two terms are the percolation versions of the connected and disconnected parts of any correlation function. Writing the contribution for each of these types separately and summing over all possible loop configurations together with their respective weights, we write Eq.~\eqref{cal-D1-BQ} as
\begin{align}
	& (2\pi)^2 \mathcal{D}_{(1,1)}(\mathbf{r}_1, \mathbf{r}_2; z)
	\nonumber\\
	&= - 16\sum_{N_{12},N_{21}} \str [B Z^{N_{12}} Q_3 Z^{N_{21}}]
	p(\mathbf{r}_1, \mathbf{r}_2; N_{12}, N_{21})
	\nonumber \\
	& \quad - 16\sum_{N,N'} \str [B Z^{N}] \str [Q_3 Z^{N'}]
	p(\mathbf{r}_1;N | \mathbf{r}_2;N')
	\nonumber\\
	& = 4 \sum_{N} (1 - z^{2N}) p(\mathbf{r}_1, \mathbf{r}_2; N)
	\nonumber \\
	&\quad + 4 \sum_{N,N'} (1 - z^{2N})(1 - z^{2N'}) p(\mathbf{r}_1;N | \mathbf{r}_2;N').
	\label{calD11-perc}
\end{align}
This is exactly the same representation of  $\mathcal{D}_{(1,1)}(\mathbf{r}_1, \mathbf{r}_2; z)$ as was given in Eqs.~\eqref{eq:D2d} and~\eqref{eq:D2d-coeff} of Sec.~\ref{sec:sqh_percolation}.   The advantage of this from in comparison with Eq.~\eqref{eq:D11} for the same correlator is that each summation over a length $N_i$ is accompanied by the corresponding suppression  factor $(1-z^{2N_i})$. As has been already emphasized in {  Sec.~\ref{sec:perc-analyt-q2}}, such representation are particularly favorable for the purpose of analytical investigation of scaling properties.

In a similar way we can obtain all other correlators $\mathcal{D}_\lambda$ with $q \leq 3$ in terms of percolation probabilities. We have verified that the results are identical with those obtained in Sec.~\ref{sec:sqh_percolation} within the alternative mapping procedure used there.

The mapping to percolation (in either form) does not permit to obtain all correlators $\mathcal{D}_\lambda$ with $q > 3$. However,  it turns out that the SUSY method is suitable to obtain exact percolation expressions for the most irrelevant scaling operators $\mathcal{P}^C_{(1^q)}$ for any integer $q$. To describe the derivation of this remarkable result, we need some additional notation.

\subsection{Young tableaux and point configurations.}
\label{sec:partitions-tableaux}

A more general and comprehensive notation for probabilities and other objects of our interest uses partitions (or Young diagrams) and Young tableaux. For more details about these objects and associated quantities see the books~\cite{Fulton-book, Fulton-Harris-book, Macdonald-book, Stanley-book-2, Stanley-book-1}.

In a general correlation function $\mathcal{D}_\lambda$ or relevant percolation probabilities, points $\mathbf{r}_i$, $i = 1,\ldots, q$, are separated into $n$ groups of sizes $q_a$, $a = 1, 2, \ldots, n$. This corresponds to the integer partition $\lambda = (q_1, q_2, \ldots, q_n)$ with  $q_1 \ge q_2 \ge \ldots \ge q_n > 0$ and $q_1 + \ldots + q_n = q$, see also
Sec.~\ref{sec:intro}. For partitions of $q$ we write $|\lambda| = q$ and, equivalently, $\lambda \vdash q$.
Another notation that we will use is $\lambda = (1^{m_1}, 2^{m_2},\dots, k^{m_k}, \ldots)$ to denote the partition that has $m_k$ copies of the integer $k$. The number
\begin{align}
	l(\lambda) = n = \sum_k m_k
	\label{partition-length}
\end{align}
is called the length of the partition. To each partition we associate a Young diagram in a standard way. Thus, the Young diagram corresponding to the partition $\lambda = (q_1, q_2, \ldots, q_n)$ consists of left-aligned rows of square boxes with the top row containing $q_1$ boxes, the next row containing $q_2$ boxes, etc. The total number of rows $n$ is the length of the partition.

We can fill in the boxes of a Young diagram $\lambda$ by points $\mathbf{r}_i$, forming a Young tableau $T$. We can re-label all points $\mathbf{r}_i$ using a superscript $(a)$ to denote points that belong to the $a$-th row of the Young tableau $T$ and a subscript $j = 1, 2, \ldots, q_a$ for all points within this group. With this notation, the Young tableau precisely specifies which points belong to which group and also gives their order (reading the points from left to right along each row), when it is important. The Young diagram $\lambda$ is called the shape of the tableau $T$.

The permutation group $S_q$ acts on tableaux with $q$ boxes by permuting the points in the boxes. If $\sigma \in S_q$, we denote by $\sigma T$ the tableau which has the point $\mathbf{r}_{\sigma(i)}$ in the box where $T$ has $\mathbf{r}_i$. The row group $R_\lambda$ of a Young tableau $T$ of shape $\lambda$ acts on $T$ by permuting {  points} only within each row:
\begin{align}
	R_\lambda &= S_{q_1} \times S_{q_2} \times \ldots \times S_{q_n}.
\end{align}
A tabloid $\{T\}$ is an equivalence class of Young tableaux of the same shape $\lambda$, two being equivalent if corresponding rows contain the same entries. So $\{T'\} = \{T\}$ exactly when $T' = \sigma T$ for some $\sigma \in R_\lambda$. The number of distinct tabloids of a given shape $\lambda$ is
\begin{align}
	c_\lambda &= \frac{q!}{\prod_{k} (k!)^{m_k} m_k! }.
	\label{c_lambda}
\end{align}

The cyclic subgroup
\begin{align}
	\mathbb{Z}_\lambda &= \mathbb{Z}_{q_1} \times \mathbb{Z}_{q_2} \times \ldots \times \mathbb{Z}_{q_n}
\end{align}
of the row group $R_\lambda$ acts on Young tableaux by cyclic permutations of points in each row. A cyclic tabloid $[T]$ is an equivalence class of Young tableaux of the same shape $\lambda$, two being equivalent if corresponding rows differ by a cyclic permutation. So $[T'] = [T]$ exactly when $T' = \sigma T$ for some $\sigma \in \mathbb{Z}_\lambda$. The number of distinct cyclic tabloids for a given shape $\lambda$ is
\begin{align}
	d_\lambda &= \frac{q!}{\prod_{k} k^{m_k} m_k!}
	= c_\lambda \prod_{k} [(k-1)!]^{m_k}.
	\label{d_lambda}
\end{align}
The coefficients $d_\lambda$ are the numbers of permutations in the conjugacy classes $C_\lambda$ of $S_q$ labeled by the Young diagrams $\lambda \vdash q$. These coefficients appear, in particular, in the RG results for pure-scaling operators, see Appendix~\ref{appendix:rg}. Specifically, the coefficients in the expansion of the most relevant pure-scaling operators $\mathcal{P}^{A}_{(q)}$ and $\mathcal{P}^{C}_{(q)}$ over basis operators are given by $d_\lambda$ both in classes A and C, see bottom rows in Eqs.~\eqref{eq:rg_A_result},~\eqref{eq:rg_C_result} and comments below these equations.   The least relevant operator $\mathcal{P}^{A}_{(1^q)}$ [top row in Eq.~\eqref{eq:rg_A_result}]
has coefficients $(-1)^{q-l(\lambda)} d_\lambda$.

In the context of our problem, we are interested in probabilities of events where points are connected by percolation hulls in various ways. Specifically, let us distinguish {\it configurations} and {\it arrangements} of points. Given a partition $\lambda = (q_1, q_2, \ldots, q_n)$, a configuration of $q$ points is simply the partitioning of the points in $n$ groups of sizes $q_a$, $a = 1, 2, \ldots, n$. All points within a group belong to one percolation hull. It is clear that configurations defined in this way are in one-to-one correspondence with tabloids $\{T\}$, and can be labeled by them. The number of distinct configurations for a given partition $\lambda$ is $c_\lambda$. On the other hand, an arrangement of points is a configuration with a specific order of points along each percolation hull. These are determined modulo cyclic permutations of points on each hull, so are labeled by the cyclic tabloids $[T]$. Correspondingly, the number of distinct arrangements for a given partition $\lambda$ is $d_\lambda$.

In a given configuration $\{T\}$ or an arrangement $[T]$ we may need to specify all or some of the lengths of segments of the hulls connecting the points in each group. Denoting the length of the segment $\mathbf{r}^{(a)}_i \leftarrow \mathbf{r}^{(a)}_j$ by $N^{(a)}_{ij}$, we can write the necessary probabilities of arrangements as
\onecolumngrid
\begin{align}
	p_{[T]}(\mathbf{r}_1,\ldots,\mathbf{r}_q) = p_{[T]}(\mathbf{r}_1^{(1)}, \ldots,\mathbf{r}_{q_1}^{(1)}; N_{12}^{(1)}, \ldots, N_{q_1,1}^{(1)}| \ldots | \mathbf{r}_1^{(n)},\ldots,\mathbf{r}_{q_n}^{(n)}; N_{12}^{(n)}, \ldots, N_{q_n,1}^{(n)}),
\end{align}
\twocolumngrid
\noindent
where the subscript $[T]$ reminds us that the points within each group can be cyclically permuted. The final forms of our percolation correlators involve the symmetrized expressions
\begin{align}
	p^{(s)}_\lambda(\mathbf{r}_1,\ldots,\mathbf{r}_q)
	&= \frac{1}{q!} \sum_{\sigma \in S_q} p_{[\sigma T]}(\mathbf{r}_1, \mathbf{r}_2, \ldots, \mathbf{r}_q).
\end{align}
The symmetrized probabilities clearly depend only of the partition of the points into groups, and so are labeled by Young diagrams $\lambda$.

In a similar way, we can denote generic averages of products of Green's functions. Let us consider the partition of points according to a Young tableau $T$ as described above. For each group of points $\mathbf{r}^{(a)}_j$ in a given row $a$ we define the following object:
\begin{align}
	D_{[(q_a)]}(\mathbf{r}_1^{(a)}, \ldots, \mathbf{r}_{q_a}^{(a)})
	&\equiv \frac{1}{(2\pi)^{q_a}} \Tr \bigg[\prod_{j=1}^{q_a}
	\Delta G\big(\mathbf{r}^{(a)}_{j}, \mathbf{r}^{(a)}_{j+1}\big) \bigg],
	\label{D_(q)}
\end{align}
where we assume $q_a+1 \equiv 1$ for the last term in the product. This trace is invariant under cyclic permutations of the points, and this is why it is labeled by the cyclic tabloid $[(q_a)]$. Then we take products of these traces for all the rows of the Young tableau $T$:
\begin{align}
	D_{[T]}(\mathbf{r}_1, \ldots, \mathbf{r}_q)
	&= \prod_{a=1}^{n} D_{[(q_a)]}\big(\mathbf{r}_1^{(a)}, \ldots, \mathbf{r}_{q_a}^{(a)}\big)
	\nonumber \\
	&= \frac{1}{(2\pi)^{q}} \prod_{a=1}^{n} \Tr \bigg[\prod_{j=1}^{q_a}
	\Delta G\big(\mathbf{r}^{(a)}_{j}, \mathbf{r}^{(a)}_{j+1}\big) \bigg].
\end{align}
We will also need the versions of these objects symmetrized over the rows of $T$:
\begin{align}
	D_{\{(q_a)\}}(\mathbf{r}_1^{(a)}, \ldots, \mathbf{r}_{q_a}^{(a)})
	&\equiv \frac{1}{q_a!} \sum_{\sigma \in S_{q_a}}
	D_{[(q_a)]}(\mathbf{r}_{\sigma(1)}^{(a)}, \ldots, \mathbf{r}_{\sigma(q_a)}^{(a)}),
	\nonumber \\
	D_{\{T\}}(\mathbf{r}_1, \ldots, \mathbf{r}_q)
	&\equiv \prod_{a=1}^{n} D_{\{(q_a)\}}(\mathbf{r}_1^{(a)}, \ldots, \mathbf{r}_{q_a}^{(a)})
	\nonumber \\
	&= \frac{1}{\prod_a q_a!}
	\sum_{\sigma \in R_\lambda} D_{[\sigma T]}(\mathbf{r}_1, \ldots, \mathbf{r}_q),
\end{align}
as well as over all permutations of the $q$ points:
\begin{align}
	D^{(s)}_\lambda(\mathbf{r}_1, \ldots, \mathbf{r}_q)
	&\equiv \frac{1}{q!} \sum_{\sigma \in S_q} D_{[\sigma T]}(\mathbf{r}_1, \ldots, \mathbf{r}_q) \label{eq:ds}.
\end{align}
These fully symmetrized expressions are completely determined by the Young diagram $\lambda$. Finally, we define the disorder averages
\begin{align}
	\mathcal{D}^{(s)}_\lambda(\mathbf{r}_1, \ldots, \mathbf{r}_q)
	&= \big\la D^{(s)}_\lambda(\mathbf{r}_1, \ldots, \mathbf{r}_q) \big\ra.
	\label{D_lambda}
\end{align}

\subsection{SUSY for the most irrelevant operators $\mathcal{P}^C_{(1^q)}$}
\label{subsec:SUSY-bottom}

The most irrelevant scaling operators $\mathcal{P}^C_{(1^q)}$ at the ``bottom'' of the generalized multifractal ``tower'' for a given $q$ are associated with the Young diagram which consists of a single vertical column of height $q$. As we have shown in Ref.~\cite{gruzberg2013classification}, in the case of class A, the multifractal scaling operators associated with such Young diagrams can be obtained within the minimal version of the SUSY formalism (with the minimal number of bosons and fermions), and also using only fermionic variables. It turns out, the same is true in class C, as we now demonstrate.

Consider the operator
\begin{align}
	\hat{A} &\equiv (\fdown - \fdup)(\fddown + \fup)
	= - 2\hat{Q}_3 - \hat{Q}_+ + \hat{Q}_-,
\end{align}
and its correlation functions $ \la \la \hat{A}(\mathbf{r}_1) \ldots \hat{A}(\mathbf{r}_q) \ra_\text{q} \ra$. We will first work out an expression for this correlator in terms of percolation probabilities, and then will find what combination of the Green's functions products $\mathcal{D}_\lambda$ it corresponds to. That combination will turn out to be exactly $\mathcal{P}^C_{(1^q)}$.

As before, to proceed with the mapping to percolation, we use the matrix
\begin{align}
	A = \begin{pmatrix}
		1 & 0 & 1 \\ 0 & 0 & 0 \\ -1 & 0 & -1
	\end{pmatrix}
\end{align}
that represents $\hat{A}$ in the fundamental representation of sl$(2|1)$. This matrix has a nice property: when it is multiplied by powers of the attenuation matrix $Z$, we get
\begin{align}
	A Z^N = \begin{pmatrix}
		1 & 0 & z^{2N} \\ 0 & 0 & 0 \\ -1 & 0 & -z^{2N}
	\end{pmatrix}.
\end{align}
It is easy to check that $AZ^{N_1} AZ^{N_2} = (1 - z^{2N_1}) AZ^{N_2}$ and, by induction,
\begin{align}
	\prod_{k=1}^{m} A Z^{N_k} &= \prod_{k=1}^{m-1} (1 - z^{2N_k}) A Z^{N_m}.
\end{align}
Taking the supertrace, we obtain
\begin{align}
	\str \prod_{k=1}^{m} A Z^{N_k} &= \prod_{k=1}^{m} (1 - z^{2N_k}).
\end{align}
Thus, the crucial suppression factors automatically appear for all percolation hulls that contribute to the correlation functions of $\hat{A}$.

Let us look at simple examples with $q \leq 3$:
\onecolumngrid
\begin{align}
	\la \la \hat{A}(\mathbf{r}) \ra_\text{q} \ra
	&= \sum_N \str [A Z^N] p(\mathbf{r},N)
	= \sum_N (1 - z^{2N}) p(\mathbf{r},N) = \pi \mathcal{D}_{(1)}(\mathbf{r})
	= \pi \mathcal{P}^C_{(1)}(\mathbf{r}),
	\label{hatA}
	\\
	\la \la \hat{A}(\mathbf{r}_1) \hat{A}(\mathbf{r}_2) \ra_\text{q} \ra
	&= \sum_{N_{12},N_{21}} \str[ A Z^{N_{12}} A Z^{N_{21}}] p(\mathbf{r}_1,\mathbf{r}_2;N_{12},N_{21})
	+ \sum_{N,N'} \str [A Z^{N}] \str [A Z^{N'}] p(\mathbf{r}_1,N|\mathbf{r}_2,N')
	\nonumber \\
	&= \sum_{N,N'} (1 - z^{2N})(1 - z^{2N'}) [p(\mathbf{r}_1, \mathbf{r}_2; N_{12}=N, N_{21}=N')
	+ p(\mathbf{r}_1;N |\mathbf{r}_2;N')]
	= \pi^2 \mathcal{P}^C_{(1,1)}(\mathbf{r}_1, \mathbf{r}_2),
	\label{hatA-hatA}
	\\
	\la \la \hat{A}(\mathbf{r}_1) \hat{A}(\mathbf{r}_2) \hat{A}(\mathbf{r}_3) \ra_\text{q} \ra
	&= \sum_{N, N', N''} (1 - z^{2N})(1 - z^{2N'})(1 - z^{2N''})
	\big[ 
	p^{(s)}_{(1^3)}(\mathbf{r}_1; N|\mathbf{r}_2; N'|\mathbf{r}_3; N'')
    \nonumber \\
	& \quad + 3 p^{(s)}_{(2,1)}(\mathbf{r}_1, \mathbf{r}_2; N_{12}=N,N_{21}=N'|\mathbf{r}_3;N'')
	\nonumber \\
	& \quad + 2 p^{(s)}_{(3)}(\mathbf{r}_1, \mathbf{r}_2, \mathbf{r}_3; N_{12}=N,N_{23}=N',N_{31}=N'') \big]
	= \pi^3 
    \mathcal{P}^C_{(1^3)}(\mathbf{r}_1, \mathbf{r}_2, \mathbf{r}_3).
	\label{hatA-hatA-hatA}
\end{align}
\twocolumngrid
This can be easily continued. The general pattern is that we get contributions from all
possible partitions $\lambda$ of $q$ points. Within each partition we get contributions of all
possible arrangements $[T]$ of points on percolation hulls, where the shape of the tableau $T$ is $\lambda$, as described in Sec.~\ref{sec:partitions-tableaux}. This ensures that all points are symmetrized in the final expression that can be written as
\begin{align}
	\Big \langle \Big\langle \prod_{k=1}^q A(\mathbf{r}_k) \Big\rangle_\text{q} \Big \rangle
	&= \sum_{\lambda \vdash q} d_\lambda \sum_{N_1,\ldots, N_q} \prod_{k=1}^q (1-z^{2N_k}) \nonumber \\
	& \times p^{(s)}_\lambda(\mathbf{r}_1,\ldots,\mathbf{r}_q).
\end{align}
Importantly, this expression contains the suppression factors $1- z^{2N_k}$ for all segments of percolation hulls involved, and we claim that this is the percolation representation for the ``bottom'' (the most irrelevant) pure scaling operator $\pi^q \mathcal{P}^C_{(1^q)}$.

To demonstrate this claim, we now work backward and determine what combinations of Green's functions are represented by the correlators of $\hat{A}$'s. To do this, let us denote
\begin{align}
	\hat{A} &= fg,
	&
	f &\equiv \fdown - \fdup,
	&
	g &\equiv \fddown + \fup.
\end{align}
It is easy to see that $\{f,f\} = \{f,g\} = \{g,g\} = 0$. We can then use Wick's theorem to compute correlators of $\hat{A} = fg$ in the free theory for a given disorder realization. The building blocks are the second quantized expectation values of all quadratic combinations of $f$ and $g$. For compactness, let us use subscripts to denote spatial points. Then we have
\begin{align}
	\la \hat{A}_1 \ra_\text{q} &= \la f_1 g_1 \ra_\text{q}
	= \la [\fdown(\mathbf{r}_1) - \fdup(\mathbf{r}_1)] [\fddown(\mathbf{r}_1) + \fup(\mathbf{r}_1)] \ra_\text{q}
	\nonumber \\
	&= \tr G_{11}(z) - 1 = \pi D_{(1)}(\mathbf{r}_1),
\end{align}
see Eq.~\eqref{D1}. Next we have
\begin{align}
	\la f_1 f_2 \ra_\text{q}
	&= G_{\down\up}(\mathbf{r}_2,\mathbf{r}_1; z) - G_{\down\up}(\mathbf{r}_1,\mathbf{r}_2; z)
	\nonumber \\
	&= G_{\down\up}(\mathbf{r}_1,\mathbf{r}_2;z^{-1}) - G_{\down\up}(\mathbf{r}_1,\mathbf{r}_2;z)
	\nonumber \\
	&= - \Delta G_{\down\up}(\mathbf{r}_1,\mathbf{r}_2)
	= \Delta G_{\down\up}(\mathbf{r}_2,\mathbf{r}_1).
\end{align}
In a similar way we obtain
\begin{align}
	\la g_1 g_2 \ra_\text{q}
	&= \Delta G_{\up\down}(\mathbf{r}_1, \mathbf{r}_2)
	= - \Delta G_{\up\down}(\mathbf{r}_2, \mathbf{r}_1),
	\nonumber \\
	\la f_1 g_2 \ra_\text{q}
	& = \Delta G_{\down\down}(\mathbf{r}_1, \mathbf{r}_2)
	= \Delta G_{\up\up}(\mathbf{r}_2,\mathbf{r}_1),
	\nonumber \\
	\la g_1 f_2 \ra_\text{q}
	&= - \Delta G_{\up\up}(\mathbf{r}_1,\mathbf{r}_2)
	= - \Delta G_{\down\down}(\mathbf{r}_2, \mathbf{r}_1).
\end{align}

Now we can compute
\begin{align}
	\la \hat{A}_1 \hat{A}_2 \ra_\text{q}
	&= \la \hat{A}_1 \ra_\text{q}  \la \hat{A}_2 \ra_\text{q}
	+ \la \hat{A}_1 \hat{A}_2 \ra_\text{q, c}.
\end{align}
Here
\begin{align}
	\la \hat{A}_1 \ra_\text{q} \la \hat{A}_2 \ra_\text{q}
	&= \pi^2 D_{(1)}(\mathbf{r}_1) D_{(1)}(\mathbf{r}_2)
	= \pi^2 D_{(1,1)}(\mathbf{r}_1, \mathbf{r}_2),
\end{align}
and in the second term the subscript ``c'' denotes the connected correlator, where the patterns of Wick contractions span all the points:
\begin{align}
	\la \hat{A}_1 \hat{A}_2 \ra_\text{q,c}
	&= \la f_1 g_1 f_2 g_2 \ra_\text{q,c}
	\nonumber \\
	&= \la f_1 g_2 \ra_\text{q} \la g_1 f_2 \ra_\text{q}
	- \la f_1 f_2 \ra_\text{q} \la g_1 g_2 \ra_\text{q}.
\end{align}
The two terms here can be written in various ways:
\begin{align}
	\la f_1 f_2 \ra_\text{q} \la g_1 g_2 \ra_\text{q}
	&= \Delta G_{\down\up}(\mathbf{r}_1,\mathbf{r}_2) \Delta G_{\up\down}(\mathbf{r}_2, \mathbf{r}_1)
	\nonumber \\
	&= \Delta G_{\up\down}(\mathbf{r}_1, \mathbf{r}_2) \Delta G_{\down\up}(\mathbf{r}_2,\mathbf{r}_1),
	\nonumber \\
	\la f_1 g_2 \ra_\text{q} \la g_1 f_2 \ra_\text{q}
	&= - \Delta G_{\up\up}(\mathbf{r}_1,\mathbf{r}_2) \Delta G_{\up\up}(\mathbf{r}_2,\mathbf{r}_1)
	\nonumber \\
	&= - \Delta G_{\down\down}(\mathbf{r}_1, \mathbf{r}_2) \Delta G_{\down\down}(\mathbf{r}_2, \mathbf{r}_1).
\end{align}
In fact, the most symmetric way is to use the average of the two expressions for each term, which gives
\begin{align}
	\la \hat{A}_1 \hat{A}_2 \ra_\text{q,c}
	&= - \frac{1}{2} \tr \Delta G_{12} \Delta G_{21}
	= - 2 \pi^2 D_{(2)}(\mathbf{r}_1, \mathbf{r}_2).
\end{align}
Finally, as expected, we get
\begin{align}
	\la \hat{A}_1 \hat{A}_2 \ra_\text{q}
	&= \pi^2 [D_{(1,1)}(\mathbf{r}_1, \mathbf{r}_2) - 2 D_{(2)}(\mathbf{r}_1, \mathbf{r}_2)]
	\nonumber \\
	&\equiv \pi^2 P^C_{(1,1)}(\mathbf{r}_1, \mathbf{r}_2).
\end{align}
After disorder average the left-hand side becomes the percolation correlator~\eqref{hatA-hatA}, while the right-hand side becomes $\pi^2 \mathcal{P}^C_{(1,1)}(\mathbf{r}_1, \mathbf{r}_2)$, see Eq.~\eqref{rg:perc}.

Let us consider $q=3$:
\onecolumngrid
\begin{align}
	\la \hat{A}_1 \hat{A}_2 \hat{A}_3 \ra_\text{q}
	&= \la \hat{A}_1 \ra_\text{q}\la \hat{A}_2 \ra_\text{q}\la \hat{A}_3 \ra_\text{q}
	+ 3\big[\la \hat{A}_1 \hat{A}_2 \ra_\text{q,c} \la \hat{A}_3 \ra_\text{q}\big]^{(s)}
	+ \la \hat{A}_1 \hat{A}_2 \hat{A}_3 \ra_\text{q,c}
	\nonumber \\
	&= \pi^3 D_{(1)}(\mathbf{r}_1) D_{(1)}(\mathbf{r}_2) D_{(1)}(\mathbf{r}_3)
	- 6 \pi^3 [D_{(2)}(\mathbf{r}_1, \mathbf{r}_2) D_{(1)}(\mathbf{r}_3)]^{(s)}
	+ \la \hat{A}_1 \hat{A}_2 \hat{A}_3 \ra_\text{q,c}
	\nonumber \\
	&= \pi^3 D_{(1,1,1)}(\mathbf{r}_1,\mathbf{r}_2,\mathbf{r}_3)
	- 6 \pi^3 D_{(2,1)}^{(s)}(\mathbf{r}_1, \mathbf{r}_2, \mathbf{r}_3)
	+ \la \hat{A}_1 \hat{A}_2 \hat{A}_3 \ra_\text{q,c}.
\end{align}
\twocolumngrid
\noindent
The new connected correlator here is $\la \hat{A}_1 \hat{A}_2 \hat{A}_3 \ra_\text{q,c} = \la f_1 g_1 f_2 g_2 f_3 g_3 \ra_\text{q,c}$. There are eight possible Wick pairings that span the three points. Each of the eight terms has two different representations in which the points follow the two possible sequences on the loop. Computing the terms as before, we notice that the sign factors work out in such a way that all representations in terms of $\Delta G$ come with a plus sign. Summing up the two possible orderings of the points on the hull, we get the traces:
\begin{align}
	\la \hat{A}_1 \hat{A}_2 \hat{A}_3 \ra_\text{q,c}
	&= \tr \Delta G_{12} \Delta G_{23} \Delta G_{31} = 8 \pi^3 D_{(3)}(\mathbf{r}_1, \mathbf{r}_2, \mathbf{r}_3)
	\nonumber \\
	&= \tr \Delta G_{13} \Delta G_{32} \Delta G_{21} = 8 \pi^3 D_{(3)}(\mathbf{r}_1, \mathbf{r}_3, \mathbf{r}_2).
\end{align}
The symmetrized form (the half-sum of the two expressions) is
\begin{align}
	\la \hat{A}_1 \hat{A}_2 \hat{A}_3 \ra_\text{q,c}
	&= 8 \pi^3 D_{(3)}^{(s)}(\mathbf{r}_1, \mathbf{r}_2, \mathbf{r}_3).
\end{align}
Finally, we get for $q=3$
\begin{align}
	\la \hat{A}_1 \hat{A}_2 \hat{A}_3 \ra_\text{q}
	&= \pi^3 [
    D^{(s)}_{(1^3)}(\mathbf{r}_1,\mathbf{r}_2,\mathbf{r}_3)
	- 6D^{(s)}_{(2,1)}(\mathbf{r}_1,\mathbf{r}_2,\mathbf{r}_3)
	\nonumber \\
	&\quad + 8 D^{(s)}_{(3)}(\mathbf{r}_1,\mathbf{r}_2,\mathbf{r}_3)]
	\equiv \pi^3 
    P^{C}_{(1^3)}(\mathbf{r}_1, \mathbf{r}_2, \mathbf{r}_3).
	\label{PC-111}
\end{align}
Again, upon averaging over disorder, the left-hand side becomes the percolation correlator~\eqref{hatA-hatA-hatA}, while the right-hand side becomes the pure scaling operator 
$\pi^3 \mathcal{P}^C_{(1^3)}(\mathbf{r}_1, \mathbf{r}_2, \mathbf{r}_3)$, 
see Eq.~\eqref{eq:q3-pure-scaling-matrix}.

This pattern continues for a general $q \geq 3$, and the counting works out in the following way. First, the connected correlator $\la \prod_{k=1}^q \hat{A}(\mathbf{r}_k) \ra_\text{q,c}$ has $2^{q-1} (q-1)!$ Wick contractions. The factor $(q-1)!$ counts the number of possible closed paths spanning all $q$ points, and the factor $2^{q-1}$ counts the number of choices of $f$ or $g$ at each point along a path (except the last point). The paths naturally break into $(q-1)!/2$ pairs, such that in each pair the two paths go through the points in exactly the opposite way. For $q=3$ there was one such pair of paths that went through the points as $1 \leftarrow 2 \leftarrow 3 \leftarrow 1$ and $1 \leftarrow 3 \leftarrow 2 \leftarrow 1$. Each of the $(q-1)!/2$ pairs gets $2^q$ contributions from different choices of $f$ and $g$, and each such contribution can be written in two ways according to the sequence of points on the corresponding loop. The signs of all contributions turn out to be $(-1)^{q-1}$, and summing them all we get (after the symmetrization over the members of the pairs)
\begin{align}
	\Big\la \prod_{k=1}^q \hat{A}(\mathbf{r}_k) \Big\ra_\text{q,c}
	&= \pi^q n_{(q)} D_{\{(q)\}}(\mathbf{r}_1,\ldots,\mathbf{r}_q),
\end{align}
where $n_{(q)} = (-2)^{q-1} d_{(q)} = (-2)^{q-1} (q-1)!$

A general correlator $\la \prod_{k=1}^q \hat{A}(\mathbf{r}_k) \ra_\text{q}$ can be written as the sum of contributions from all point configurations labelled by Young tabloids $\{T\}$:
\onecolumngrid
\begin{align}
	\Big\la \prod_{k=1}^q \hat{A}(\mathbf{r}_k) \Big\ra_{\text{q},\{T\}}
	&= \prod_{a=1}^{n} \big\la \hat{A}\big(\mathbf{r}_1^{(a)}\big) \ldots \hat{A}\big(\mathbf{r}_{q_a}^{(a)}\big) \big\ra_\text{q,c}
	= \prod_{a=1}^{n} \pi^{q_a} n_{(q_a)} D_{\{(q_a)\}}\big(\mathbf{r}_1^{(a)}, \ldots,\mathbf{r}_{q_a}^{(a)} \big)
	= \pi^q n_\lambda D_{\{T\}}(\mathbf{r}_1, \ldots, \mathbf{r}_q),
\end{align}
where $\lambda$ is the shape of the tableaux $T$, and
\begin{align}
	n_\lambda &= \prod_{a=1}^{n} n_{(q_a)}
	= (-2)^{q - l(\lambda)} \prod_{a=1}^{n} (q_a - 1)!
	= (-2)^{q - l(\lambda)} \prod_k [(k-1)!]^{m_k}.
\end{align}
We can group all such contributions by the partitions $\lambda$ of $q$. For each such partition there are $c_\lambda$ Young tabloids, and the summation over them leads to the symmetrization of all points, so that we get
\begin{align}
	\Big\la \prod_{k=1}^q \hat{A}(\mathbf{r}_k) \Big\ra_\text{q}
	&= \sum_{\lambda \vdash q} c_\lambda \Big\la \prod_{k=1}^q \hat{A}(\mathbf{r}_k) \Big\ra_{\text{q},\lambda}^{(s)}
	= \pi^q \sum_{\lambda \vdash q} c_\lambda n_\lambda D_\lambda^{(s)}(\mathbf{r}_1,\ldots,\mathbf{r}_q)
	\nonumber \\
	& = \pi^q \sum_{\lambda \vdash q} (-2)^{q - l(\lambda)} d_\lambda D_\lambda^{(s)}(\mathbf{r}_1,\ldots,\mathbf{r}_q)
	\equiv \pi^q P^{C}_{(1^q)}(\mathbf{r}_1,\ldots,\mathbf{r}_q).
	\label{eq:prod-A-q}
\end{align}
The disorder average of the right-hand side is exactly $\pi^q$ times the pure scaling operator $\mathcal{P}^{C}_{(1^q)}$, and the final result can be written as
\begin{align}
	\mathcal{P}^{C}_{(1^q)}(\mathbf{r}_1,\ldots,\mathbf{r}_q)
	&\equiv \sum_{\lambda \vdash q} (-2)^{q - l(\lambda)} d_\lambda \mathcal{D}_\lambda^{(s)}(\mathbf{r}_1,\ldots,\mathbf{r}_q)
	= \frac{1}{\pi^q} \sum_{\lambda \vdash q} d_\lambda \sum_{N_1,\ldots, N_q}
	\prod_{k=1}^q (1-z^{2N_k}) p^{(s)}_\lambda(\mathbf{r}_1,\ldots,\mathbf{r}_q)
	\label{eq:pc}.
\end{align}
\twocolumngrid
In full consistency with Eq.~\eqref{eq:pc}, the coefficients $(-2)^{q - l(\lambda)} d_\lambda$ appear in the RG result for the class-C pure-scaling operators, see Eq.~\eqref{eq:rg_C_result} in Appendix~\ref{appendix:rg}. The operator $\mathcal{P}^{C}_{(1^q)}$, in its NL$\sigma$M form, is given by the top row of the coefficient matrices appearing there. The functions $\mathcal{D}_\lambda^{(s)}$ from Eq.~\eqref{eq:pc} correspond to the $K-$invariant operators $O_\lambda$, Eq.~\eqref{eq:o_lambda}, in the NL$\sigma$M language.

Let us now analyze the scaling of the percolation expression in the right-hand-side of Eq.~\eqref{eq:pc}. Each factor $1-z^{2N_k}$ is close to unity for $ N_k \gg \gamma^{-1}$ and is suppressed for $ N_k \ll \gamma^{-1}$. Using Eq.~\eqref{eq:scaling}, we find that the contribution of the region where all $N_k$ are of the order of $\gamma^{-1}$ is governed by the scaling exponent $x_q^{\rm h}$, in full analogy with the cases $q=2$ ans $q=3$ analyzed in Sections~\ref{sec:perc-analyt-q2} and~\ref{sec:perc-analyt-q3}. One might, however, expect also contributions coming from the lower limits of sums over $N_k$ (i.e. $N_k \sim r^{\frac74}$) for some of $N_k$. These contributions would then form a hierarchy with scaling exponents $x_{q-j}^{\rm h}+i_j\frac74$, where $j$ and $i_j$ are positive integers and $i_j \ge j$, as can be obtained by expanding $1-z^{2N_k} \approx 2N_k\gamma + \ldots$ for $N_k \ll \gamma^{-1}$. We have already discussed in Sections~\ref{sec:perc-analyt-q2} and~\ref{sec:perc-analyt-q3}  that these contributions should in fact cancel, since there is no room for such additional exponents in the generalized-multifractality spectrum of SQH transition and since $\mathcal{P}^{C}_{(1^q)}$ should be a pure-scaling operator.

For $q > 3$, the situation is even more intriguing, since some of these additional contributions, if present, would dominate over the $x_q^{\rm h}$ term. Indeed, for $q>3$ we have $x_{q-1}^{\rm h} + \frac74 < x_q^{\rm h}$, so that the contribution from the operator $x_{q-1}^{\rm h}+\frac74$ would be more relevant than $x_q^{\rm h}$. This can be thought of as the $(q-1)$-hull operator
fusing with the energy operator with dimension $\frac74$. However, we argue again that all these additional contributions cancel out, as suggested by the analogy with the cases $q=2$ and $q=3$ and by the pure-scaling character of $\mathcal{P}^{C}_{(1^q)}$ known form the sigma-model RG. The numerical studies presented in the next section indeed confirm this expectation, by showing numerical evidence for $x_{(1^4)} = x_4^{\rm h}$ and $x_{(1^5)} = x_5^{\rm h}$. We thus argue that
\begin{align}
	x_{(1^q)} = x_q^{\rm h}.
\end{align}

By virtue of the Weyl symmetry, this result further has the implication that all of the $2^q q!$ scaling dimensions in the Weyl orbit of $x_{(1^q)}$ are equal to $x_q^{\rm h}$. In particular, this means that $x_{(2,1^{q-1})} = x_q^{\rm h}$, which can be nicely observed both in the classical analysis of Sections~\ref{sec:sqh_Pana} and~\ref{sec:sqh_Pnumerics} and in the quantum numerics in Sec.~\ref{sec:sqh_numerics}.

\section{Pure-scaling wave-function combinations}
\label{sec:scaling}

In order to determine the generalized-multifractality exponents numerically, one needs to construct wave-function observables that exhibit the corresponding scaling.
While the latter goal has been achieved in Ref.~\cite{karcher2021generalized}, the solution was not optimal. Specifically, the pure-scaling eigenfunction combinations found in Ref.~\cite{karcher2021generalized} did not have a definite sign before averaging, which required a very large number of disorder realizations
in order to get a meaningful result for the average.

Alternative paths towards a numerical determination of the exponents (see Sec.~\ref{sec:intro} for more details) have helped to partly resolve this complication and to find numerical values of several exponents. However, the ultimate goal in this context---a general construction of strictly positive pure-scaling wave-function combinations for an arbitrary Young diagram $\lambda=(q_1,q_2, \ldots, q_n)$---has not been achieved in Ref.~\cite{karcher2021generalized}. In this section, we present a solution for this problem.
In fact, $\lambda$ does not need to be a Young tableau in the conventional sense, as $q_j$ in the construction below do not need to be integer (in fact, they can even be complex).

The logic and the structure of this section are as follows. We consider first the case of the least relevant operator $\lambda = (1^q)$ in each order $q$. In Sec.~\ref{sec:scal_pos} we present strictly positive eigenfunction combinations for such $\lambda$, which have the form $P^C_{(1^q)}[\psi] = \det(M_q[\psi])$, with the matrix $M_q$ introduced below in Eq.~\eqref{eq:det}. Further, in Sec.~\ref{sec:scal_pure} we show that their disorder averages map to pure-scaling operators $\mathcal{P}^C_{(1^q)}[Q]$ satisfying Abelian fusion in the non-linear $\sigma$ model (NL$\sigma$M) framework,
\begin{align}
	\big\langle P^C_{(1^q)}[\psi] \big\rangle \equiv \mathcal{P}^C_{(1^q)}[\psi]
	\ \longleftrightarrow \  \mathcal{P}^C_{(1^q)}[Q].
	\label{eq:corr}
\end{align}
These NL$\sigma$M operators $\mathcal{P}^C_{(1^q)}[Q]$ are given by the Pfaffians of the $2q\times 2q$ replica sub-block of the advanced-advanced bosonic part of the rotated $Q$-field $\mathcal{Q}$ of the NL$\sigma$M
\begin{align}
	\mathcal{P}^C_{(1^q)}[Q] &= \mathrm{Pf}\begin{pmatrix}
		\left.\mathcal{Q}^{01}\right|_{q\times q} & \left.\mathcal{Q}^{00}\right|_{q\times q}\\
		\left.\mathcal{Q}^{11}\right|_{q\times q} &  \left.\mathcal{Q}^{10}\right|_{q\times q}
	\end{pmatrix}
	\equiv \mathrm{Pf}\big(\mathcal{Q}\big|_{2q\times 2q}\big);
	\label{eq:P1qQ-Pfaffian}
\end{align}
we refer the reader to Ref.~\cite{karcher2021generalized} for details. (See, in particular,  Eq. (313) of Ref.~\cite{karcher2021generalized}  and the text around it.)

A complementary proof of the pure-scaling character of the wave-function observables $P^C_{(1^q)}[\psi]$ constructed in Sec.~\ref{sec:scal_pos} is presented in Sec.~\ref{sec:scal_conn} where we demonstrate a direct correspondence between $P^C_{(1^q)}[\psi]$ and the pure-scaling combinations $P^C_{(1^q)}[G]$ of Green's functions  from Sec.~\ref{sec:susy} [see Eq.~\eqref{eq:prod-A-q}],
\begin{align}
	P^C_{(1^q)}[G] &= \sum_{\mu \vdash q} (-2)^{q - l(\mu)} d_\mu D_\mu^{(s)}(\mathbf{r}_1,\ldots,\mathbf{r}_q).
	\label{eq:green_scaling}
\end{align}

In Sec.~\ref{sec:scal_gen} the construction is extended to a generic $\lambda$, with $\lambda = (1^q)$ serving as building blocks.
The sigma-model operators~\eqref{eq:P1qQ-Pfaffian} satisfy the Abelian fusion property,
\begin{align}
	\mathcal{P}^C_{\lambda+\mu}[Q] &\sim \mathcal{P}^C_{\lambda}[Q] \mathcal{P}^C_{\mu}[Q] \,.
	\label{eq:general}
\end{align}
which allows one to construct~\cite{karcher2021generalized} the eigenoperators $\mathcal{P}^C_{\lambda}[Q]$ for any $\lambda$. Using the one-to-one correspondence between the eigenfunction and sigma-model observables, Eq.~\eqref{eq:corr}, we can obtain pure-scaling positive wave-function observables $P^C_{(1^q)}[\psi]$ for arbitrary $\lambda$. This construction bears analogy with that for class A that was developed in Ref.~\cite{gruzberg2013classification}.

\subsection{Scaling combinations $P^C_{(1^q)}[\psi]$ and positivity}
\label{sec:scal_pos}

In analogy with the class-A construction~\cite{gruzberg2013classification}, the $\lambda = (1^q)$ wave-function combination can be written as a Slater determinant (we denote the corresponding matrix with entries given by eigenfunction amplitudes by $M_q[\psi]$):
\onecolumngrid
\begin{align}
	P^C_{(1^q)}[\psi] = \det
	\left(M_q[\psi]\right)
	&  = \det
	\left(\begin{array}{c|c}
		(\psi_{i,\uparrow}(\RR{j}))_{q\times q} & (\psi_{-i,\uparrow}(\RR{j}))_{q\times q}\\
		\hline
		(\psi_{i,\downarrow}(\RR{j}))_{q\times q} & (\psi_{-i,\downarrow}(\RR{j}))_{q\times q}
	\end{array}\right) \nonumber\\[0.2cm]
	& \equiv \det
	\left(\begin{array}{ccc|ccc}
		\psi_{1,\uparrow}(\RR1) & \psi_{2,\uparrow}(\RR1) & \ldots & \psi_{-1,\uparrow}(\RR1) & \psi_{-2,\uparrow}(\RR1) & \ldots\\
		\psi_{1,\uparrow}(\RR2) & \psi_{2,\uparrow}(\RR2) & \ldots &
		\psi_{-1,\uparrow}(\RR2) & \psi_{-2,\uparrow}(\RR2) & \ldots\\
		\vdots & \vdots & \ddots &\vdots &\vdots & \ddots \\
		\hline
		\psi_{1,\downarrow}(\RR1) & \psi_{2,\downarrow}(\RR1) & \ldots &
		\psi_{-1,\downarrow}(\RR1) & \psi_{-2,\downarrow}(\RR1) & \ldots\\
		\psi_{1,\downarrow}(\RR2) & \psi_{2,\downarrow}(\RR2) & \ldots &
		\psi_{-1,\downarrow}(\RR2) & \psi_{-2,\downarrow}(\RR2) & \ldots\\
		\vdots & \vdots & \ddots &\vdots &\vdots & \ddots \\
	\end{array}\right).
	\label{eq:det}
\end{align}
\twocolumngrid
\noindent
Here $\psi_i$ with $i=1,2, \ldots,q$ are $q$ distinct eigenfunctions with positive energies. The energies should be close to zero to study the system at criticality;  a good choice is  to take wave functions corresponding to $q$ lowest positive eigenenergies. Further, $\RR{i}$ with $i=1,2, \ldots,q$ are $q$ distinct spatial points, while the subscripts $\uparrow$ and $\downarrow$ label the corresponding spin components of the wave function. The eigenfunction $\psi_{-i}$ is the partner of $\psi_i$ related by particle-hole conjugation, which implies
\begin{align}
	-\psi_{i,\downarrow}^*(\RR{} ) &= \psi_{-i,\uparrow}(\RR{} ), &
	\psi_{i,\uparrow}^*(\RR{} ) &= \psi_{-i,\downarrow}(\RR{} ).
	\label{eq:ph}
\end{align}
The energies of the wave functions $\psi_i$ and $\psi_{-i}$ are related via $\epsilon_{-i} = - \epsilon_i$.  Using Eq.~\eqref{eq:ph}, we can rewrite Eq.~\eqref{eq:det} in the form
\onecolumngrid
\begin{align}
	&P^C_{(1^q)}[\psi]
	=  \det
	\left(\begin{array}{c|c}
		(\psi_{i,\uparrow}(\RR{j}))_{q\times q} & (-\psi^*_{i,\downarrow}(\RR{j}))_{q\times q}\\
		\hline
		(\psi_{i,\downarrow}(\RR{j}))_{q\times q} & (\psi_{i,\uparrow}^*(\RR{j}))_{q\times q}
	\end{array}\right)
	=\det
	\left(\begin{array}{ccc|ccc}
		\psi_{1,\uparrow}(\RR1) & \psi_{2,\uparrow}(\RR1) & \ldots & -\psi_{1,\downarrow}^*(\RR1) & -\psi_{2,\downarrow}^*(\RR1) & \ldots\\
		\psi_{1,\uparrow}(\RR2) & \psi_{2,\uparrow}(\RR2) & \ldots &
		-\psi_{1,\downarrow}^*(\RR2) & -\psi_{2,\downarrow}^*(\RR2) & \ldots\\
		\vdots & \vdots & \ddots &\vdots &\vdots & \ddots \\
		\hline
		\psi_{1,\downarrow}(\RR1) & \psi_{2,\downarrow}(\RR1) & \ldots &
		\psi_{1,\uparrow}^*(\RR1) & \psi_{2,\uparrow}^*(\RR1) & \ldots\\
		\psi_{1,\downarrow}(\RR2) & \psi_{2,\downarrow}(\RR2) & \ldots &
		\psi_{1,\uparrow}^*(\RR2) & \psi_{2,\uparrow}^*(\RR2) & \ldots\\
		\vdots & \vdots & \ddots &\vdots &\vdots & \ddots \\
	\end{array}\right).
	\label{eq:det1}
\end{align}
\twocolumngrid
\noindent
It follows from the representation~\eqref{eq:det1} that these  determinants are real and positive for any disorder realization. Indeed, the matrix $M_q[\psi]$ of wave-function amplitudes in Eq.~\eqref{eq:det1}  inherits the particle-hole symmetry~\eqref{eq:ph}, yielding
\begin{equation}
	\sigma_2 M_q[\psi]^*\sigma_2 = M_q[\psi].
\end{equation}
This implies that eigenvalues of $M_q$ come in complex conjugate pairs $z_i, z_i^*$. Therefore, the determinant is given by $\det(M_q)=\prod_i |z_i|^2$ and is thus real and positive.

Let us discuss in more detail analogies and differences between this construction and that for class A developed in Ref.~\cite{gruzberg2013classification}. In the case of class A, the $\lambda = (1^q)$ pure-scaling wave-function observables read $P^A_{(1^q)}[\psi] = |\det M^A_q[\psi] |^2$, where $\det M^A_q[\psi]$ is the Slater determinant of a $q\times q$ matrix $M^A_q[\psi]$ built on $q$ wave functions at $q$ distinct points. This determinant is a complex number, and its absolute value squared yields the sought $\lambda = (1^q)$ observable.  On the other hand, in class C, the matrix $M_q$ has the size $2q \times 2q$; the doubling reflects the presence of the spin index and of particle-hole partner wave functions. Now, the Slater determinant itself has a real positive value and provides the $\lambda = (1^q)$ wave-function observable. In both cases of class A and class C, the Slater determinant structure ensures the full antisymmetrization corresponding to the $\lambda = (1^q)$ Young diagram.

\subsection{Pure-scaling property of $P^C_{(1^q)}[\psi]$:  Mapping to the sigma model}
\label{sec:scal_pure}

Now we provide a formal proof of the pure $(1^q)$ scaling character of the wave-function combination $P^C_{(1^q)}[\psi]$ defined above. For this purpose, we expand the determinant~\eqref{eq:det} using the Leibnitz formula and group the resulting $(2q)!$ terms into $(2q)!!=2^qq!$ Pfaffians of $2q\times 2q$ matrices $A_q$ defined below. From Ref.~\cite{karcher2021generalized}, we know that such Pfaffians satisfy pure $(1^q)$ scaling.

According to the Leibnitz formula, we  have the following expansion of Eq.~\eqref{eq:det}:
\begin{align}
	P^C_{(1^q)}[\psi]
	&= \sum_{\pi\in S_{2q}} \mathrm{sign}(\pi)
	\prod_{i=1}^{q} \psi_{\pi(i),\uparrow}(\mathbf{r}_i)\psi_{\pi(-i),\downarrow}(\mathbf{r}_i),
	\label{eq:det2}
\end{align}
where $S_{2q}$ is the permutation group for the set of integers $-q,\ldots, -1, 1,\ldots ,q$.
We define the bilinears $\mathcal{A}_{i,j}(\RR1)$ via
\begin{align}
	&\mathcal{A}_{i,j}(\RR1) = \psi_{i,\uparrow}(\RR1) \psi_{j,\downarrow}(\RR1).
\end{align}
In view of the particle-hole conjugation symmetry~\eqref{eq:ph}, these bilinears satisfy the symmetry relation
\begin{align}
	&\mathcal{A}_{i,j}(\RR1)  = -\mathcal{A}_{-j,-i}^*(\RR1).
	\label{eq:A-ph-symmetry}
\end{align}

We define now a $2q \times 2q$ matrix $A_q^{+\cdots +}[\psi]$  built out of bilinears $\mathcal{A}_{i,j}$:
\onecolumngrid
\begin{align}
	A_q^{+\cdots +}[\psi] &=
	\begin{pmatrix}
		A_{q,+-}& A_{q,++}\\
		A_{q,--}& A_{q,-+}
	\end{pmatrix}
	\quad
	\begin{array}{c c c} (A_{q,++})_{ij}=\mathcal{A}_{i,q-j+1},  \ & (A_{q,+-})_{i<j}=\mathcal{A}_{i,-j},  \ & (A_{q,-+})_{i<j}=\mathcal{A}_{-i,j},\\
		(A_{q,--})_{ij}=\mathcal{A}_{q-j+1,i}^*,  \ & (A_{q,+-})_{i>j}=\mathcal{A}_{j,-i}^*,  \ & (A_{q,-+})_{i>j}=\mathcal{A}_{-j,i}^*.
		\label{A_q^++matrix}
	\end{array}
\end{align}
This matrix has thus the following form:
\begin{align}
	A_q^{+\cdots +}[\psi] &= \left(\begin{array}{cccc|cccc}
		0 & \mathcal{A}_{1,-2} & \mathcal{A}_{1,-3} & \ldots & \ldots &\mathcal{A}_{1,3} & \mathcal{A}_{1,2} & \mathcal{A}_{1,1}\\
		\mathcal{A}^*_{2,-1} & 0 & \mathcal{A}_{2,-3} & \ldots & \ldots &
		\mathcal{A}_{2,3} &
		\mathcal{A}_{2,2} & \mathcal{A}_{2,1}\\
		\mathcal{A}^*_{3,-1} & \mathcal{A}^*_{3,-2} & 0& \ldots & \ldots &
		\mathcal{A}_{3,3} &
		\mathcal{A}_{3,2} & \mathcal{A}_{3,1}\\
		\vdots & \vdots & \vdots & \ddots & \iddots &\vdots & \vdots & \vdots \\
		\hline
		\vdots & \vdots & \vdots & \iddots & \ddots &\vdots & \vdots & \vdots \\
		\mathcal{A}_{-3,-1}^* & \mathcal{A}_{-3,-2}^* & \mathcal{A}_{-3,-3}^* & \ldots &
		\ldots & 0 & \mathcal{A}_{-3,2} & \mathcal{A}_{-3,1}\\
		\mathcal{A}_{-2,-1}^* & \mathcal{A}_{-2,-2}^* & \mathcal{A}_{-2,-3}^* & \ldots &
		\ldots& \mathcal{A}^*_{-2,3} & 0 & \mathcal{A}_{-2,1}\\
		\mathcal{A}_{-1,-1}^* & \mathcal{A}_{-1,-2}^* & \mathcal{A}_{-1,-3}^* & \ldots  & \ldots &
		\mathcal{A}_{-1,3}^* & \mathcal{A}_{-1,2}^* & 0\\
	\end{array}\right).
	\label{A_q^++matrix-explicit}
\end{align}
\twocolumngrid
\noindent
Here we do not specify the coordinates (this will be done below), i.e., all $\mathcal{A}_{i,j}$ are understood at this stage as functions
$\mathcal{A}_{i,j}: \mathbb{R}^2\rightarrow\mathbb{C}$. In view of Eq.~\eqref{eq:A-ph-symmetry}, the matrix $A_q^{+\cdots +}[\psi]$ is antisymmetric.
Further, we define a more generic matrix $A_q^{s_1\cdots s_q}[\psi]$, with $s_i  = +, -$. It is obtained from $A_q^{+\cdots +}[\psi]$ by flipping the sign of the index $i$ (i.e., exchanging $i\leftrightarrow -i$) in the right-hand side of Eq.~\eqref{A_q^++matrix-explicit} each time that we have $s_i = -$.
The Pfaffian of $A_q^{s_1\cdots s_q}[\psi]$ can be written as a homogeneous polynomial of degree $q$ with respect to the entries $\mathcal{A}_{i,j}$. In view of the antisymmetry property of $A_q^{s_1\cdots s_q}[\psi]$, the Pfaffian can be expressed using only the entries in the upper triangle of $A_q^{s_1\cdots s_q}[\psi]$. Since all $(2q-1)!!$ terms in the Pfaffian are of degree $q$, we can now assign to them the spatial arguments $\RR1, \ldots, \RR{q}$. The freedom that appears in this assignment is not essential, since we will be only interested in the Pfaffian symmetrized with respect to permutations of the coordinates.
The determinant in Eq.~\eqref{eq:det2} can be now written as a sum of Pfaffians of matrices $A_q^{s_1\cdots s_q}[\psi]$:
\begin{align}
	P^C_{(1^q)}[\psi] &=  \! \! \sum_{\sigma\in S_q}\sum_{s_1,\ldots, s_q=\pm} \!\! \mathrm{Pf} \left(A_q^{s_1\ldots s_q}[\psi]\right) (\RR{\sigma(1)},\ldots, \RR{\sigma(q)}).
	\label{eq:PC-1q-Pfaffian}
\end{align}
The summation here goes over $2^q$ possible choices of the $q$ signs $s_i$ and over $q!$ permutations of the positions $\RR{i}$. Since $2^q q! (2q-1)!! = (2q)!$, we have exactly the correct number of terms (2q)! in the expansion of the determinant. For $q=1,2$, the identity~\eqref{eq:PC-1q-Pfaffian} is easily checked by an explicit calculation. For arbitrary $q$, it can be proved by induction using the row/column expansion of the Pfaffian and the determinant.

We use now the correspondence between the disorder-averaged wave-function combinations and NL$\sigma$M operators derived in Ref.~\cite{karcher2021generalized}; see Eqs.~(275)--(281) there. Spe\-ci\-fically, the local bilinears $\mathcal{A}_{i,j}$ are in one-to-one correspondence with the components $\mathcal{Q}_{ij}$ of the rotated NL$\sigma$M field $\mathcal{Q}$. Using this ``translation dictionary'', we find that each of the Pfaffians of $A_q$ in Eq.~\eqref{eq:PC-1q-Pfaffian} translates to a pure-scaling Pfaffian of the rotated NL$\sigma$M field $\mathcal{Q}$,
Eq.~\eqref{eq:P1qQ-Pfaffian}.
This completes the proof of the pure-scaling character of $P^C_{(1^q)}[\psi]$.

\subsection{Pure-scaling property of $P^C_{(1^q)}[\psi]$:  Connection to the correlators $D^{(s)}_{(1^q)}$}
\label{sec:scal_conn}

The Leibnitz representation~\eqref{eq:det2} can also be used to establish a connection to the pure-scaling $(1^q)$ Green's-function observables~\eqref{eq:green_scaling} derived in Sec.~\ref{sec:susy} using the supersymmetry formalism. We introduce bilinears corresponding to Green's functions (without the energy denominator):
\begin{align}
	\mathcal{G}_{\alpha,\beta}(\RR{i},\RR{j}; a)
	&\equiv \psi_{a,\alpha}(\RR{i}) \psi_{a,\beta}^*(\RR{j})
	\nonumber\\
	&= (-1)^{\alpha+\beta}\psi_{-a,-\alpha}^*(\RR{i}) \psi_{-a,-\beta}(\RR{j})
	\nonumber\\
	&= (-1)^{\alpha+\beta}\mathcal{G}_{-\beta,-\alpha}(\RR{j},\RR{i}; -a).
	\label{G-symmetry}
\end{align}
In analogy with Sec.~\ref{sec:scal_pure}  where it was convenient to consider the bilinears $\mathcal{A}_{i,j}$ as functions $\mathbb{R}^2\rightarrow\mathbb{C}$ (i.e., without specifying the spatial coordinate), here it is convenient to consider bilinears $\mathcal{G}_{\sigma,\sigma'}(\RR{i},\RR{j};\cdot)$ with given coordinates but without specifying the wave-function (i.e., the energy) index, i.e. as functions $\{1,\ldots q\}\rightarrow\mathbb{C}$.  We define an antisymmetric matrix of such bilinears:
\onecolumngrid
\begin{align}
	G_q^{+\cdots +}[\psi] = \left(\begin{array}{c c}
		G_{q,\uparrow\downarrow} & G_{q,\uparrow\uparrow} \\
		G_{q,\downarrow\downarrow} & G_{q,\downarrow\uparrow}
	\end{array}\right),
	&&
	\begin{array}{l l}
		(G_{q,\uparrow\uparrow})_{ij} = \mathcal{G}_{\uparrow\uparrow}(\RR{i},\RR{q-j+1};\cdot),
		&
		(G_{q,\downarrow\downarrow})_{ij} = -\mathcal{G}_{\downarrow\downarrow}(\RR{q-j+1},\RR{i};-\cdot),
		\\
		(G_{q,\uparrow\downarrow})_{i<j} = \mathcal{G}_{\uparrow\downarrow}(\RR{i},\RR{j};\cdot),
		&
		(G_{q,\uparrow\downarrow})_{i>j} = \mathcal{G}_{\uparrow\downarrow}(\RR{j},\RR{i};-\cdot),
		\\
		(G_{q,\downarrow\uparrow})_{i<j} = \mathcal{G}_{\downarrow\uparrow}(\RR{i},\RR{j};\cdot),
		&
		(G_{q,\downarrow\uparrow})_{i>j} = \mathcal{G}_{\downarrow\uparrow}(\RR{j},\RR{i};-\cdot).
	\end{array}
\end{align}
\twocolumngrid
The Pfaffian of $ \mathrm{Pf}  \left( G_q^{+\cdots +} \right) $ is a homogeneous polynomial of order $q$ that can be expressed using only the entries in the upper triangle of $G_q$. Considering all possible assignments of $q$ energy indices in each of the $(2q-1)!!$  terms of the Pfaffian, we will get a function $(\{1,\ldots, q\})^q\rightarrow\mathbb{C}$. The relevant assignments are $q!$ permutations of $(1, \ldots, q)$; in analogy with permutations of coordinates in Sec.~\ref{sec:scal_pure}, we will be interested in the Pfaffian symmetrized over all of them. Thus, again in analogy with Sec.~\ref{sec:scal_pure}, the freedom in assignment of the energy indices is immaterial. The determinant in Eq.~\eqref{eq:det2} can be written as a sum of $2^q q!$ Pfaffians of matrices $G_q^{s_1\cdots s_q}[\psi]$ over the $q!$ permutations $\sigma\in S_q$ of the energy indices $1,\ldots, q$ and over $q$ spin indices $s_i = \pm$:
\begin{align}
	P^C_{(1^q)}[\psi] &= \!\! \sum_{\sigma\in S_q}\sum_{s_1,\ldots, s_q=\pm} \!\! \mathrm{Pf} \left(G_q^{s_1\cdots s_q} [\psi]\right) (\sigma(1),\ldots, \sigma(q)).
	\label{eq:PC-sum-Pf}
\end{align}
Here $G_q^{s_1\cdots s_q}[\psi]$ is obtained from $G_q^{+\cdots +}[\psi]$ by flipping the spins at $\RR{i}$  whenever $s_i=-$.

We can now classify the terms in the Pfaffian and regroup them to recover expression structurally analogous to $D^{s}_\lambda$. (It is worth mentioning that $D^{s}_\lambda$ are composite objects of Green's functions and contain energy denominators not present here. This difference however plays no role for the symmetry classification.)
{  We begin by focussing on the term with the spin multiindex $s_1\ldots s_q$ equal to $+\ldots+$; other terms are treated analogously and will be included afterwards.}
First, note that among the $(2q-1)!!$ terms in the Pfaffian, there are $q!$ terms forming the determinant of the upper right sub-block $G_{q,\uparrow\uparrow}$. These $q!$ terms {  have all } spins $s_1,\ldots, s_q$ {  up} at positions $\RR{1},\ldots,\RR{q}$, and each of them is associated, by virtue of the Leibnitz formula, with a certain permutation $\tau \in S_q$:
\begin{align}
	\det(G_{q,\uparrow\uparrow}) &= (-1)^{q+1}\sum_{\tau\in S_q}\mathrm{sign}(\tau) \prod_{i=1}^q(G_{q,\uparrow\uparrow})_{q-i+1,\tau(i)}.
	\label{eq:Pfaffian-det}
\end{align}
Let $\mu=(\mu_1,\ldots,\mu_n)$ denote the cycle lengths of $\tau$; the number of cycles is $n \equiv l(\mu)$. We can transform the sum into that over the
partitions $\mu\vdash q$ labeling the cycle decomposition of $\tau$. In the determinant part of the Pfaffian, Eq.~\eqref{eq:Pfaffian-det}, there are $d_\mu$ terms corresponding to each $\mu$, with $d_\mu$ given by Eq.~\eqref{d_lambda}.

Now we include other terms from the Pfaffian, which are not included in Eq.~\eqref{eq:Pfaffian-det}. The essential point is the $\mu$-dependent counting factor. Specifically, for each permutation $\tau$ with $l(\mu)$ cycles, there are in total $2^{q-l(\mu)}$ terms in the Pfaffian that differ in spin structure only, as we are going to explain. Starting with the determinant term, we can obtain other terms by flipping some of $\mu_k-1$ spins in the cycle $k$. To clarify this, we use the row/column expansion of the Pfaffian of an antisymmetric $2q\times 2q$ matrix $A$,
\begin{align}
	\mathrm{Pf}(A)
	&= \sum_{\substack{j=1 \\ j\neq i}}^{2q}
	(-1)^{i+j+1+\Theta(i-j)}A_{ij}\mathrm{Pf}(A_{\hat{\imath}\hat{\jmath}}),
	\label{eq:pf}
\end{align}
where the index $i$ is an arbitrary integer between $1$ and $2q$ and $\Theta(\cdot)$ is the Heaviside step function. In each term in the sum over $j$, we have to remove both $i$-th and $j$-th rows and columns; the resulting $2(q-1) \times 2(q-1)$ matrix is denoted by $A_{\hat{\imath}\hat{\jmath}}$.  Using Eq.~\eqref{eq:pf} recursively, one obtains an expansion of the Pfaffian analogous to the Leibnitz formula for the determinant.

Let $\tau\in S_q$ be a permutation with at least one non-trivial cycle (of length larger than unity). Picking $q$ pairs of integers $i_k = k$,  $j_k = 2q-\tau(k)$  (with $k=1,\ldots, q$) in the expansion  of the Pfaffian of $A=G_q^{+\cdots +}$ generated by Eq.~\eqref{eq:pf}, we arrive at the term in determinant~\eqref{eq:Pfaffian-det} associated to $\tau$. The same permutation generates in total $2^{q-l(\mu)}$ terms in the Pfaffian, where $\mu$ is the cycle class of $\tau$.
To make this clear, let us focus on a segment $s\rightarrow t\rightarrow u$ of one of the cycles,  which means that $\tau(s)= t$ and $\tau^2(s)= u$. For example, a cycle of length two would have $u=\tau^2(s)=s$.
In the above expansion of the determinant part of the Pfaffian, we will have the corresponding factors $A_{i_s j_s} A_{i_{\tau(s)} j_{\tau(s)}}$, with
$i_s = s$, $j_s = 2q-\tau(s)$, and $i_{\tau(s)}= \tau(s)$, $ j_{\tau(s)}= 2q-\tau^2(s)$.  If we start the recursion~\eqref{eq:pf}
with these two steps, we will thus get a factor $A_{s,2q-t}A_{t,2q-u}$ (a product of two entries from the upper right block of the matrix) and would be left with $A_{\hat{s}\hat{t}\widehat{2q-u}\widehat{2q-t}}$ to continue the recursion.

Alternatively, we can take $i_s = s$, $j_s = \tau(s)$, and $i_{\tau(s)}= 2q-\tau(s)$, $ j_{\tau(s)}= 2q-\tau^2(s)$, giving us the factor $A_{s,t}A_{2q-t,2q-u}$, i.e., a product of two matrix elements from the diagonal blocks. In either case, the rows and columns $s,t, 2q-t,2q-u$ are deleted, so that we continue the recursion with the matrix $A_{\hat{s}\hat{t}\widehat{2q-u}\widehat{2q-t}}$. The second option corresponds to picking the terms $\mathcal{G}_{\uparrow\downarrow}(\RR{s},\RR{t};\cdot) \, \mathcal{G}_{\downarrow\uparrow}(\RR{t},\RR{u};\cdot)$ instead of $\mathcal{G}_{\uparrow\uparrow}(\RR{s},\RR{t};\cdot) \, \mathcal{G}_{\uparrow\uparrow}(\RR{t},\RR{u};\cdot)$, which effectively flips the spin at position $t$. It is a simple exercise to verify that the total sign picked up in the recursion is equal to $\mathrm{sign}(\tau) =(-1)^{q-l(\lambda)}$ in any path associated to $\tau$ chosen through the recursion.

What we have just discussed is an elementary building block for the procedure that allows us to construct the expansion for the Pfaffian from that for the determinant of the upper right block of the matrix. For each cycle, there are $2^{\mu_k-1}$ choices analogous to the above in the expansion, giving  in total $2^{q-l(\mu)}$ terms with different spin configurations in the cycles originating from a permutation $\tau$ with cycle class $\mu$, as was stated above.

Hence, after the additional symmetrization over positions and spin configurations [i.e. taking all $A=G_q^{s_1\cdots s_q}$ into account, see Eq.~\eqref{eq:PC-sum-Pf}], there are $m_\mu = 2^{q-l(\mu)} d_\mu$ terms, each of which contributes an analog of $D^{(s)}_\mu(\mathbf{r}_1,\ldots,\mathbf{r}_q)$ from~\eqref{eq:ds} but without energy denominators [we denote it by $\tilde{D}_\mu^{(s)}(\mathbf{r}_1,\ldots,\mathbf{r}_q)$].
Thus, we obtain
\begin{align}
	P^C_{(1^q)}[\psi] &= \sum_{\mu \vdash q} (-2)^{q - l(\mu)} d_\mu \tilde{D}_\mu^{(s)}(\mathbf{r}_1,\ldots,\mathbf{r}_q).
	\label{eq:green_scaling2}
\end{align}
This is the direct analog of Eq.~\eqref{eq:prod-A-q} from Sec.~\ref{sec:susy}.
As was already pointed out below Eq.~\eqref{eq:pc}, the coefficients $(-2)^{q - l(\lambda)} d_\lambda$ appear also in NL$\sigma$M  RG analysis in Appendix~\ref{appendix:rg}. The operator $\mathcal{P}^{C}_{(1^q)}[Q]$, which is the NL$\sigma$M counterpart of $P^C_{(1^q)}[\psi]$, is given by the top row of the coefficient matrices in Eq.~\eqref{eq:rg_C_result}. The functions $\tilde{D}_\mu^{(s)}$ from Eq.~\eqref{eq:green_scaling2} map to the $K-$invariant operators $O_\mu$, Eq.~\eqref{eq:o_lambda}, in the NL$\sigma$M formalism.

In summary, the $(2q)!$ terms in the sum over $S_{2q}$ in the left-hand-side of Eq.~\eqref{eq:green_scaling2} [the Leibnitz expansion of the determinant, Eq.~\eqref{eq:det2}]  are regrouped in the right-hand side in accordance with the decomposition $(2q)! = q! \times 2^q \times (2q-1)!!$. Here the factor $q!$ comes from the position symmetrization, the factor $2^q$ from the spin traces, and the factor $(2q-1)!!$ is the total number of terms, $\sum_\mu m_\mu = (2q-1)!!$, in all partitions $\mu\vdash q$.

\subsection{Pure scaling combinations for generic $\lambda$}
\label{sec:scal_gen}

In Sec.~\ref{sec:scal_pos}, we have defined
wave-function observables $P^C_{(1^q)}[\psi]$, Eq.~\eqref{eq:det}, for the
``vertical'' partitions
$\lambda = (1^q)$.  To demonstrate that they indeed exhibit the $\lambda = (1^q)$ pure-scaling, we used  in Sec.~\ref{sec:scal_pure} the NL$\sigma$M mapping. Now we extend
this construction to generic representations characterized by
abritrary (complex) highest weights
$\lambda = (q_1,q_2,\ldots,q_n)$. To do this, we use the one-to-one mapping between the
wave-function and the NL$\sigma$M observables,
Sec.~\ref{sec:scal_pure}, as well as the Abelian fusion property~\eqref{eq:general} of the NL$\sigma$M observables that have the form of Pfaffians of the rotated NL$\sigma$M field $\mathcal{Q}$, as  obtained in the Iwasawa decomposition construction of Ref.~\cite{karcher2021generalized}.

For convenience, we introduce a short-hand notation for the determinants~\eqref{eq:det}:
\begin{align}
	f_q & =\det(M_q[\psi]).
\end{align}
We remind that these determinants are real and positive.
Then, using the Abelian fusion property~\eqref{eq:general}, we obtain the following
pure-scaling wave-function observable for an arbitrary weight
(an $n$-tuple of complex numbers) $\lambda= (q_1,\ldots, q_n)$:
\begin{align}
	P^C_\lambda[\psi] &= f_1^{q_1-q_2} f_2^{q_2-q_3}\cdots f_{n-1}^{q_{n-1}-q_n}  f_{n}^{q_n}.
	\label{eq:obs}
\end{align}
To construct the observable~\eqref{eq:obs}, we need $n$ spatial points $\RR1, \ldots, \RR{n}$ and $n$ wave functions $\psi_j$ together with their particle-hole-conjugation partners $\psi_{-j}$, with $j = 1, \ldots, n$. The determinant $f_q$ with $q \le n$ is built using the first $q$ wave functions ($j=1, \ldots, q$) and the first $q$ spatial points  $\RR1, \ldots, \RR{q}$.

The formula~\eqref{eq:obs} has the same form as an analogous expression for class A (derived in Ref.~\cite{gruzberg2013classification}), with the difference in the building block $f_q$ [corresponding to $\lambda = (1^q)$]  discussed in the end of Sec.~\ref{sec:scal_pos}. Our preliminary results indicate that, when taken together, two constructions (that for class A from Ref.~\cite{gruzberg2013classification}) and the one for class C derived here) permit to describe pure-scaling eigenfunction observables in all symmetry classes. Specifically,  for five spinless classes A, AI, AIII, BDI and D, the class-A construction from Ref.~~\cite{gruzberg2013classification} applies. On the other hand,
the other five classes AII, C, CII, CI, DIII possess a (pseudo)spin  in the sense that they obey either the time-reversal symmetry $\mathcal{T}$ with $\mathcal{T}^2=-1$ or the particle-hole symmetry $\mathcal{P}$ with $\mathcal{P}^2=-1$ and, therefore, exhibit Kramers degeneracy (at zero energy). In this situation, the pure-scaling wave-function construction developed here is applicable. We leave a detailed investigation of other symmetry classes to a future work, see also the outlook in Sec.~\ref{sec:conclusions}.

In Sec.~\ref{sec:sqh_numerics}, we will verify the construction of pure-scaling wave-function observables developed above by means of numerical simulations on an SU(2) version of the CCN model. This will allow us to efficiently determine numerical values of the generalized-multifractality exponents at the SQH transition.

\section{SU(2) Chalker-Coddington network numerics}
\label{sec:sqh_numerics}

In this section, we determine the generalized-multifracality exponents at the SQH transition numerically, using strictly positive pure-scaling wave-function observables
constructed in Sec.~\ref{sec:scaling}, see Eqs.~\eqref{eq:det} and~\eqref{eq:obs}. For numerical simulations, we use the  SU(2) version of the Chalker-Coddington network (CCN)~\cite{chalker1988percolation}. Network models of the CCN type  have been efficiently exploited for numerical studies of 2D localization transitions of various symmetry classes, see the reviews~\cite{kramer2005random, evers08}. We study  ensembles of systems of linear sizes $L=64,96,\ldots, 1024$, with $10^4$ disorder realizations for each linear system size $L$. All observables are averaged both spatially over all points of the system ($\sim L^2$) and over all disorder realizations in the ensemble ($10^4$). Thus, the total number of values over which the average is taken ranges from $\sim 10^8$ for smaller systems to $\sim 10^{10}$ for the largest systems.

\begin{figure*}
	\noindent\begin{center}
		\includegraphics[width=0.42\linewidth]{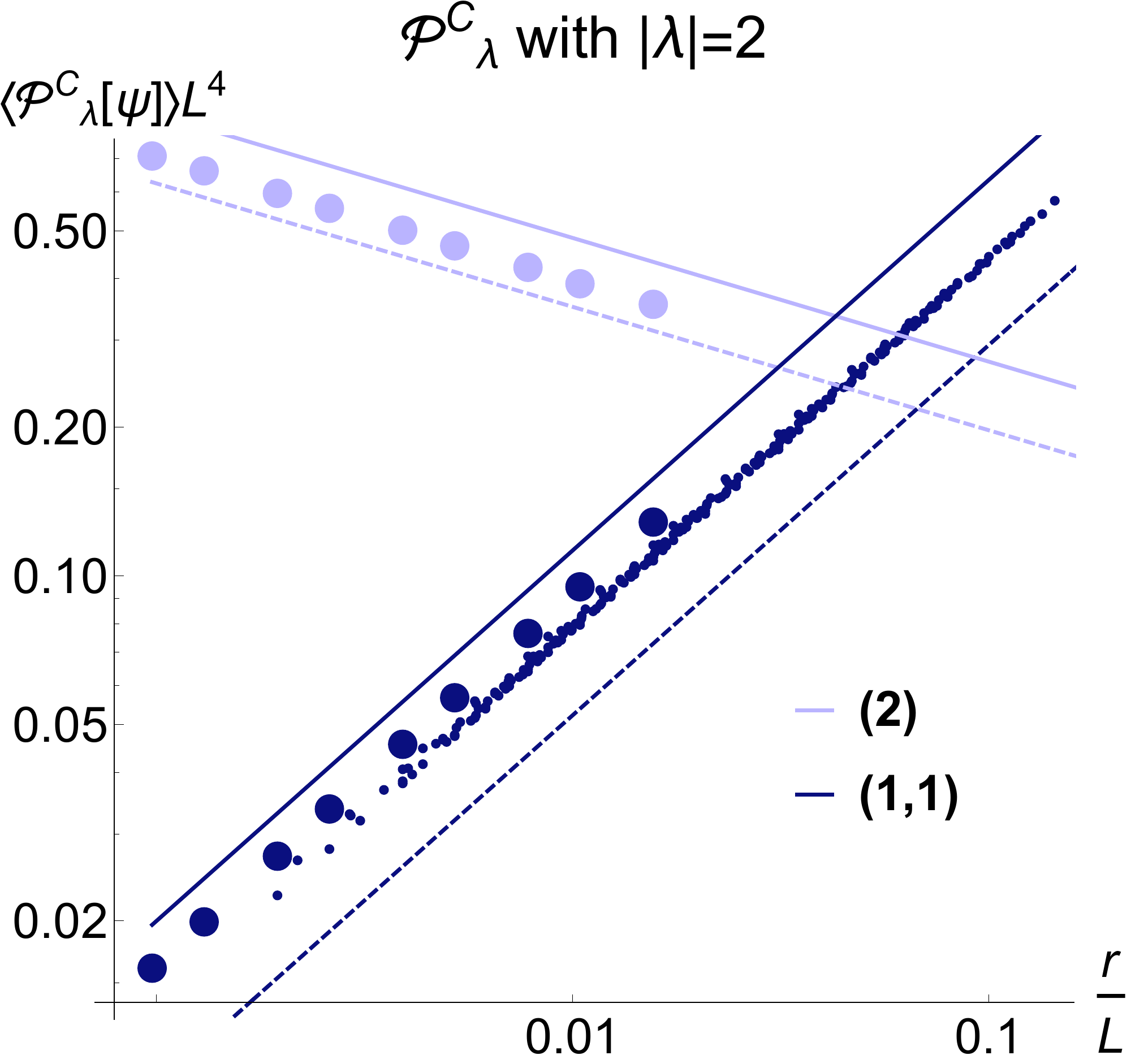}
		\includegraphics[width=0.42\linewidth]{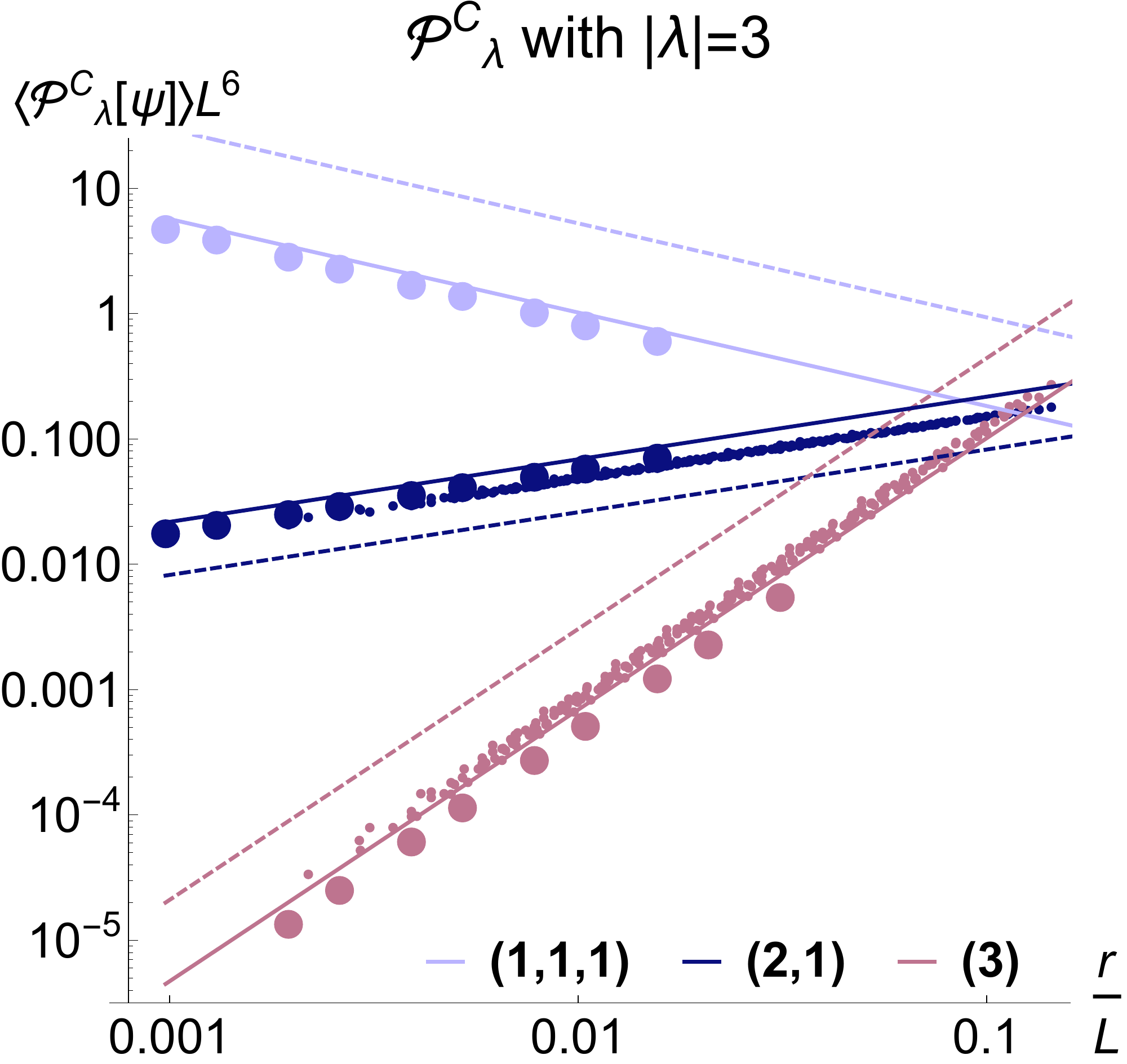}
	\end{center}
	\noindent\begin{center}
		\includegraphics[width=0.42\linewidth]{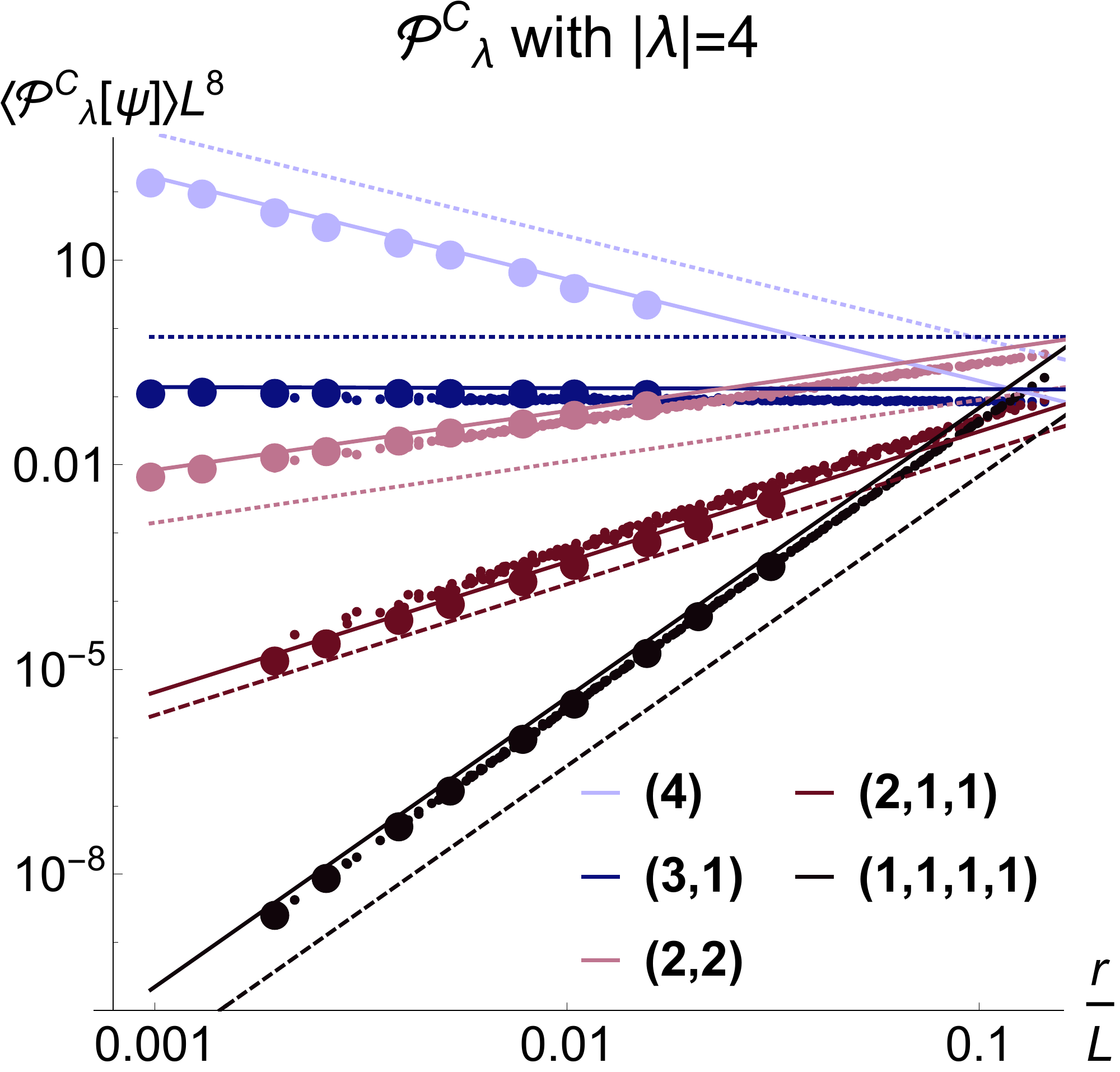}
		\includegraphics[width=0.42\linewidth]{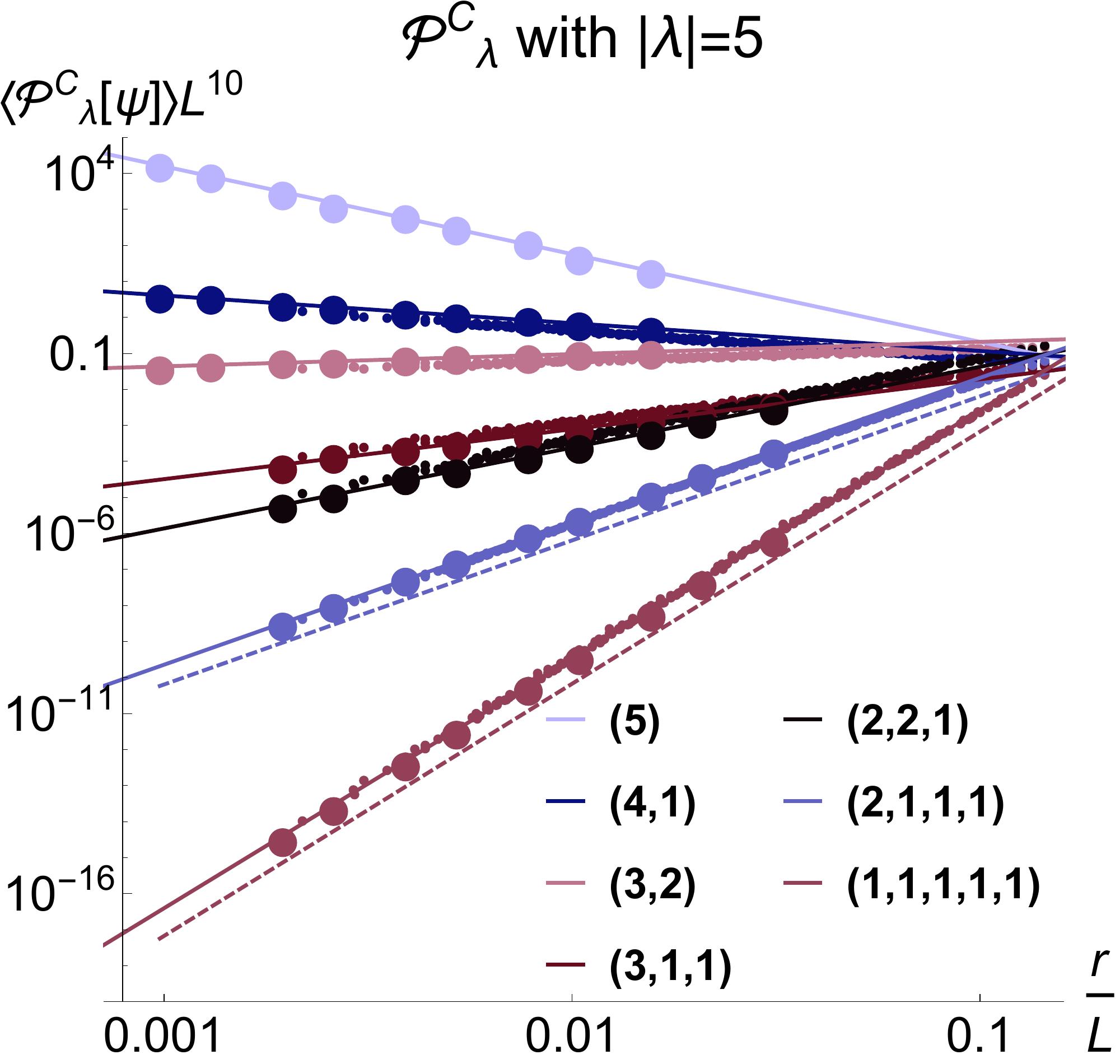}
	\end{center}
	\noindent\caption{Numerical determination of the generalized-multifractality exponents at the SQH transition by SU(2) Chalker-Coddington network (CCN) simulations of pure scaling wave-function observables, Eqs.~\eqref{eq:det}  and~\eqref{eq:obs}. Ensembles of systems of linear sizes $L=64,96,\ldots, 1024$ with $10^4$ disorder configurations for each length $L$ are studied. All observables are averaged over $\sim L^2$ spatial points and over $10^4$ disorder realizations.  Each of the four panels shows data for all  Young diagrams $\lambda$ in a given order $q=|\lambda|$, including $q=2$ (upper left),  $q=3$ (upper right), $q=4$ (bottom left), and $q=5$ (bottom right). The spatial arguments $\RR{i}$ in the wave-function observable are chosen to be separated by approximately the same distance $r$. The data are scaled by a factor $r^{\Delta_{q_1} + \ldots +  \Delta_{q_n}}$, which ensures the collapse of the data with different $r$ onto a single line $\sim (r/L)^{\Delta_\lambda}$.	
		The data corresponding to the smallest distance $r$ for each combination is highlighted as bold data points. These bold data points are used in the fit to extract the scaling exponent.  Solid lines are fits to the highlighted small-$r$ data points, whereas dashed lines are predictions from the percolation theory (whenever available). For $q=4$, the dotted lines represent numerical results obtained earlier in Ref.~\cite{karcher2021generalized} by an alternative (less general) numerical approach. Scaling exponents obtained numerically in the present work are summarized  in Table~\ref{tab:lC}, see the column $x_\lambda^{\rm qn}$.
		They are also compared there with analytical predictions of the percolation theory, with percolation numerics, and with previous numerical results of Ref.~\cite{karcher2021generalized} (whenever available).
	}
	\label{fig:scaling}
\end{figure*}

\begin{table}[t!]
	\centering
	\begin{tabular}{c|cccccc|c}
		& $\lambda$ & $x_\lambda^{\rm perc}$ & $x_\lambda^{\rm cn}$ & $x_\lambda^{\rm qn}$ & $x_\lambda^{\rm qn,A\uparrow}$ & $x_\lambda^{\rm qn,|\psi|}$ & $x^{\rm para}_\lambda $\\[-1pt]
		&&&&& {\scriptsize \ (from} & {\scriptsize Ref. 5)} & \\[5pt]
		\hline
		\hline
		&&&&&&&\\[-5pt]
		$q=1$ & (1) & $x_1^{\rm h}=\frac14$ & $0.25$ & --- & --- & ---  & $\frac14$
		\\[5pt]
		\hline
		&&&&&&\\[-5pt]
		$q=2$ & (2) & $x_1^{\rm h}=\frac14$ & $0.25$  & $0.249$ & $0.25$ & $0.25$  & $\frac14$\\[3pt]
		& (1,1) & $x_2^{\rm h}=\frac54$ & $1.22$ & $1.251$ &  $1.29$ & $1.24$ & $1$
		\\[5pt]
		\hline
		&&&&&&\\[-5pt]
		$q=3$ & (3)  & $0$ & --- & $0.004$ & $0.00$ & $0.00$ & $0$\\[3pt]
		& (2,1) & $x_2^{\rm h}=\frac{5}{4}$ & $1.23$ & $1.249$ & $1.25$ & $1.24$ & $1$\\[3pt]
		& 
        $(1^3)$
        & $x_3^{\rm h}=\frac{35}{12}$ & --- & $2.915$ & --- & --- & $\frac{9}{4}$
		\\[5pt]
		\hline
		&&&&&&\\[-5pt]
		$q=4$ & (4) & --- & --- & $-0.49$ & $-0.5$ & $-0.5$& $-\frac{1}{2}$\\[3pt]
		& (3,1)  & --- & --- & $0.99$ & $0.98$ & $0.99$ & $\frac34$\\[3pt]
		& (2,2)  & --- & --- & $1.87$ & $1.86$ & $1.91$ & $\frac{3}{2}$\\[3pt]
		& (2,1,1) & $x_3^{\rm h}=\frac{35}{12}$ & --- & $2.911$ & --- & ---  & $\frac{9}{4}$\\[3pt]
		& 
        $(1^4)$
        & $x_4^{\rm h}=\frac{21} 4$  & --- & $5.242$ & --- & ---  & $4$
		\\[5pt]
		\hline
		&&&&&&\\[-5pt]
		$q=5$ & (5)  & --- & --- & $-1.19$ & --- & --- & -$\frac{5}{4}$\\[3pt]
		& (4,1)  & --- & --- & $0.48$ & --- & --- & $\frac14$\\[3pt]
		& (3,2)  & --- & --- & $1.59$ & --- & --- & $\frac{5}{4}$\\[3pt]
		& (3,1,1)  & --- & --- & $2.64$ & --- & --- & 2\\[3pt]
		& (2,2,1) & --- & --- & $3.50$ & --- & ---  &  $\frac{11}{4}$\\[3pt]
		& 
        $(2,1^3)$
        &$x_4^{\rm h}=\frac{21} 4$  & --- & $5.23$ & --- & ---  &  $4$\\[3pt]
		& 
        $(1^5)$
        & $x_5^{\rm h}=\frac{33} 4$  & --- & $8.16$ & --- & ---  & $\frac{25}{4}$
	\end{tabular}
	\caption{Scaling exponents $x_\lambda$ of generalized multifractality at the SQH transition for wave-function observables with $q\equiv |\lambda|\leq 5$.
		The column $x_\lambda^{\rm perc}$ represents exact analytical values obtained in Sec.~\ref{sec:sqh_percolation} and
		Sec.~\ref{sec:susy} via the mapping to percolation theory.  The values $x_\lambda^{\rm cn}$ are obtained by numerical evaluation of the corresponding percolation probabilities in Sec.~\ref{sec:sqh_percolation}.  In the column $x_\lambda^{\rm qn}$,
		results of quantum numerics using the SU(2) CCN model (Sec.~\ref{sec:sqh_numerics}) are presented.  {  See text and Appendix~\ref{appendix:error-bars} for statistical error bars of the exponents $x_\lambda^{\rm qn}$. The values $x_\lambda^{\rm qn}$ are given with three digits after the decimal point in those cases when the standard deviation $\sigma_\lambda$ satisfies $2\sigma_\lambda \le 0.01$. }
		Whenever the analytical results are available, they are in excellent agreement with numerical values.  For completeness, we also include numerical results obtained in
		Ref.~\cite{karcher2021generalized} by two complementary (less general) approaches that also used the SU(2) CCN model. Specifically,
		$x_\lambda^{\rm qn,|\psi|}$ used observables involving the total density of spin up and down,  $|\psi|=\sqrt{|\psi_\uparrow|^2 + |\psi_\downarrow|^2}$, while $x_\lambda^{\rm qn,|\psi|}$  used pure-scaling combinations of class A. The symbol ``---'' means that the exponent was not determined by the corresponding approach.
		Finally, the last column $x^{\rm para}$ presents the values given by the generalized-parabolicity ansatz~\eqref{eq:gener-parab}	 with $b = \frac18$.
		The violation of generalized parabolicity is evident.
	}
	\label{tab:lC}
\end{table}

In Fig.~\ref{fig:scaling}, we show the wave-function pure-scaling observables given by Eq.~\eqref{eq:det}  and~\eqref{eq:obs} as functions of $r/L$, where $r$ is the distance between the points and $L$ the system size. The corresponding slopes (in the double-logarithmic scale) determine the generalized-multifractality scaling dimensions $x_\lambda$ of the SQH transition. The four panels display the data for Young diagrams $\lambda$ with $q=|\lambda|$ equal to 2 (top left), 3 (top right), 4 (bottom left), and 5 (bottom right).

For an observable of order $q$, we need up to $q$ distinct wave functions  {  $\psi_i$ (with positive energies)} at $q$ spatial positions $\RR1,\ldots \RR{q}$.
{  (Since $q \le 5$ in our simulations, we need at most five eigenstates.) We choose these eigenstates to be the ones with (positive) energies closest to zero.}
The configurations of points $\RR{i}$ are chosen in such a way that all pairwise distances are approximately the same, $ \sim r$. In the case $q=2$, we have two points $\RR1$ and $\RR2$ with $|\RR1-\RR2|=r$. In the case of  higher $q$, there are many possible choices;  we stick to the following geometries. For $q=3$, we choose the points to form approximately equilateral triangles (the approximation is necessary because the links of the network model form a square lattice), while for $q=4$ we use corners of squares as the four points. In the $q=5$ case, the $\RR1,\ldots \RR4$ are chosen to form a square with $\RR5$ located in the center of that square.
{  We restrict the distances to $r < 10$. In combination with the choice of the eigenstates as those with the energies closest to zero, this ensures that we stay at criticality.}

The data points shown in Fig.~\ref{fig:scaling} are scaled by a factor $r^{\Delta_{q_1} + \ldots +  \Delta_{q_n}}$. As expected, upon this rescaling, there is an excellent collapse of data with different $r$ for every $\lambda$ onto a single line corresponding to a power-law scaling with $r/L$. The slope of this line is $\Delta_\lambda$; from it we obtain $x_\lambda$ according to Eq.~\eqref{eq:Delta_lambda}.

The data for smallest possible distance $r \sim 1$ for each $\lambda$ is highlighted as bold data points. These data points are used for the fit (solid lines) to extract the scaling exponent. The dashed lines represent slopes predicted by  the mapping to percolation (whenever it applies) and the analysis of scaling of the resulting classical expressions, see Sec.~\ref{sec:sqh_Pana}.  An excellent agreement between the analytical and numerical results is evident.

The results for the  generalized-multifractality exponents at the SQH transition for wave-function observables with $q\equiv |\lambda|\leq 5$, as obtained by various methods, are summarized in Table~\ref{tab:lC}.  In the column $x_\lambda^{\rm qn}$ the values found by the CCN numerics are shown (here the superscript ``qn'' stands for ``quantum numerics'').  The column $x_\lambda^{\rm perc}$ presents the exact analytical results obtained by mapping to percolation in Sec.~\ref{sec:sqh_Pana} (for those $\lambda$, for which this mapping can be constructed), and the column $x_\lambda^{\rm cn}$ the corresponding numerics from Sec.~\ref{sec:sqh_Pnumerics} (with ``cn'' standing for ``classical numerics''). The agreement between the CCN numerics and the percolation predictions is very impressive, especially taking into  account that we proceed up to such large $q$ as $q=5$.

{  To characterize the precision of our CCN numerics, we have determined statistical error bars for the exponents $x_\lambda^{\rm qn}$, see Appendix~\ref{appendix:error-bars} for detail. The obtained standard deviation $\sigma_\lambda$ of $x_\lambda^{\rm qn}$ is as small as $\sigma_\lambda \approx 0.001$ for $\lambda = (1,1)$, (2), and 
$(1^3)$.
For the other two $q=3$ exponents, we found an excellent precision as well: $\sigma_{(2,1)} \approx 0.002$ and $\sigma_{(3)} \approx 0.004$. For $q=4$ exponents, the standard deviation ranges from 
$\sigma_{(1^4)} \approx 0.005$
to $\sigma_{(4)} \approx 0.02$, and for $q=5$ exponents from 
$\sigma_{(1^5)} \approx 0.01$
to $\sigma_{(5)} \approx 0.04$.  In nearly all the cases when there are exact analytical results for the exponents $x_\lambda$ (column $x_\lambda^{\rm perc}$ in Table \ref{tab:lC}), the numerical values $x_\lambda^{\rm qn}$ agree with $x_\lambda^{\rm perc}$ within the $2\sigma_\lambda$ interval. Such an excellent agreement is rather remarkable, especially in view of the small values of the statistical uncertainty $\sigma_\lambda$. The only exception when the deviation is a few times larger then $2\sigma_\lambda$ is the exponent 
$x_{(1^5)}$.
We attribute this to somewhat enhanced systematic errors, presumably related to very small values of the corresponding correlation function ($ \sim 10^{-15}$ for the largest $L$ in Fig.~\ref{fig:scaling}). Let us note that also for this exponent a relative deviation  from the exact result is very small: 
$x^{\rm qn}_{(1^5)} = 8.16$
instead of
$x_{(1^5)}^{\rm perc} = 33/4 = 8.25$,
i.e., just a 1\% deviation.
}

We have also included in Table~\ref{tab:lC} results of quantum numerical simulations performed in Ref.~\cite{karcher2021generalized}. There, two complementary numerical approaches were used; one using class-A combinations formed with one spin component $\psi_\uparrow$ and another one using observables involving the total density of spin up and down, $|\psi|=\sqrt{|\psi_\uparrow|^2 + |\psi_\downarrow|^2}$. Both these approaches yield strictly positive observables; however, they are not of pure-scaling nature and allowed us to obtain only a subclass of the scaling exponents (see Eq.~\eqref{eq:rg_C_to_A} in Appendix~\ref{appendix:rg} and a discussion below it for explanations with respect to the first of these approaches).
The corresponding columns in Table~\ref{tab:lC} are denoted by $x_\lambda^{\rm qn,A\uparrow}$ and $x_\lambda^{\rm qn,|\psi|}$, respectively.  There is a good agreement between these numerical results (when available) and those of the present work.  The CCN numerics of the present work ($x_\lambda^{\rm qn}$), which is based on the construction of Sec.~\ref{sec:scaling}, is clearly superior in comparison with the approaches of Ref.~\cite{karcher2021generalized} since it permits to study the scaling of pure-scaling observables for any representation $\lambda$ and with a high numerical precision.

Summarizing, all our numerical and analytical results for the exponents are in excellent agreement with each other.
It is also worth emphasizing that the obtained  values of the exponents fully agree with the exact Weyl symmetry relations:
$x_{(1)}= x_{(2)}$, $x_{(3)}= x_{(0)}\equiv 0$, $x_{(1,1)}= x_{(2,1)}$, 
$x_{(1^3)}= x_{(2,1,1)}$,
and 
$x_{(1^4)}= x_{(2,1^3)}$.

Finally, let us discuss a comparison between the generalized-multifractality spectrum of the SQH transition and the parabolic ansatz
\begin{equation}
	x_\lambda^{\rm para} = b \left[ q_1(3-q_1) + q_2(7-q_2) + q_3 (11-q_3) + \ldots \right].
	\label{eq:gener-parab}
\end{equation}
The analytically known values of the exponents $x_{(1)} = x_{(2)}= \frac14$ and $x_{(3)} = 0$ agree with $x_\lambda^{\rm para}$ with $b= \frac 18$.  The corresponding ``parabolic'' values $x_\lambda^{\rm para}$ are given in the last column of Table~\ref{tab:lC}. Numerical studies of the leading multifracatlity spectrum $x_{(q)}$ in Ref.~\cite{mirlin2003wavefunction}, and more recently in Ref.~\cite{puschmann2021quartic}, showed relatively small but significant deviations from parabolicity.  In Ref.~\cite{karcher2021generalized}, we have presented  numerical evidence of a strong violation of generalized parabolicity in the subleading multifractal spectrum (in particular, in the value of $x_{(1,1)}$).  Here we have shown that the strong violation of generalized parabolicity is manifest in the analytical results (within the percolation mapping) and nicely confirmed by an efficient quantum SU(2) CCN numerics as well as by the classical percolation numerics.  In particular, we have derived an exact analytical result $x_{(1,1)}= \frac 54$, whereas the parabolicity assumption would give $x_{(1,1)}= 1$. Similarly, we have an exact analytical value 
$x_{(1^3)}= \frac {35}{12}$,
while the ``parabolic'' value would be 
$x_{(1^3)}^{\rm para} = \frac 94$,
and so on. The violation of the generalized parabolicity implies~\cite{karcher2021generalized} that the local conformal invariance does not hold at the SQH critical point.

\section{Conclusions}
\label{sec:conclusions}

In this paper, we have studied the generalized multifractality at the SQH transition, which is a counterpart of the conventional integer QH transition for superconducting systems.
The key findings of the paper are as follows:

\begin{enumerate}
	
	\item
	By using two complementary approaches, we have developed in  Sec.~\ref{sec:sqh_mapping} and Sec.~\ref {sec:susy} a percolation mapping for a subset of eigenfunction correlation functions. This includes all correlation functions  up to the order $q=3$  as well as the ``most subleading'' 
$\lambda = (1^q)$
observables for any integer $q=|\lambda|$.
Employing this mapping, we have found analytically the SQH generalized-multifractality exponents $x_{(1)}$, $x_{(2)}$, $x_{(1,1)}$, $x_{(3)}$, $x_{(2,1)}$, and 
$x_{(1^3)}$
in Sec.~\ref{sec:sqh_Pana}, as well as 
$x_{(1^q)}$
in Sec.~\ref{subsec:SUSY-bottom}. They are all expressed in terms of the scaling dimensions $x_n^{\rm h}$ of $n$-hull operators known from classical percolation theory. These analytical results are supported  by numerical simulations of percolation expressions in Sec.~\ref{sec:sqh_Pnumerics}.
	
	\item
	In Sec.~\ref{sec:scaling}, we have derived a general form of wave-function observables that are positive in any disorder realization and exhibit the pure scaling upon disorder averaging. This construction holds for a generic representation $\lambda = (q_1, q_2, \ldots, q_n)$, with arbitrary $q_i$ (that do not need to be integer, and in general can be also complex) and permits an efficient numerical evaluation of exponents $x_\lambda$. Using a network model in class C, we have numerically determined in Sec.~\ref{sec:scaling} all exponents corresponding to polynomial pure-scaling observables up to the order $q=5$. An excellent agreement between the numerical and the analytical results has been found (for those $\lambda$ for which the analytical results via the percolation mapping were derived). The obtained analytical and numerical values of the generalized-multifractality exponents are summarized in Table~\ref{tab:lC}.
	
\end{enumerate}

Numerical simulations in Ref.~\cite{karcher2021generalized} provided evidence of a strong violation of generalized parabolicity of the generalized-multifractality spectrum of the SQH transition.  Here, the analytical results based on the percolation mapping unambiguously demonstrate this violation.  This conclusion is also fully supported by the advanced numerical simulations.  Therefore, local conformal invariance is violated in SQH critical systems. This excludes any theories with local conformal invariance---in particular,
Wess-Zumino-Novikov-Witten models---as candidates for the fixed-point theory of the SQH transition.

The following question may arise at this point. The 2D percolation-transition critical point obeys the local conformal invariance. So, is the mapping to percolation not in conflict with the  violation of local conformal invariance at the SQH criticality?  The answer becomes clear if one recalls the status of this mapping.  Indeed, the mapping to percolation holds only for a discrete subset of multi-indices $\lambda$ characterizing the observables. These correlation functions should thus exhibit local conformal invariance. However, they represent only a tiny sector of the whole SQH critical theory, for which the local conformal invariance does not hold.

Another interesting question is whether our results have any implications for the conjecture~\cite{ghorashi2018critical,karcher2021how} that the energy-perturbed Wess-Zumino-Novikov-Witten theory describing finite-energy states on a surface of a class-CI topological insulator flows into the fixed point of the SQH transition. Our findings do not invalidate this conjecture, since the energy perturbation explicitly breaks the conformal invariance of the Wess-Zumino-Novikov-Witten model.

Before closing, we briefly mention a few prospects for future research;  the work in some of these directions is currently underway.

\begin{enumerate}
	
	\item {  The analysis of generalized multifractality in this paper (both analytical and numerical) can be extended to surfaces of SQH critical systems. The surface multifractality in the conventional setting (LDOS moments) at this critical point has been studied in Ref.~\cite{subramaniam2008boundary}.  Interestingly, it was found that the surface multifractality is affected by the presence of topological boundary modes~\cite{bondesan2012exact}. We expect that the surface generalized multiftactality may serve as an additional useful hallmark of the critical theory.}
			
	\item The construction of pure-scaling eigenfunction observables developed in Sec.~\ref{sec:scaling} can be extended to other symmetry classes; see a discussion at the end of Sec.~\ref{sec:scaling}. This will permit a numerical investigation of the generalized multifractality at other Anderson-localization critical points. {  One of  intriguing questions in this context is whether the violation of the generalized parabolicity  (and thus of the local conformal invariance) applies also to other 2D critical points. In particular, we envision the investigation of generalized multifractality at the following 2D critical points:
	\begin{enumerate}
	
		\item Classes AII, DIII, and D exhibit weak antilocalization and, as a result, host in two dimensions extended metallic phases as well as metal-insulator transitions~\cite{asada2002anderson, Obuse-Multifractality-2007, obuse2007two-dimensional, fulga2012thermal, chalker2001thermal, mildenberger2007density, medvedyeva2010effective, wang_multicriticality_2021}.
	
		\item In the chiral classes AIII, BDI and CII, there is the Gade-Wegner critical-metal phase~\cite{gade1991the, gade1993anderson} characterized by strong multifractality. Furthermore, these systems undergo a metal-insulator transition~\cite{motrunich2002particle-hole, bocquet2003network, koenig2012metal-insulator}.
	
		\item It was recently discovered that stacks of critical states generically emerge at finite energy on surfaces of topological superconductors~\cite{sbierski2020spectrum-wide, ghorashi2020criticality, ghorashi2018critical}.
		
	\end{enumerate}}
	
	\item The percolation mapping has allowed us to get exact values of the exponents characterizing the scaling of the observables with $q= |\lambda| \le 3$. While we know from the SQH sigma-model analysis that these are pure-scaling observables, it might be interesting to have a proof of this based solely on the percolation mapping, see discussions in the end of Sec.~\ref{sec:perc-analyt-q2}  and Sec.~\ref{sec:perc-analyt-q3}. The full spectrum of operators and their fusion rules for the CFT of two-dimensional percolation are not known at present, though very promising results have been recently obtained in Refs.~\cite{Grans-Samuelsson-The-action-2020} and~\cite{Grans-Samuelsson-Global-2021}.
	Future analyses may give complete fusion rules for the hull operators and, thereby, the full scaling content of the percolation probabilities beyond the leading scaling. Perhaps, the RG analysis of the pure scaling operators that can be expressed in terms of percolation probabilities, done in this paper, can help with this effort.
	
\end{enumerate}

\section{Acknowledgments}

We thank R.~Ziff for a discussion of corrections to leading scaling of classical percolation probabilities, J.~L.~Jacobsen for a discussion of fusion rules in the CFT for percolation,  and I.~Burmistrov and S.~Babkin for a collaboration on a related project on surface multifractality. {  Further, we thank H.~Obuse, N.~Read, and M.~R.~Zirnbauer for comments on the manuscript. } J.F.K. and A.D.M. acknowledge support by the Deutsche Forschungsgemeinschaft (DFG) via the grant MI 658/14-1.

\appendix

\onecolumngrid
\section{Details of the percolation mapping}
\label{appendix-mapping_}

{  
\subsection{Mapping for $q=2$.}
\label{appendix-mapping2}

\subsubsection{Correlator $\mathcal{D}_{(2)}$.}
\label{appendix-mapping2-2}

One of the contributions entering Eq.~\eqref{eq:F} in Sec.~\ref{sec:sqh_pnot} (with $z=w$) is straightforwardly evaluated within the percolation mapping:
\begin{align}
& \big\langle \Tr(G(\RR1,\RR2;z)G(\RR2,\RR1;z))\big\rangle = -2\sum_{N=1}^\infty p(\RR1,\RR2;N)z^{2N}.
\label{eq:D2-term1}
\end{align}
Since $\RR1,\RR2$ are distinct, there is no contribution from the unit operator as above. The factor 2 is because there are two pairs of paths that contribute: either one path goes $\RR1\leftarrow \RR2 \leftarrow \RR1\leftarrow \RR2$ picking up the factor $z^{2N - N_{\RR2\RR1}}$ and the second path $\RR2\leftarrow \RR1$ picking up $z^{N_{\RR2\RR1}}$, or vice versa, see the upper configuration in Fig.~\ref{fig:paths-12}c.  Further, using Eq.~\eqref{eq:inv}, we get
\begin{align}
& \big\langle \Tr(G(\RR1,\RR2;z^{-1})G(\RR2,\RR1;z^{-1}))\big\rangle = -2\sum_{N=1}^\infty p(\RR1,\RR2;N)z^{2N}.
\label{eq:D2-term2}
\end{align}
Finally, the remaining terms yield
\begin{align}
& \big\langle \Tr(G(\RR1,\RR2;z)G(\RR2,\RR1;z^{-1})) \big\rangle = -2\sum_{N=1}^\infty p_1(\RR1,\RR2;N) z^{2N},\nonumber\\
& \big\langle \Tr(G(\RR1,\RR2;z^{-1})G(\RR2,\RR1;z)) \big\rangle = -2\sum_{N=1}^\infty p_1(\RR1,\RR2;N) z^{2N},
\label{eq:tr_Gz_Gz-1}
\end{align}
and an analogous term with interchanged $\RR1$ and $\RR2$, as illustrated by the lower configuration in Fig.~\ref{fig:paths-12}c. To prove Eq.~\eqref{eq:tr_Gz_Gz-1}, we use, following Ref.~\cite{mirlin2003wavefunction}, an invariance argument based on explicit averages over the link SU(2) marices. For this purpose, consider first
\begin{align}
&\big\langle \Tr(G(\RR1,\RR2;z)G(\RR1,\RR2;z))\big\rangle
=\int_{\rm SU(2)} d\mu(U_{\RR1}) d\mu(U_{\RR2})
\big\langle \Tr(U^\dagger_{\RR1} G(\RR1,\RR2;z)U_{\RR2}U^\dagger_{\RR1}G(\RR1,\RR2;z)U_{\RR2}) \big\rangle.
\end{align}
To perform the averaging over the SU(2)  random variables $U_{\RR1}$ and $U_{\RR2}$, we
use the identity
\begin{align}
\int_{\rm SU(2)} d\mu(U) U_{\alpha\beta} U_{\gamma\delta}
&= \frac12\delta_{\alpha\beta} \delta_{\gamma\delta}-\frac12 \delta_{\alpha\delta} \delta_{\gamma\beta}
\label{eq:su2avg}.
\end{align}
This yields
\begin{align}
&\big\langle \Tr [G(\RR1,\RR2;z)G(\RR1,\RR2;z)] \big\rangle = -\big\langle  \Tr G(\RR1,\RR2;z) \Tr G(\RR1,\RR2;z) \big\rangle.
\label{eq:identity_tr_GG}
\end{align}
Further, we can express $G(\RR2,\RR1,z^{-1})$ in terms of $G(\RR1,\RR2,z)$ using Eqs.~\eqref{eq:inv} and~\eqref{eq:Csym}:
\begin{align}
&G(\RR2,\RR1,z^{-1}) = \mathds{1}(\RR2,\RR1) - \left[G(\RR2,\RR1,z^*)\right]^\dagger = \mathds{1}(\RR2,\RR1) + (i\sigma_2) \left[G(\RR1,\RR2,z^*)\right]^T (i\sigma_2).
\end{align}
This implies (we can drop the identity $\mathds{1}$, since $\RR1,\RR2$ are distinct)
\begin{align}
&\big\langle  \Tr \left[G(\RR1,\RR2,z)  G(\RR2,\RR1,z^{-1})\right]\big\rangle
= \big\langle  \Tr \left \{G(\RR1,\RR2,z) (i\sigma_2) [G(\RR1,\RR2,z)]^T (i\sigma_2)\right\} \big\rangle
\nonumber\\
&= - \big\langle \Tr[G(\RR1,\RR2;z)G(\RR1,\RR2;z)] - \left[\Tr G(\RR1,\RR2;z))\right]^2 \big\rangle
= -2\big\langle \Tr(G(\RR1,\RR2;z)G(\RR1,\RR2;z))\big\rangle.
\end{align}
Here we used the identity
\begin{align}
(i\sigma_2)_{ij}
(i\sigma_2)_{kl} &= \delta_{ij}\delta_{kl}-\delta_{il}\delta_{kj},
\end{align}
and Eq.~\eqref{eq:identity_tr_GG}. This proves Eq.~\eqref{eq:tr_Gz_Gz-1}.

Combining the individual contributions from Eqs.~\eqref{eq:D2-term1},~\eqref{eq:tr_Gz_Gz-1} and~\eqref{eq:D2-term2}, we obtain Eq.~\eqref{eq:D2} of Sec.~\ref{sec:mapping-q2}.

\subsubsection{Correlator $\mathcal{D}_{(1,1)}$.}
\label{appendix-mapping2-11}

The results for the individual terms read (the corresponding path configurations are illustrated in Fig.~\ref{fig:paths-12}b):
\begin{align}
\big\langle \Tr G(\RR1,\RR1;z) \Tr G(\RR2,\RR2;z) \big\rangle
&= 4 - 4\sum_{N=1}^\infty p^{(s)}(\RR1;N) z^{2N}
+ \sum_{N,N'=1}^{\infty}p(\RR1;N|\RR2;N')z^{2(N+N')}+\sum_{N}p(\RR1,\RR2;N)z^{2N},
\nonumber\\
\big\langle \Tr G(\RR1,\RR1;z) \Tr G(\RR2,\RR2;z^{-1}) \big\rangle
&= 2\sum_{N=1}^\infty p(\RR2;N)z^{2N}
-\sum_{N,N'=1}^{\infty}p(\RR1;N|\RR2;N')z^{2(N+N')}-\sum_{N}p(\RR1,\RR2;N)z^{2N},
\nonumber\\
\big\langle \Tr G(\RR1,\RR1;z^{-1})\Tr G(\RR2,\RR2;z^{-1}) \big\rangle
&= \sum_{N}p(\RR1,\RR2;N)z^{2N} +\sum_{N,N'=1}^{\infty}p(\RR1;N|\RR2;N')z^{2(N+N')}.
\label{eq:D11-perc-terms}
\end{align}
(There is also a counterpart of the second term in Eq.~\eqref{eq:D11-perc-terms} with interchanged coordinates $\RR1$ and $\RR2$.)
Here the superscript $(s)$ again denotes symmetrization with respect to the links $\RR1, \RR2$:
\begin{align}
2p^{(s)}(\RR1;N) &= p(\RR1;N) + p(\RR2;N).
\end{align}
It is easy to see that all terms in the right-hand-side of equations~\eqref{eq:D11-perc-terms} are symmetric with respect to the interchange $\RR1 \leftrightarrow \RR2$, so that
the full percolation expression for $\mathcal{D}_{(1,1)}$ explicitly possesses this symmetry (which immediately follows from the definition of $\mathcal{D}_{(1,1)}$).

Combining the individual contributions from Eq.~\eqref{eq:D11-perc-terms}, we obtain Eq.~\eqref{eq:D11} of Sec.~\ref{sec:mapping-q2}.}

\subsection{Mapping for $q=3$.}
\label{appendix-mapping}

In this appendix, we provide intermediate formulas for the percolation mapping of $q=3$ correlation functions, Sec.~\ref{sec:perc-mapping-q3}.
Specifically, we present the results of the percolation mapping for individual terms resulting from the definition of the correlation functions.
Configurations of paths yielding these contributions are shown in Fig.~\ref{fig:paths-3}.

\subsubsection{Correlator $\mathcal{D}_{(3)}$.}
\label{appendix-mapping3-3}

\begin{figure}
	\includegraphics[width=0.95\textwidth]{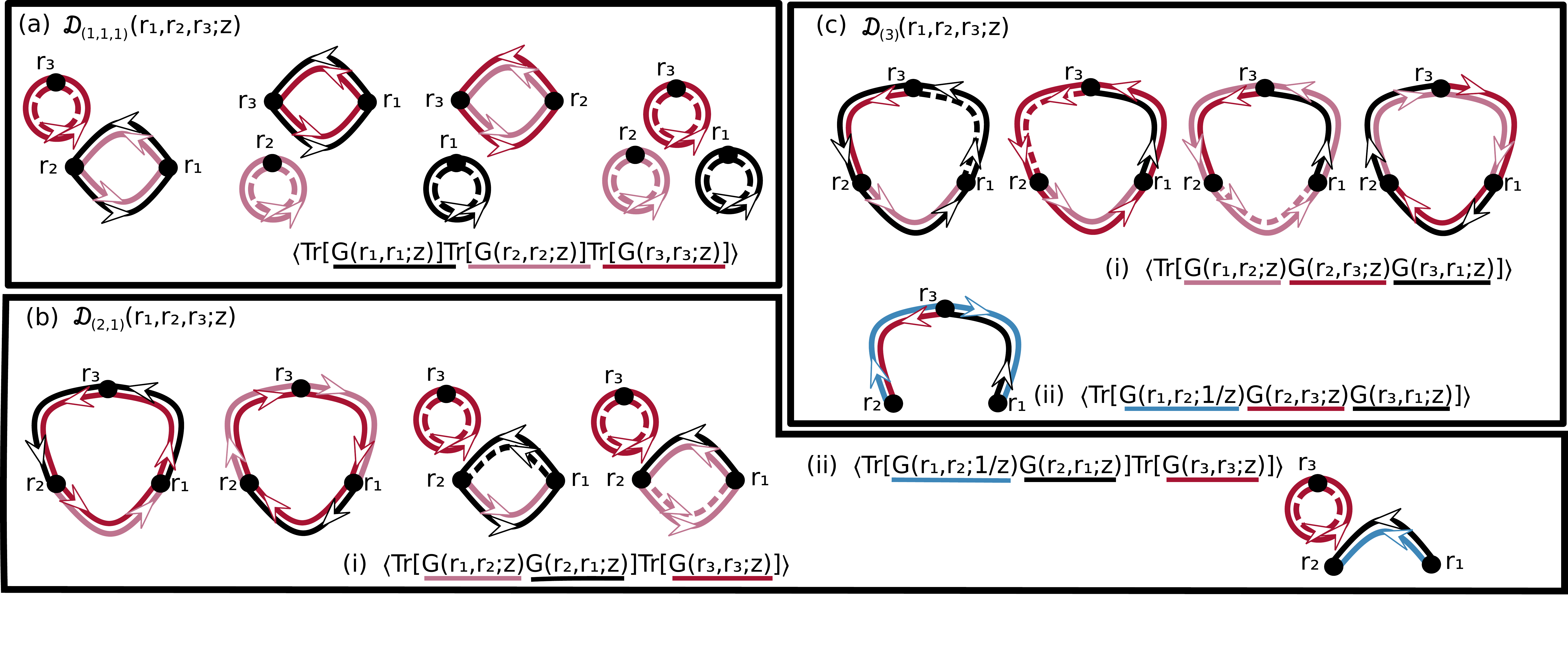}
	\vspace{-0.5cm}
	\caption{Schematic representation of path configurations yielding individual contributions within the percolation mapping for the $q=3$ correlation functions presented in Appendix~\ref{appendix-mapping} [Eqs.~\eqref{eq:d3ic},~\eqref{eq:d21ic} and ~\eqref{eq:tr111}].
		Paths corresponding to each Green's function are shown by the color corresponding to the line underlining this Green's function. When a path traverses a segment of the loop for the second time, it is shown by a dashed line.  Only contributions containing explicitly all three points $\RR1$, $\RR2$, and $\RR3$ are shown. In addition, there are contributions originating from unity terms in the expansion of some of the Green's functions $G(\RR{i},\RR{i};z)$, see Eqs.~\eqref{eq:d21ic} and ~\eqref{eq:tr111}.}
	\label{fig:paths-3}
\end{figure}

For the individual terms entering Eq.~\eqref{eq:dg3}, the percolation mapping yields:
\begin{align}
	\left\langle\Tr\left [G(\RR1,\RR2;z)G(\RR2,\RR3;z)G(\RR3,\RR1;z) \right ] \right\rangle &= -\sum_{N} \left[3p(\RR1,\RR2,\RR3;N)+p(\RR3,\RR2,\RR1;N)\right]z^{2N}, \nonumber\\
	\left\langle\Tr \left[ G(\RR1,\RR2;z^{-1})G(\RR2,\RR3;z^{-1})G(\RR3,\RR1;z^{-1}) \right]\right\rangle &= \sum_{N} \left[p(\RR1,\RR2,\RR3;N)+3p(\RR3,\RR2,\RR1;N)\right]z^{2N}, \nonumber\\
	\left\langle\Tr \left[ G(\RR1,\RR2;z)G(\RR2,\RR3;z)G(\RR3,\RR1;z^{-1}) \right] \right\rangle &= -2\sum_{N} p_1(\RR1,\RR2,\RR3;N)z^{2N},\nonumber\\
	\left\langle\Tr(G(\RR1,\RR2;z)G(\RR2,\RR3;z^{-1})G(\RR3,\RR1;z))\right\rangle &= -2\sum_{N} p_1(\RR3,\RR1,\RR2;N)z^{2N},\nonumber\\
	\left\langle\Tr \left[ G(\RR1,\RR2;z^{-1})G(\RR2,\RR3;z)G(\RR3,\RR1;z) \right] \right\rangle &= -2\sum_{N} p_1(\RR2,\RR3,\RR1;N)z^{2N},\nonumber\\
	\left\langle\Tr \left[ G(\RR1,\RR2;z^{-1})G(\RR2,\RR3;z^{-1})G(\RR3,\RR1;z) \right] \right\rangle &= 2\sum_{N} p_1(\RR3,\RR2,\RR1;N)z^{2N},\nonumber\\
	\left\langle\Tr \left[ G(\RR1,\RR2;z^{-1})G(\RR2,\RR3;z)G(\RR3,\RR1;z^{-1}) \right] \right\rangle &= 2\sum_{N} p_1(\RR2,\RR1,\RR3;N)z^{2N},\nonumber\\
	\left\langle\Tr \left[ G(\RR1,\RR2;z)G(\RR2,\RR3;z^{-1})G(\RR3,\RR1;z^{-1}) \right] \right\rangle &= 2\sum_{N}
	p_1(\RR1,\RR3,\RR2;N)z^{2N}.
	\label{eq:d3ic}
\end{align}
The first two equations correspond to the diagrams in the upper panel (i) of Fig.~\ref{fig:paths-3}c. The factor of three in front of $p(\RR1,\RR2,\RR3;N)z^{2N}$ in the first term is related to three distinct ways one can traverse a loop $\RR1\leftarrow\RR2\leftarrow\RR3\leftarrow\RR1$ exactly twice, as shown in the figure. The rightmost diagram gives the contribution $p(\RR1,\RR3,\RR2;N)z^{2N}$ originating from a loop  with a reverse orientation, $\RR1\leftarrow\RR3\leftarrow\RR2\leftarrow\RR1$. The remaining six equations in Eq.~\eqref{eq:d3ic} correspond to the diagram shown in the lower panel (ii) of Fig.~\ref{fig:paths-3}c. Here, only a part of the loop between $\RR2\leftarrow\RR3\leftarrow\RR1$ is traversed, which is related to the probability $p_1(\RR1,\RR2,\RR3;N)z^{2N}$ specifying the length $N_{\RR1,\RR3}=N$ of that segment.
Adding up the individual contributions from Eq.~\eqref{eq:d3ic}, we get  Eq.~\eqref{eq:D3-perc} of Sec.~\ref{sec:perc-mapping-q3}.


\subsubsection{Correlator $\mathcal{D}_{(2,1)}$.}
\label{appendix-mapping3-21}

For the correlation function~\eqref{eq:dg21}, the individual terms are mapped onto percolation expressions as follows:
\begin{align}
	\left\langle\Tr \left[ G(\RR1,\RR2;z)G(\RR2,\RR1;z) \right] \Tr G(\RR3,\RR3;z) \right\rangle
	& = \sum_{N} \left[p(\RR1,\RR2,\RR3;N)+p(\RR1,\RR3,\RR2;N)\right] z^{2N}
	\nonumber \\
	&+2\sum_{N,N''} p(\RR1,\RR2;N|\RR3;N^{\prime\prime})z^{2N}z^{2N''} -4\sum_N p(\RR1,\RR2;N)z^{2N}, \nonumber\\
	\left\langle\Tr \left[ G(\RR1,\RR2;z^{-1})G(\RR2,\RR1;z^{-1}) \right] \Tr G(\RR3,\RR3;z) \right\rangle & = \sum_{N} \left[p(\RR1,\RR2,\RR3;N)+p(\RR1,\RR3,\RR2;N)\right] z^{2N}\nonumber \\
	& +2\sum_{N,N''} p(\RR1,\RR2;N|\RR3;N^{\prime\prime})z^{2N}z^{2N''} -4\sum_N p(\RR1,\RR2;N)z^{2N}, \nonumber\\
	\left\langle\Tr \left[ G(\RR1,\RR2;z)G(\RR2,\RR1;z) \right] \Tr G(\RR3,\RR3;z^{-1})\right\rangle
	& = -\sum_{N} \left[p(\RR1,\RR2,\RR3;N)+p(\RR1,\RR3,\RR2;N)\right] z^{2N}\nonumber\\
	&-2\sum_{N,N''} p(\RR1,\RR2;N|\RR3;N^{\prime\prime})z^{2N}z^{2N''}, \nonumber\\
	\left\langle\Tr \left[ G(\RR1,\RR2;z^{-1})G(\RR2,\RR1;z^{-1}) \right] \Tr G(\RR3,\RR3;z^{-1})\right\rangle
	& = -\sum_{N} \left[p(\RR1,\RR2,\RR3;N)+p(\RR1,\RR3,\RR2;N)\right] z^{2N}\nonumber\\
	&-2\sum_{N,N''} p(\RR1,\RR2;N|\RR3;N^{\prime\prime})z^{2N}z^{2N''},\nonumber\\
	\left\langle\Tr \left[ G(\RR1,\RR2;z)G(\RR2,\RR1;z^{-1}) \right] \Tr G(\RR3,\RR3;z )\right\rangle
	& = 2\sum_{N,N''} p_1(\RR1,\RR2;N|\RR3;N^{\prime\prime})z^{2N}z^{2N''} -4\sum_N p_1(\RR1,\RR2;N)z^{2N},\nonumber\\
	\left\langle\Tr \left[ G(\RR1,\RR2;z^{-1})G(\RR2,\RR1;z) \right] \Tr G(\RR3,\RR3;z)\right\rangle
	& = 2\sum_{N,N''} p_1(\RR2,\RR1;N|\RR3;N^{\prime\prime})z^{2N}z^{2N''} -4\sum_N p_1(\RR2,\RR1;N)z^{2N},\nonumber\\
	\left\langle\Tr \left[ G(\RR1,\RR2;z)G(\RR2,\RR1;z^{-1}) \right] \Tr G(\RR3,\RR3;z^{-1}) \right\rangle &= -2\sum_{N,N''} p_1(\RR1,\RR2;N|\RR3;N^{\prime\prime})z^{2N}z^{2N''},\nonumber\\
	\left\langle\Tr \left[ G(\RR1,\RR2;z^{-1})G(\RR2,\RR1;z) \right] \Tr G(\RR3,\RR3;z^{-1})\right\rangle &=-2\sum_{N,N''} p_1(\RR2,\RR1;N|\RR3;N^{\prime\prime})z^{2N}z^{2N''}.\label{eq:d21ic}
\end{align}
The contributions in the first four equations correspond to the diagrams shown in the panel (i) of Fig.~\ref{fig:paths-3}b. There are two types of these diagrams. One possibility is that all points belong to the same loop, yielding the term $p(\RR1,\RR2,\RR3;N)z^{2N}$ for the loop orientation $\RR1\leftarrow\RR2\leftarrow\RR3\leftarrow \RR1$ and the term $p(\RR1,\RR3,\RR2;N)z^{2N}$ for the opposite orientation, $\RR1\leftarrow\RR3\leftarrow\RR2\leftarrow \RR1$. Alternatively, the points $\RR1,\RR2$ may belong to one loop,
and $\RR3$ to a distinct loop. In that case, we get two contributions giving $p(\RR1,\RR2;N|\RR3;N'')z^{2N}z^{2N''}$ each.

For the last four equations of Eq.~\eqref{eq:d21ic}, the diagrams of relevant paths are shown in the panel (ii) of  Fig.~\ref{fig:paths-3}b. Here, configurations with a loop through $\RR1,\RR2$ and a distinct loop containing $\RR3$ contribute. Since only a part of the $\RR1,\RR2$-loop is traversed, the probabilities $p_1(\RR2,\RR1;N|\RR3;N'')z^{2N}z^{2N''}$  and $p_1(\RR1,\RR2;N|\RR3;N'')z^{2N}z^{2N''}$ appear under the sum.

For those correlation functions in Eq.~\eqref{eq:d21ic} that contain $G(\RR3,\RR3;z)$, there are additional contributions in the right-hand-side that do not contain explicitly $\RR3$. These contributions originate from the unity term in the expansion of the Green's function $G(\RR3,\RR3;z)$, in analogy with  Eq.~\eqref{eq:D11-perc-terms} for $\mathcal{D}_{(1,1)}$.

Combining the individual contributions from Eq.~\eqref{eq:d21ic}, we obtain Eq.~\eqref{eq:D21-perc} of Sec.~\ref{sec:perc-mapping-q3}.


\subsubsection{Correlator 
$\mathcal{D}_{(1^3)}$.
}
\label{appendix-mapping3-111}

For the correlation function  ~\eqref{eq:dg111}, the percolation mapping yields for individual terms:
\begin{align}
	&\left\langle\Tr G(\RR1,\RR1;z) \Tr G(\RR2,\RR2;z) \Tr G(\RR3,\RR3;z)  \right\rangle = 2^3-2^2\sum_N \left[p(\RR1;N)+p(\RR2;N)+p(\RR3;N)\right]z^{2N}\nonumber\\
	&+2\sum_{NN'} \left[p(\RR1;N|\RR2;N')+p(\RR1;N|\RR3;N')+p(\RR2;N|\RR3;N')\right]z^{2N}z^{2N'}\nonumber\\
	&+2\sum_{N} \left[p(\RR1,\RR2;N)+p(\RR2,\RR3;N)+p(\RR1,\RR3;N)\right]z^{2N}-\sum_{NN'N^{\prime\prime}}p(\RR1;N|\RR2;N'|\RR3;N^{\prime\prime})z^{2N}z^{2N'}z^{2N^{\prime\prime}}\nonumber\\
	&-\sum_{NN'}\left[p(\RR1,\RR2;N|\RR3;N')+ p(\RR1,\RR3;N|\RR2;N')+ p(\RR2,\RR3;N|\RR1;N')\right]z^{2N}z^{2N'},\nonumber\\
	&\left\langle\Tr G(\RR1,\RR1;z) \Tr G(\RR2,\RR2;z) \Tr G(\RR3,\RR3;z^{-1}) \right\rangle = 2^2\sum_N \left[p(\RR3;N)\right]z^{2N}\nonumber\\
	&-2\sum_{NN'} \left[p(\RR1;N|\RR3;N')+p(\RR2;N|\RR3;N')\right]z^{2N}z^{2N'}\nonumber\\
	&-2\sum_{N} \left[p(\RR1,\RR2;N)+p(\RR1,\RR3;N)+p(\RR1,\RR3;N)\right]z^{2N}+\sum_{NN'N^{\prime\prime}}p(\RR1;N|\RR2;N'|\RR3;N^{\prime\prime})z^{2N}z^{2N'}z^{2N^{\prime\prime}}\nonumber\\
	&+\sum_{NN'}\left[p(\RR1,\RR2;N|\RR3;N')+ p(\RR1,\RR3;N|\RR2;N')+ p(\RR2,\RR3;N|\RR1;N')\right]z^{2N}z^{2N'},\nonumber\\
	&\left\langle\Tr G(\RR1,\RR1;z) \Tr G(\RR2,\RR2;z^{-1}) \Tr G(\RR3,\RR3;z^{-1})\right\rangle =
	2\sum_{NN'} p(\RR2;N|\RR3;N')z^{2N}z^{2N'}\nonumber\\
	&+2\sum_{N}p(\RR2,\RR3;N)z^{2N}-\sum_{NN'N^{\prime\prime}}p(\RR1;N|\RR2;N'|\RR3;N^{\prime\prime})z^{2N}z^{2N'}z^{2N^{\prime\prime}}
	\nonumber\\
	&-\sum_{NN'}\left[p(\RR1,\RR2;N|\RR3;N')+ p(\RR1,\RR3;N|\RR2;N')+ p(\RR2,\RR3;N|\RR1;N')\right]z^{2N}z^{2N'},\nonumber\\
	&\left\langle\Tr G(\RR1,\RR1;z^{-1}) \Tr G(\RR2,\RR2;z^{-1}) \Tr G(\RR3,\RR3;z^{-1}) \right\rangle = \sum_{NN'N^{\prime\prime}}p(\RR1;N|\RR2;N'|\RR3;N^{\prime\prime})z^{2N}z^{2N'}z^{2N^{\prime\prime}}\nonumber\\
	&+\sum_{NN'}\left[p(\RR1,\RR2;N|\RR3;N')+ p(\RR1,\RR3;N|\RR2;N')+ p(\RR2,\RR3;N|\RR1;N')\right]z^{2N}z^{2N'}.
	\label{eq:tr111}
\end{align}
Diagrams showing the corresponding paths are presented in Fig.~\ref{fig:paths-3}a. There are two types of these diagrams. One possibility is that all three points are in distinct loops loop, yielding $p(\RR1;N|\RR2;N'|\RR3;N^{\prime\prime})z^{2N}z^{2N'}z^{2N^{\prime\prime}}$ (rightmost diagram in this panel). Alternatively, there are three partitions of the points in two distinct loops (other diagrams in this panel). In that case, we get three contributions giving $p(\RR1,\RR2;N|\RR3;N'')z^{2N}z^{2N''}$ and cyclic permutations thereof.

In addition, the right-hand-sides of the formulas in Eq.~\eqref{eq:tr111} contain contributions that do not contain explicitly one or several of the arguments $\RR1$, $\RR2$, and $\RR3$. These contributions originate from the unity term in the expansions of the Greeen's functions $G(\RR1,\RR1;z)$, $G(\RR2,\RR2;z)$, and/or $G(\RR3,\RR3;z)$.

Adding up the individual contributions from Eq.~\eqref{eq:tr111},  we obtain Eq.~\eqref{eq:D111-perc} of Sec.~\ref{sec:perc-mapping-q3}.

\section{NL$\sigma$M RG in $K$-invariant basis}
\label{appendix:rg}

In this appendix, we collect some of key results of the RG analysis within the NL$\sigma$M framework from Ref.~\cite{karcher2021generalized}, which are used in the present paper or have a direct connection to it. In Ref.~\cite{karcher2021generalized}, we have studied  the RG flow of  composite operators defined on NL$\sigma$M symmetric spaces $G/K$ corresponding to symmetry classes A and C. These RG equations can be conveniently written in a $K$-invariant basis. For both classes A and C, this basis can be labeled by integer partitions $\lambda=(q_1,\ldots,q_n)$  of $q$ with $q_1\geq\ldots\geq q_n$, where $q_i$ are positive integers and $ \sum_i q_i =q$. Elements of this basis have the form
\begin{align}
	O_\lambda \equiv O_{(q_1,\ldots, q_n)} &= \prod_{i=1}^n \mathrm{tr} \left((\Lambda Q)^{q_i}\right),
	\label{eq:o_lambda}
\end{align}
where $Q$ is NL$\sigma$M field, $\Lambda$ is the ``origin'' of the NL$\sigma$M manifold, and $\mathrm{tr}$ is the trace over all indices of $Q$.
The integers $q_j$ label the lengths of the $n$ cycles of $\Lambda Q$-strings over which the traces are taken.

For $q=2$ there are two basis operators, with $\lambda = (1,1)$ and (2).  In a one-loop RG operation $\delta_{A/C}$ (integrating out a fast momentum shell) two fast fields originating from either different $Q$-fields or the same $Q$-field are contracted. This contraction gives a loop integral factor $(-2I_f)$ and can glue the $\Lambda Q$-strings together or cut them apart:
\begin{align}
	\delta_A \!\!
	\begin{pmatrix}
		{\rm tr}(AQ) {\rm tr}(BQ)\\
		{\rm tr}(AQ BQ)
	\end{pmatrix}
	\!\!
	&=- 2I_f
	\!\!
	\underbrace{\begin{pmatrix*}[r]
			0 & -1 \\
			-1 & 0
	\end{pmatrix*}}_{=:M_2^A}
	\!\!
	\begin{pmatrix}
		{\rm tr}(AQ) {\rm tr}(BQ)\\
		{\rm tr}(AQ BQ)
	\end{pmatrix}\!\!,
	\nonumber\\
	\delta_C \!\!
	\begin{pmatrix}
		{\rm tr}(AQ) {\rm tr}(BQ)\\
		{\rm tr}(AQ BQ)
	\end{pmatrix}
	\!\!
	&=- 2I_f
	\!\!
	\underbrace{\begin{pmatrix*}[r]
			2 & - 2 \\
			- 1 & 3
	\end{pmatrix*}}_{=:M_2^C}
	\!\!
	\begin{pmatrix}
		{\rm tr}(AQ) {\rm tr}(BQ)\\
		{\rm tr}(AQ BQ)
	\end{pmatrix}\!\!.
	\label{eq:rgres}
\end{align}
Setting here $A = B = \Lambda$ yields the RG flow for the $K$-invariant operators~\eqref{eq:o_lambda} for $q=2$.

It was further shown in Ref.~\cite{karcher2021generalized} how the one-loop RG flow for polynomial operators of degree $q$ can be obtained from the $q=2$ flow, with Eq.~\eqref{eq:rgres} used as a building block.  A convenient formalism  to extract the action RG on a generic polynomial operator is as follows. We identify the $K$-invariant operators $O_\lambda$ with polynomials, by rewriting $\lambda = (1^{m_1}, 2^{m_2},\dots, k^{m_k}, \ldots)$  in terms of cycle lengths $k$ and multiplicities $m_k$. The monomial associated to $O_\lambda$ is then  $X_\lambda = \prod_k x_k^{m_k}$.
The one loop RG operators have a very simple form in this basis:
\begin{align}
	\mathcal{D}_A &= \sum_{j< i} jx_{i-j} x_j \partial_i + \frac12\sum_{i,j} ijx_{i+j} \partial_i\partial_j,
	\label{eq:rg_operators-A}
	\\
	\mathcal{D}_C &= \sum_{j< i} jx_{i-j} x_j \partial_i + \sum_{i,j} ijx_{i+j} \partial_i\partial_j-\sum_i \dfrac{i(i+1)}{2}x_i\partial_i,
	\label{eq:rg_operators-C}
\end{align}
where $\partial_i \equiv \partial/\partial x_i$.
Below, we briefly comment on the origin of different terms in these differential operators.

\paragraph{Class A.} The first term in Eq.~\eqref{eq:rg_operators-A} describes cutting a cycle of length $i$  into two cycles of length $j$ and $i-j$.  Here, the derivative removes one factor $x_i$ and yields a factor $m_i$, corresponding to the fact that this can happen to any of the $m_i$ cycles of length $i$.
The multiplication by $x_j x_{i-j}$ corresponds to the appearance of two cycles with the lengths $j$ and $i-j$. In total, there are $i=j+(i-j)$ realizations of such a cut. The second (quadratic with respect to the derivatives) term in Eq.~\eqref{eq:rg_operators-A} describes the fusion of cycles of length $i$ and $j$ into a cycle of length $i+j$. Here, the derivatives remove one cycle of length $i$ and one of length $j$, while the multiplication by $x_{i+j}$ adds one cycle of the corresponding length. In total, there are $ij$ channels for this process: the first fast field can come from each of the $i$ $Q$-fields in the cycle of length $i$,
and the second one from each of the $j$ $Q$-fields in the cycle of length $j$.

\paragraph{Class C.} Again, the first term in Eq.~\eqref{eq:rg_operators-C} describes cutting a cycle of length $i$ into two cycles of lengths $j$ and $i-j$. In total, there are $i=j+(i-j)$ realizations of such a cut. The second (quadratic with respect to the derivatives) term in Eq.~\eqref{eq:rg_operators-C} describes the fusion of cycles of length $i$ and $j$ into one of length $i+j$. There is an additional factor of two here in comparison to class A, as is clear from the inspection of the formulas~\eqref{eq:rg_operators-A} and~\eqref{eq:rg_operators-C} for the $q=2$ case. Finally, the last term in Eq.~\eqref{eq:rg_operators-C} is due to contractions of fast fields originating from (i) the same $Q$ (as in the renormalization of the average LDOS) and (ii)  from distinct $Q$ fields in a cycle of length $i$ preserving the cycle.

We note that both differential operators $\mathcal{D}_A$ and $\mathcal{D}_C$ preserve the degree $q = \sum_{k} k m_k$ of the composite operator. We can therefore restrict them to a sector of the theory with a given $q=|\lambda|$. Then we have
\begin{align}
	\mathcal{D}_A \sum_\lambda a_\lambda X_\lambda &= \sum_{\lambda,\mu} a_\lambda  (M^A_q)_{\lambda,\mu} X_\mu, \
	\nonumber\\
	\mathcal{D}_C \sum_\lambda a_\lambda X_\lambda &= \sum_{\lambda,\mu} a_\lambda  (M^C_q)_{\lambda,\mu} X_\mu,
	\label{eq:rg_operator_action}
\end{align}
with matrices $M^A_q$ and $M^C_q$ describing the renormalization of operators of degree $q$.  If we consider~\eqref{eq:rg_operator_action} as equations describing the action of the RG operators $\mathcal{D}_{A/C}$ on vectors of the coefficients $a_\lambda$, this action is clearly characterized by the transposed matrix $(M^{A/C}_q)^T$.

Once  the matrices $(M_q^A)^T$ are found with this procedure, one can determine their eigenvectors and assign to them a Young label $\lambda\vdash q$ by identifying the eigenvalues with those of the Laplacian on the NL$\sigma$M manifold. The results for $q = 2, 3$, and 4 are:
\onecolumngrid
\begin{align}
	\begin{pmatrix}
		\mathcal{P}_{(1,1)}^A[Q]\\
		\mathcal{P}_{(2)}^A[Q]
	\end{pmatrix}
	=
	\begin{pmatrix*}[r]
		1 & -1 \\
		1 & 1
	\end{pmatrix*}
	\begin{pmatrix}
		\mathrm{tr}(\Lambda Q)
		\mathrm{tr}(\Lambda Q)\\
		\mathrm{tr}(\Lambda Q\Lambda Q)
	\end{pmatrix},
	&&
	\begin{pmatrix}
        \mathcal{P}_{(1^3)}^A[Q] \\
		\mathcal{P}_{(2,1)}^A[Q]\\
		\mathcal{P}_{(3)}^A[Q]
	\end{pmatrix}
	=
	\begin{pmatrix*}[r]
		1 & -3 & 2 \\
		1 & 0 & -1 \\
		1 & 3 & 2
	\end{pmatrix*}
	\begin{pmatrix}
		\mathrm{tr}(\Lambda Q)
		\mathrm{tr}(\Lambda Q)
		\mathrm{tr}(\Lambda Q)\\
		\mathrm{tr}(\Lambda Q\Lambda Q)
		\mathrm{tr}(\Lambda Q)\\
		\mathrm{tr}(\Lambda Q\Lambda Q\Lambda Q)
	\end{pmatrix},
	\nonumber\\
	\begin{pmatrix}
        \mathcal{P}_{(1^4)}^A[Q] \\
		\mathcal{P}_{(2,1,1)}^A[Q]\\
		\mathcal{P}_{(2,2)}^A[Q]\\
		\mathcal{P}_{(3,1)}^A[Q]\\
		\mathcal{P}_{(4)}^A[Q]
	\end{pmatrix}
	=
	\begin{pmatrix*}[r]
		1 & -6 & 3 & 8 & -6 \\
		1 & -2 & -1 & 0 & 2 \\
		1 & 0 & 3 & -4 & 0 \\
		1 & 2 & -1 & 0 & -2 \\
		1 & 6 & 3 & 8 & 6
	\end{pmatrix*}
	\begin{pmatrix}
		\mathrm{tr}(\Lambda Q)
		\mathrm{tr}(\Lambda Q)
		\mathrm{tr}(\Lambda Q)
		\mathrm{tr}(\Lambda Q)\\
		\mathrm{tr}(\Lambda Q\Lambda Q)
		\mathrm{tr}(\Lambda Q)
		\mathrm{tr}(\Lambda Q)\\
		\mathrm{tr}(\Lambda Q\Lambda Q)\mathrm{tr}(\Lambda Q\Lambda Q)\\
		\mathrm{tr}(\Lambda Q\Lambda Q\Lambda Q)
		\mathrm{tr}(\Lambda Q)\\
		\mathrm{tr}(\Lambda Q\Lambda Q\Lambda Q\Lambda Q)
	\end{pmatrix}.\hspace{-6cm}
	\label{eq:rg_A_result}
\end{align}
Note that the last row in each of the matrices in Eq.~\eqref{eq:rg_A_result}, which corresponds to the most relevant operator $(q)$, is formed by the coefficients $d_\lambda$ defined in Eq.~\eqref{d_lambda}  and used in Sections~\ref{sec:susy} and~\ref{sec:scaling} of the present paper. Further, the first row, which contains the coefficients of the least relevant scaling operator $(1^q)$, has entries $(-1)^{q-l(\lambda)}d_\lambda$ in class A.

For class C, the pure-scaling operators are obtained in an analogous way, with the following results:
\begin{align}
	\begin{pmatrix}
		\mathcal{P}_{(1,1)}^C[Q]\\
		\mathcal{P}_{(2)}^C[Q]
	\end{pmatrix}
	=
	\begin{pmatrix*}[r]
		1 & -2 \\
		1 & 1
	\end{pmatrix*}
	\begin{pmatrix}
		\mathrm{tr}(\Lambda Q)
		\mathrm{tr}(\Lambda Q)\\
		\mathrm{tr}(\Lambda Q\Lambda Q)
	\end{pmatrix},
	&&
	\begin{pmatrix}
        \mathcal{P}_{(1^3)}^C[Q] \\
		\mathcal{P}_{(2,1)}^C[Q]\\
		\mathcal{P}_{(3)}^C[Q]
	\end{pmatrix}
	=
	\begin{pmatrix*}[r]
		1 & -6 & 8 \\
		1 & -1 & -2 \\
		1 & 3 & 2
	\end{pmatrix*}
	\begin{pmatrix}
		\mathrm{tr}(\Lambda Q)
		\mathrm{tr}(\Lambda Q)
		\mathrm{tr}(\Lambda Q)\\
		\mathrm{tr}(\Lambda Q\Lambda Q)
		\mathrm{tr}(\Lambda Q)\\
		\mathrm{tr}(\Lambda Q\Lambda Q\Lambda Q)
	\end{pmatrix},
	\nonumber\\
	\begin{pmatrix}
        \mathcal{P}_{(1^4)}^C[Q] \\
		\mathcal{P}_{(2,1,1)}^C[Q]\\
		\mathcal{P}_{(2,2)}^C[Q]\\
		\mathcal{P}_{(3,1)}^C[Q]\\
		\mathcal{P}_{(4)}^C[Q]
	\end{pmatrix}
	=
	\begin{pmatrix*}[r]
		1 & -12 & 12 & 32 & -48 \\
		1 & -5 & -2 & 4 & 8 \\
		1 & -2 & 7 & -8 & 2 \\
		1 & 1 & -2 & -2 & -4 \\
		1 & 6 & 3 & 8 & 6
	\end{pmatrix*}
	\begin{pmatrix}
		\mathrm{tr}(\Lambda Q)
		\mathrm{tr}(\Lambda Q)
		\mathrm{tr}(\Lambda Q)
		\mathrm{tr}(\Lambda Q)\\
		\mathrm{tr}(\Lambda Q\Lambda Q)
		\mathrm{tr}(\Lambda Q)
		\mathrm{tr}(\Lambda Q)\\
		\mathrm{tr}(\Lambda Q\Lambda Q)\mathrm{tr}(\Lambda Q\Lambda Q)\\
		\mathrm{tr}(\Lambda Q\Lambda Q\Lambda Q)
		\mathrm{tr}(\Lambda Q)\\
		\mathrm{tr}(\Lambda Q\Lambda Q\Lambda Q\Lambda Q)
	\end{pmatrix}.
	\hspace{-6cm}
	\label{eq:rg_C_result}
\end{align}
For each $q$, the last row in the matrix corresponds to the most relevant operator $(q)$ describing the averaged $q$-th moment of the LDOS. The corresponding coefficients are again given by $d_\lambda$, as in class A. For the least relevant operator $(1^q)$ (the first row), the coefficients are $(-2)^{q-l(\lambda)}d_\lambda$ in accordance with Eq.~\eqref{eq:prod-A-q} of Sec.~\ref{sec:susy}.   Note that in Eq.~\eqref{eq:rg_C_result} we have corrected sign typos that appeared in the last two columns of the $q=4$ matrix in Eq.~(251) of  Ref.~\cite{karcher2021generalized}.

Quite remarkably, the relation between class-A and class-C pure-scaling operators has a nontrivial pattern. Specifically, ordering the Young labels as in Eqs.~\eqref{eq:rg_A_result}  and~\eqref{eq:rg_C_result}, we obtain
\begin{align}
	\begin{pmatrix}
		\mathcal{P}_{(1,1)}^C[Q]\\
		\mathcal{P}_{(2)}^C[Q]
	\end{pmatrix}
	=
	\begin{pmatrix*}[r]
		\frac{3}{2} & -\frac{1}{2} \\
		0 & 1
	\end{pmatrix*}
	\begin{pmatrix}
		\mathcal{P}_{(1,1)}^A[Q]\\
		\mathcal{P}_{(2)}^A[Q]
	\end{pmatrix},
	&&
	\begin{pmatrix}
        \mathcal{P}_{(1^3)}^C[Q] \\
		\mathcal{P}_{(2,1)}^C[Q]\\
		\mathcal{P}_{(3)}^C[Q]
	\end{pmatrix}
	=
	\begin{pmatrix*}[r]
		\frac{5}{2} & -2 & \frac{1}{2} \\
		0 & \frac{4}{3} & -\frac{1}{3} \\
		0 & 0 & 1
	\end{pmatrix*}
	\begin{pmatrix}
        \mathcal{P}_{(1^3)}^A[Q] \\
		\mathcal{P}_{(2,1)}^A[Q]\\
		\mathcal{P}_{(3)}^A[Q]
	\end{pmatrix},
	\nonumber\\
	\begin{pmatrix}
        \mathcal{P}_{(1^4)}^C[Q] \\
		\mathcal{P}_{(2,1,1)}^C[Q]\\
		\mathcal{P}_{(2,2)}^C[Q]\\
		\mathcal{P}_{(3,1)}^C[Q]\\
		\mathcal{P}_{(4)}^C[Q]
	\end{pmatrix}
	=
	\begin{pmatrix*}[r]
		\frac{35}{8} & -\frac{45}{8} & -\frac{1}{2} & \frac{27}{8} & -\frac{5}{8} \\
		0 & \frac{9}{4} & -\frac{1}{2} & -1 & \frac{1}{4} \\
		0 & 0 & 2 & -1 & 0 \\
		0 & 0 & 0 & \frac{5}{4} & -\frac{1}{4} \\
		0 & 0 & 0 & 0 & 1
	\end{pmatrix*}
	\begin{pmatrix}
        \mathcal{P}_{(1^4)}^A[Q] \\ 
		\mathcal{P}_{(2,1,1)}^A[Q]\\
		\mathcal{P}_{(2,2)}^A[Q]\\
		\mathcal{P}_{(3,1)}^A[Q]\\
		\mathcal{P}_{(4)}^A[Q]
	\end{pmatrix}.
	\hspace{-6cm}
	\label{eq:rg_A_to_C}
\end{align}
This corresponds to Eq. (293) of Ref.~\cite{karcher2021generalized}, with the difference that there the operators $\mathcal{P}_\lambda^C[Q]$
were defined in such a way
that the matrices relating pure-scaling class-A and class-C operators had unit entries on the diagonal. Inverting these relations gives:
\begin{align}
	\begin{pmatrix}
		\mathcal{P}_{(1,1)}^A[Q]\\
		\mathcal{P}_{(2)}^A[Q]
	\end{pmatrix} = \left(
	\begin{array}{cc}
		\frac{2}{3} & \frac{1}{3} \\
		0 & 1 \\
	\end{array}
	\right) \begin{pmatrix}
		\mathcal{P}_{(1,1)}^C[Q]\\
		\mathcal{P}_{(2)}^C[Q]
	\end{pmatrix}; &&
	\begin{pmatrix}
        \mathcal{P}_{(1^3)}^A[Q] \\
		\mathcal{P}_{(2,1)}^A[Q]\\
		\mathcal{P}_{(3)}^A[Q]
	\end{pmatrix} = \left(
	\begin{array}{ccc}
		\frac{2}{5} & \frac{3}{5} & 0 \\
		0 & \frac{3}{4} & \frac{1}{4} \\
		0 & 0 & 1 \\
	\end{array}
	\right)\begin{pmatrix}
        \mathcal{P}_{(1^3)}^C[Q] \\
		\mathcal{P}_{(2,1)}^C[Q]\\
		\mathcal{P}_{(3)}^C[Q]
	\end{pmatrix};
	\nonumber\\
	\begin{pmatrix}
        \mathcal{P}_{(1^4)}^A[Q] \\
		\mathcal{P}_{(2,1,1)}^A[Q]\\
		\mathcal{P}_{(2,2)}^A[Q]\\
		\mathcal{P}_{(3,1)}^A[Q]\\
		\mathcal{P}_{(4)}^A[Q]
	\end{pmatrix}  = \left(
	\begin{array}{ccccc}
		\frac{8}{35} & \frac{4}{7} & \frac{1}{5} & 0 & 0 \\
		0 & \frac{4}{9} & \frac{1}{9} & \frac{4}{9} & 0 \\
		0 & 0 & \frac{1}{2} & \frac{2}{5} & \frac{1}{10} \\
		0 & 0 & 0 & \frac{4}{5} & \frac{1}{5} \\
		0 & 0 & 0 & 0 & 1 \\
	\end{array}
	\right)\begin{pmatrix}
        \mathcal{P}_{(1^4)}^C[Q] \\
		\mathcal{P}_{(2,1,1)}^C[Q]\\
		\mathcal{P}_{(2,2)}^C[Q]\\
		\mathcal{P}_{(3,1)}^C[Q]\\
		\mathcal{P}_{(4)}^C[Q]
	\end{pmatrix}.\hspace{-6cm}
	\label{eq:rg_C_to_A}
\end{align}
The matrices in Eqs.~\eqref{eq:rg_A_to_C} and ~\eqref{eq:rg_C_to_A} have an upper triangular structure. Furthermore, for $q > 2$ the matrices in Eq.~\eqref{eq:rg_C_to_A} have zeros around the upper-right corner. This implies that pure-scaling observables of class A can be used to extract some of the class-C exponents. Specifically, for $q =2$, 3, and 4, the following class-C exponents are accessible in this way: $(2), (3), (2,1), (4), (3,1)$ and $(2,2)$. This method was used in Ref.~\cite{karcher2021generalized} to obtain the corresponding class-C exponents. For completeness, the corresponding exponents are included in Table~\ref{tab:lC}, see column $x_\lambda^{\rm qn,A\uparrow}$. While the method developed in the present work which allows us to get positive pure-scaling observables of class C for all $\lambda$ (see Sec.~\ref{sec:scaling} and the numerics in Sec.~\ref{sec:sqh_numerics} presented as $x_\lambda^{\rm qn}$ in Table~\ref{tab:lC}), is clearly superior to the numerical approach of Ref.~\cite{karcher2021generalized} described above, the values $x_\lambda^{\rm qn,A\uparrow}$ are in a very good agreement with $x_\lambda^{\rm qn}$.

{ 
\section{Error bars for the generalized multifractality exponents $x_\lambda^{\rm qn}$ obtained by CCN simulations}
\label{appendix:error-bars}

We obtain statistical errors for the fitted exponents $x_\lambda^{\rm qn}$ by partitioning the $10^4$ disorder configurations in $s$ data sets $\mathcal{S}_i$ with $i\in\{1,\ldots, s\}$. For each data set $\mathcal{S}_i$, we compute the observable $\langle \mathcal{P}^C_\lambda[\psi]\rangle_i$ by averaging over the configurations in $\mathcal{S}_i$. We then perform the scaling analysis described in the main text (Sec. \ref{sec:sqh_numerics}) and obtain $s$ sets of exponents $x_{\lambda,i}^{\rm qn} (s)$. Their mean value is denoted by $x_{\lambda}^{\rm qn} (s)$, and their standard deviation divided by $\sqrt{s}$ serves as a measure for the statistical error (standard deviation of the mean) $\sigma_\lambda(s)$.

The results for $s\in\{25,16,5\}$ are shown in Tab. \ref{tab:exp}, see also a discussion in Sec. \ref{sec:sqh_numerics}. We observe that the obtained error $\sigma_\lambda(s)$ is essentially the same in the considered range of $s$, which confirms the consistency of the numerical procedure. The found $\sigma_\lambda$ thus serves as the standard deviation of $x_\lambda^{\rm qn}$ obtained by full ensemble averaging over $10^4$ configurations (no partitioning, $s=1$) and given in Tab. \ref{tab:lC} in Sec. \ref{sec:sqh_numerics}. We also note that the average values $x_\lambda^{\rm qn}(s)$ with $s = 25, 16, 5$ are all in agreement with each other and with $x_\lambda^{\rm qn}$ within $\approx 2\sigma_\lambda$.

The statistical error is the largest in the case of $x^{\rm qn}_{(5)}$, and also shows a sizeable variation with $s$. This is an indication of the fact that the requirement on the size of the statistical ensemble becomes more stringent when one studies eigenfunction moments $\langle |\psi_{i,\alpha}|^{2q} \rangle$ with high $q$. In principle, one can improve the accuracy of determination of the ``standard multifractality'' spectrum $x_{(q)}$ at high $q$ by using a considerably larger number of disorder realizations that $10^4$ used here. This is, however, outside the scope of the present work, which focusses on generalized-multifractality exponents.

\hspace*{3cm}

\begin{table}
	\begin{tabular}{cc|cccc|c}
		& & $x_\lambda^{\rm qn}(s=25)$ & $x_\lambda^{\rm qn}(s=16)$ & $x_\lambda^{\rm qn}(s=5)$ & $x_\lambda^{\rm qn}$ & $x_\lambda^{\rm perc}$\\[5pt]
		\hline\hline
		&&&&&&\\[-5pt]
	&(2) & $0.249\pm 0.001$ & $0.249\pm 0.001$ & $0.249\pm 0.001$ & $0.249$ \
	& $\frac{1}{4} = 0.25$\\[3pt]
	&(1, 1) & $1.251\pm 0.001$ & $1.251\pm 0.001$ & $1.251\pm 0.001$ & \
	$1.251$ & $\frac{5}{4} = 1.25$\\[5pt]
	&&&&&&\\[-5pt]
	\hline
	&(3) & $0.004\pm 0.004$ & $0.004\pm 0.003$ & $0.004\pm 0.004$ & $0.004$ \
	& $0$\\[3pt]
	&(2, 1) & $1.249\pm 0.002$ & $1.249\pm 0.001$ & $1.249\pm 0.002$ & \
	$1.249$ & $\frac{5}{4} = 1.25$\\[3pt]
	&
    $(1^3)$
    & $2.916\pm 0.002$ & $2.915\pm 0.002$ & $2.915\pm 0.002$ & \
	$2.915$ & $\frac{35}{12} \approx 2.917$\\[5pt]
	&&&&&&\\[-5pt]
	\hline
	&(4) & $-0.480\pm 0.016$ & $-0.483\pm 0.016$ & $-0.488\pm 0.021$ & \
	$-0.492$ & \\[3pt]
	&(3, 1) & $0.986\pm 0.007$ & $0.986\pm 0.006$ & $0.985\pm 0.007$ & \
	$0.985$ & \\[3pt]
	&(2, 2) & $1.867\pm 0.006$ & $1.866\pm 0.005$ & $1.866\pm 0.006$ & \
	$1.865$ & \\[3pt]
	&(2, 1, 1) & $2.911\pm 0.004$ & $2.911\pm 0.004$ & $2.911\pm 0.005$ & \
	$2.911$ & $\frac{35}{12} \approx 2.917$\\[3pt]
	&
    $(1^4)$
    & $5.242\pm 0.004$ & $5.242\pm 0.004$ & $5.242\pm \
	0.004$ & $5.242$ & $\frac{21}{4} = 5.25$\\[5pt]
	&&&&&&\\[-5pt]
	\hline
	&(5) & $-1.104\pm 0.038$ & $-1.118\pm 0.041$ & $-1.151\pm 0.061$ & \
	$-1.186$ & \\[3pt]
	&(4, 1) & $0.500\pm 0.023$ & $0.497\pm 0.025$ & $0.485\pm 0.025$ & \
	$0.482$ & \\[3pt]
	&(3, 2) & $1.602\pm 0.016$ & $1.598\pm 0.017$ & $1.594\pm 0.019$ & \
	$1.592$ & \\[3pt]
	&(3, 1, 1) & $2.648\pm 0.013$ & $2.645\pm 0.014$ & $2.640\pm 0.019$ & \
	$2.636$ & \\[3pt]
	&(2, 2, 1) & $3.506\pm 0.012$ & $3.502\pm 0.012$ & $3.499\pm 0.017$ & \
	$3.495$ & \\[3pt]
	&
    $(2,1^3)$
    & $5.233\pm 0.008$ & $5.233\pm 0.008$ & $5.231\pm \
	0.009$ & $5.231$ & $\frac{21}{4} = 5.25$\\[3pt]
	&
    $(1^5)$
    & $8.164\pm 0.008$ & $8.164\pm 0.008$ & $8.163\pm \
	0.007$ & $8.163$ & $\frac{33}{4} = 8.25$\\[3pt]
	\end{tabular}
\caption{Error bars for the generalized multifractality exponents $x_\lambda^{\rm qn}$ obtained by CCN simulations. To find the statistical error, the total number of $10^4$ disorder configurations is partitioned in $s$ data sets with $s = 25$, 16, and 5, see text for detail. The obtained error $\sigma_\lambda$ is nearly independent on $s$. The only case when a somewhat stronger dependence is obtained is $x_{(5)}$, indicating that requirements on statistical ensemble become more stringent for $\lambda = ( q)$ with large $q$.
The column $x_\lambda^{\rm qn}$ presents the exponents obtained by averaging over the full ensemble,  which are also given in Table \ref{tab:lC} of Sec. \ref{sec:sqh_numerics}. For convenience, we also included (the last column) exact analytical values $x_\lambda^{\rm perc}$ (when available). }
\label{tab:exp}
\end{table}

}

\twocolumngrid

\bibliography{class-C}

\end{document}